\documentclass[aps,prb,10pt,twocolumn,amsmath,amssymb,longbibliography,superscriptaddress,nofootinbib]{revtex4-2}
\usepackage{amsmath,amssymb,amsfonts} 	
\usepackage{graphicx}
\usepackage{hhline}
\usepackage[latin1]{inputenc}
\usepackage{subfigure}
\usepackage[normalem]{ulem}
\usepackage{xcolor}
\begin{document}

\title{Dynamic transverse magnetic susceptibility in the projector augmented-wave method. Application to Fe, Ni, and Co.}

\author{Thorbj\o rn Skovhus}
\affiliation{CAMD, Department of Physics, Technical University of Denmark, 2820 Kgs. Lyngby Denmark}
\author{Thomas Olsen}
\email{tolsen@fysik.dtu.dk}
\affiliation{CAMD, Department of Physics, Technical University of Denmark, 2820 Kgs. Lyngby Denmark}

\begin{abstract}
We present a first principles implementation of the dynamic transverse magnetic susceptibility in the framework of linear response time-dependent density functional theory. The dynamic susceptibility allows one to obtain the magnon dispersion as well as magnon lifetimes for a particular material, which strongly facilitates the interpretation of inelastic neutron scattering experiments as well as other spectroscopic techniques. We apply the method to Fe, Ni, and Co and perform a thorough convergence analysis with respect the basis set size, $k$-point sampling, spectral smearing and unoccupied bands.
In particular, it is shown that while the gap error (acoustic magnon energy at $\mathbf{q}=\mathbf{0}$) is highly challenging to converge, the spin-wave stiffness and the dispersion relation itself are much less sensitive to convergence parameters. Our final results agrees well with experimentally extracted magnon dispersion relations except for Ni, where it is well-known that the exchange splitting energy is poorly represented in the local density approximation. We also find good agreement with previous first principles calculations and explain how differences in the calculated dispersion relations can arise from subtle differences in computational approaches.
\end{abstract}
\maketitle

\section{Introduction}\label{sec:intro}
The dynamic transverse magnetic susceptibility is a central object of interest in the study of magnetic excitations. It is a fundamental material property giving the induced transverse magnetization in response to external perturbations such as transverse magnetic fields. In particular, the susceptibility has poles at frequencies corresponding to the magnon quasi-particle excitations of the material. 
Magnons are relevant both for theoretical development and technological applications. They have been proposed to play a role in the pairing mechanism of certain classes of high-temperature superconductors\cite{Monthoux2007,Scalapino2012} and may possibly be used as a medium for data communication and processing in future magnonics-based information technology devices\cite{Neusser2009}. Moreover, a wide range of thermodynamical properties, such as the heat capacity and Curie/N\'{e}el temperature, are directly related to the temperature dependence of the susceptibility\cite{Moriya1985}.

Experimentally, the transverse magnetic susceptibility can be directly probed by, or at least inferred from, a wide range of different spectroscopic techniques including inelastic neutron scattering (INS)\cite{VanHove1954,Jensen1991}, spin-polarized electron energy loss spectroscopy (SPEELS)\cite{Qin2015,Zakeri2014}, inelastic scanning tunneling spectroscopy (ISTS)\cite{Hirjibehedin2006,Balashov2006,Balashov2009} and resonant inelastic x-ray spectroscopy (RIXS)\cite{Brookes2020}. From the measured magnon dispersion, it is possible to extract valuable information about the underlying quantum system. The interpretation and analysis needed to accomplish this often relies on theoretical calculations - either based on models or a first principles treatment.

From a computational point of view, calculating the magnon dispersion poses a major challenge due to the many-body nature of collective magnetic excitations. 
For first principles calculations there essentially exists two different approaches for obtaining the linear dynamic susceptibility. 1) Many-body perturbation theory where the susceptibility is obtained by solving a Bethe-Salpeter equation\cite{Aryasetiawan1999,Karlsson2000,SasIoglu2010,Friedrich2014,Muller2016,Friedrich2020}. 2) Time-dependent density functional theory (TDDFT)\cite{HohenbergP.1973,Kohn1965,Runge1984}, which (although exact in principle) is limited by approximations for the applied exchange-correlation kernel. 
Both of these methods are restricted to $T=0$ and 
thermodynamical properties are currently inaccessible by direct {\it ab initio} methods. Nevertheless, the $T=0$ limit of the susceptibility provides fundamental insight into the magnetic properties of a given material and one can directly extract the magnon spectrum from it. 
In this paper, we present an implementation of the transverse magnetic susceptibility within linear response time-dependent density functional theory (LR-TDDFT)\cite{HohenbergP.1973,Kohn1965,Runge1984,Gross1985} in the projected augmented wave method (PAW)\cite{Blochl1994}. 
Applying the adiabatic local density approximation (ALDA) for the exchange-correlation kernel, we study the magnon spectrum of itinerant ferromagnets iron, nickel and cobalt. The extracted magnon dispersions agree well with experimental results, except for the case of fcc-Ni, where LDA is known to overestimate the exchange splitting energy by a factor of two\cite{SasIoglu2010}.

Through a rigorous convergence analysis, we address some of the general computational challenges in performing theoretical magnon spectroscopy on itinerant ferromagnets. We neglect spin-obit effects in our calculations, which implies the existence of a gapless acoustic magnon mode with $\omega_{\mathbf{q}=\mathbf{0}}=0$. 
The gapless mode is fundamentally protected by symmetry, but in a numerical treatment the vanishing gap is not protected against numerical inconsistencies or general numerical limitations such as truncation of basis sets or electronic bands. Through a systematic convergence analysis, we pinpoint contributions to the gap error from different computational parameters and show that the problem can be effectively overcome by applying a gap error correction procedure. This conclusion validates the common practise in literature\cite{Buczek2011b,Lounis2011,Rousseau2012,Singh2018}. 
Furthermore, we discuss the convergence of magnon modes inside the Stoner continuum, 
the transverse magnetic continuum of single-particle excitations. Overlap with the Stoner continuum gives rise to Landau damping of the collective magnon modes, which manifests itself as a broadening in the magnon lineshape. From a numerical perspective, the treatment of Landau damped magnons is particularly challenging 
as they require a good continuum description of the low frequency Stoner excitations. 
In this regard, we present an empirical convergence parameter, 
which directly allows one to extract the minimal spectral broadening required to smoothen out the low frequency Stoner excitations of a given $k$-point sampling.

The paper is organized as follows. In section \ref{sec:theoretical magnon spectroscopy}, the dynamic transverse magnetic susceptibility is formally introduced and its relation to quasi-particle excitations discussed. 
The LR-TDDFT methodology is presented and it is shown how one can compute the dynamic transverse magnetic susceptibility within the ALDA. 
In sections \ref{sec:PAW for pw susc}-\ref{sec:numerical details}, the technical details of the implementation within the PAW method are given, and in sections \ref{sec:sum rule}-\ref{sec:dispersion convergence}, the convergence analysis of the implementation is provided. The converged transverse magnetic excitation spectra of bcc-Fe, fcc-Ni, fcc-Co and hcp-Co are presented and discussed in section \ref{sec:results}. Finally, a summary and outlook is given in section \ref{sec:summary}. 
The general theoretical framework applied throughout the paper is complemented by Appendix \ref{sec:theory}, which provides a self-contained presentation of the \textit{Kubo theory} for spectroscopy in periodic crystals.

\section{Theoretical Magnon Spectroscopy}\label{sec:theoretical magnon spectroscopy}
In this section, the fundamentals of theoretical magnon spectroscopy are presented. The transverse magnetic plane wave susceptibility is introduced as the central macroscopic quantity of interest, its connection with magnon quasi-particles is discussed and it is shown how to compute it within LR-TDDFT. Finally, the Goldstone theorem and sum rules are discussed.

Throughout the main body of the paper, the Born-Oppenheimer approximation is employed and only the linear response in electronic coordinates is considered. Furthermore, zero temperature is assumed and contributions from the orbital magnetization are neglected.


\subsection{The four-component susceptibility tensor}
%

For an electronic Hamiltonian, $\hat{H}_0$, 
the magnetic response (neglecting contributions from orbital magnetization) may be described in terms of the four-component electron density operator
\begin{align}
    \hat{n}^{\mu}(\mathbf{r}) = \sum_{s, s'}\sigma^\mu_{s s'}\, \hat\psi^\dag_{s}(\mathbf{r})\hat\psi_{s'}(\mathbf{r}),
    \label{eq:n_mu}
\end{align}
%
with $\mu\in\{0, x, y, z\}$. The index $s$ indicates the spin-projection, $\uparrow$ or $\downarrow$, and $\sigma^\mu=(\sigma^0, \sigma^x, \sigma^y, \sigma^z)$ is composed of the Pauli matrices augmented by the $2\times2$ identity matrix $\sigma^0$. 
%
%
The electron density degrees of freedom are perturbed by an external (classical) electromagnetic field:
\begin{subequations}
    \begin{equation}
        \hat H_\mathrm{ext}(t)=\sum_\mu\int d\mathbf{r}\, \hat n^\mu(\mathbf{r}) W^\mu_\mathrm{ext}(\mathbf{r}, t),
        \label{eq:dH}
    \end{equation}
    \begin{align}
        \big(W_{\mathrm{ext}}^{\mu}(\mathbf{r}, t)\big)
        &= 
        \big(V_{\mathrm{ext}}^{\mu}(\mathbf{r}, t), \mathbf{W}_{\mathrm{ext}}^{\mu}(\mathbf{r}, t)\big) 
        \nonumber \\
        &= 
        \big(-e \phi_{\mathrm{ext}}(\mathbf{r}, t),\, 
        \mu_{\mathrm{B}} \mathbf{B}_{\mathrm{ext}}(\mathbf{r}, t)\big),
        \label{eq:Zeeman field}
    \end{align}
    \label{eq:electromagnetic field interaction}
\end{subequations}
where $-e$ is the electron charge, $\mu_{\mathrm{B}}$ is the Bohr magneton, while $\phi_{\mathrm{ext}}(\mathbf{r}, t)$ and $\mathbf{B}_{\mathrm{ext}}(\mathbf{r}, t)$ are the external scalar potential and magnetic field respectively. 
%
The response to the perturbation \eqref{eq:electromagnetic field interaction} may be quantified in terms of the change in four-component density, $\delta n^{\mu}(\mathbf{r}, t) = \langle \hat{n}^{\mu}(\mathbf{r}, t) \rangle - \langle \hat{n}^{\mu}(\mathbf{r}) \rangle_0$, where $\langle \cdot \rangle_0$ denotes the expectation value with respect to the unperturbed ground state (see also Eqs. \eqref{eq:linear response H} and \eqref{eq:linear response def.}). 
To linear order in the perturbing field, the induced density can be written formally as
\begin{equation}\label{eq:dn_mu}
    \delta n^{\mu}(\mathbf{r}, t) = \sum_{\nu}\int_{-\infty}^{\infty}dt'\int d\mathbf{r}' \, \chi^{\mu\nu}(\mathbf{r}, \mathbf{r}', t-t') W_{\mathrm{ext}}^{\nu}(\mathbf{r}', t').
\end{equation}
This equation defines the retarded four-component susceptibility tensor $\chi^{\mu\nu}$, which fully characterizes the linear response of the system. 
%

The susceptibility may be calculated from the \textit{Kubo formula} (Eq. \eqref{eq:Kubo formula}):
\begin{equation}
    \chi^{\mu\nu}(\mathbf{r}, \mathbf{r}', t-t') = - \frac{i}{\hbar} \theta(t-t') \langle\, [\hat{n}^{\mu}_0(\mathbf{r}, t), \hat{n}^{\nu}_0(\mathbf{r'}, t')] \,\rangle_0,
    \label{eq:four component susc tens def}
\end{equation}
%
in which the four-component density operators carry the time-dependence of the interaction picture, $\hat n_0^\mu(\mathbf{r},t) \equiv e^{i\hat H_0t/\hbar} \, \hat n^\mu(\mathbf{r}) \, e^{-i\hat H_0t/\hbar}$. 
In the frequency domain, one may express the susceptibility in terms of the system eigenstates, $\hat{H}_0 |\alpha\rangle = E_{\alpha} |\alpha\rangle$, that is within the Lehmann representation (see Eqs. \eqref{eq:lehmann} and \eqref{eq:lehmann T=0})
\begin{align}
    \chi^{\mu\nu}(\mathbf{r},\mathbf{r}',\omega) = \lim_{\eta \rightarrow 0^+} \sum_{\alpha \neq \alpha_0}\bigg( &\frac{n^\mu_{0\alpha}(\mathbf{r}) n^\nu_{\alpha0}(\mathbf{r}')}{\hbar \omega - (E_{\alpha}-E_{0}) + i \hbar \eta}
    \notag \\
    -& \frac{n^\nu_{0\alpha}(\mathbf{r}') n^\mu_{\alpha0}(\mathbf{r})}{\hbar \omega + (E_{\alpha}-E_{0}) + i \hbar \eta}\bigg).
    \label{eq:lehmann_r}
\end{align}
Here $|\alpha_0\rangle$ and $E_0$ denote the ground state and ground state energy respectively. 
Thus, the dynamic four-component susceptibility tensor is comprised of simple 
poles at excitation energies $\hbar \omega = E_{\alpha}-E_{0}$, each weighted by the transition matrix elements $n^\mu_{0\alpha}(\mathbf{r})=\langle \alpha_0|\hat n^\mu(\mathbf{r})|\alpha\rangle$ and $n^\nu_{\alpha0}(\mathbf{r}') = \langle \alpha|\hat n^\nu(\mathbf{r}')|\alpha_0\rangle$. 

In order to further illustrate the physics embedded in the four-component susceptibility tensor, a single frequency component is considered, $W^\mu_\mathrm{ext}(\mathbf{r},t)= W^\mu_\mathrm{ext}(\mathbf{r})\cos(\omega_0t)$. In this case, the real and imaginary parts of the dynamic susceptibility determine the in- and out-of-phase response respectively (see Eq. \eqref{eq:in and out of phase response}):
\begin{align}
    \delta n^{\mu}(\mathbf{r}, t) = &\sum_{\nu} \int d\mathbf{r}'\, \Big[ \mathrm{Re}\left\{ \chi^{\mu\nu}(\mathbf{r}, \mathbf{r}', \omega_0) \right\} \cos(\omega_0 t)
    \nonumber \\
    &+ \mathrm{Im}\left\{ \chi^{\mu\nu}(\mathbf{r}, \mathbf{r}', \omega_0) \right\} \sin(\omega_0 t) \Big] W_{\mathrm{ext}}^{\nu}(\mathbf{r}').
    \label{eq:four-comp. harmonic response}
\end{align}
Here it was used that the four-component density operator is Hermitian, $\hat{n}^{\mu}(\mathbf{r})^{\dagger}=\hat{n}^{\mu}(\mathbf{r})$, such that $\chi^{\mu\nu}(\mathbf{r},\mathbf{r}',-\omega)=\chi^{\mu\nu *}(\mathbf{r},\mathbf{r}',\omega)$ (see Eq. \eqref{eq:relation daggers cc}). The rate of energy absorption into the system under the perturbation \eqref{eq:electromagnetic field interaction} is given by $Q=d\langle \hat{H}\rangle/dt$ and from \eqref{eq:four-comp. harmonic response} it then follows, that only the out-of-phase response contributes to the energy dissipation on average (see Eq. \eqref{eq:mean energy dissipation}):
%
\begin{align}
    \bar{Q}=-\frac{\omega_0}{2}\sum_{\mu, \nu}
    \iint &d\mathbf{r}d\mathbf{r}'\, W^{\mu}_\mathrm{ext}(\mathbf{r}) 
    \nonumber \\
    &\times \mathrm{Im}\left\{\chi^{\mu\nu}(\mathbf{r},\mathbf{r}',\omega_0)\right\}  W^{\nu}_\mathrm{ext}(\mathbf{r}').
    \label{eq:four-component energy dissipation Imchi}
\end{align}
Now, instead of using $\chi^{\mu\nu}(\mathbf{r},\mathbf{r}',-\omega)=\chi^{\mu\nu *}(\mathbf{r},\mathbf{r}',\omega)$ to express the mean rate of energy absorption 
in terms of the imaginary part of $\chi^{\mu\nu}$, 
one may instead interchange summation and integration variables, such that it becomes expressed in terms of the dissipative (anti-symmetric) part instead. 
%
%
This is advantageous, 
because the dissipative part of $\chi^{\mu\nu}$ (defined in Eq. \eqref{eq:dissipative part}) is proportional to the spectral function of induced excitations (see Eqs. \eqref{eq:connection to spectral functions} and \eqref{eq:spectral function def.})
\begin{subequations}
    \begin{align}
        S^{\mu\nu}(\mathbf{r}, \mathbf{r}', \omega) 
        &\equiv 
        - \frac{1}{2\pi i}\left\{\chi^{\mu\nu}(\mathbf{r},\mathbf{r}',\omega) - \chi^{\nu\mu}(\mathbf{r}',\mathbf{r}, -\omega)\right\}
        \\
        &= 
        A^{\mu\nu}(\mathbf{r}, \mathbf{r}', \omega) - A^{\nu\mu}(\mathbf{r}', \mathbf{r}, -\omega),
    \end{align}
    \label{eq:spectrum of density excitations}
\end{subequations}
where
\begin{equation}
    A^{\mu\nu}(\mathbf{r}, \mathbf{r}', \omega) \equiv \sum_{\alpha\neq\alpha_0} n^\mu_{0\alpha}(\mathbf{r}) n^\nu_{\alpha0}(\mathbf{r}') \, \delta\big(\hbar \omega - (E_{\alpha} - E_{0})\big).
    \label{eq:single frequency density spectral functions}
\end{equation}
Using these definitions,
\begin{align}
    \bar{Q}=\frac{\pi \omega_0}{2}\sum_{\mu, \nu}
    \iint d\mathbf{r}d\mathbf{r}'\, &W^{\mu}_\mathrm{ext}(\mathbf{r})
    S^{\mu\nu}(\mathbf{r},\mathbf{r}', \omega_0)  W^{\nu}_\mathrm{ext}(\mathbf{r}').
    \label{eq:energy dissipation density response chi''}
\end{align}
In this way, Eqs. \eqref{eq:spectrum of density excitations}, \eqref{eq:single frequency density spectral functions} and \eqref{eq:energy dissipation density response chi''} comprise the linear response formulation of the fact, that energy dissipation is directly governed by the spectrum of induced excitations. This also illustrates the direct connection to Fermi's golden rule. 

\subsection{The four-component susceptibility tensor in circular coordinates}\label{sec:circular coordinates}
%
In a collinear description, magnons are collective quasi-particles carrying a unit of spin angular momentum. With the ground state magnetization aligned along the $z$-axis ($\mathbf{m}(\mathbf{r}) = \langle \hat{n}^z(\mathbf{r}) \rangle_0 \, \mathbf{e}_z$), they are generated by the 
spin-raising and spin-lowering operators,
\begin{align}
    &\hat{n}^+(\mathbf{r}) = \frac{1}{2}\Big(\hat{n}^x(\mathbf{r})+ i\hat{n}^y(\mathbf{r}\Big)=\hat{\psi}^{\dagger}_{\uparrow}(\mathbf{r}) \hat{\psi}_{\downarrow}(\mathbf{r}),\label{eq:n+}\\
    &\hat{n}^-(\mathbf{r}) = \frac{1}{2}\Big(\hat{n}^x(\mathbf{r})- i\hat{n}^y(\mathbf{r})\Big)=\hat{\psi}^{\dagger}_{\downarrow}(\mathbf{r})         \hat{\psi}_{\uparrow}(\mathbf{r}),\label{eq:n-}
\end{align}
which flip the spin of a spin-down and a spin-up electron at position $\mathbf{r}$ respectively. 
In terms of the external electromagnetic field, spin-raising and spin-lowering excitations are induced by the circular components
\begin{equation}
    W_{\mathrm{ext}}^{\pm}(\mathbf{r}, t) = W_{\mathrm{ext}}^x(\mathbf{r}, t) \pm i\, W_{\mathrm{ext}}^y(\mathbf{r}, t),
\end{equation}
such that the perturbation from Eq. \eqref{eq:electromagnetic field interaction} can be written
\begin{align}
    \hat{H}_{\mathrm{ext}}(t) &= \int d\mathbf{r} \, \Big[\hat{n}(\mathbf{r}) V_{\mathrm{ext}}(\mathbf{r}, t) + \hat{n}^{+}(\mathbf{r}) W_{\mathrm{ext}}^{-}(\mathbf{r}, t) 
    \nonumber \\
    &\hspace{43pt}+ \hat{n}^{-}(\mathbf{r}) W_{\mathrm{ext}}^{+}(\mathbf{r}, t) + \hat{\sigma}^{z}(\mathbf{r}) W_{\mathrm{ext}}^{z}(\mathbf{r}, t) \Big]
    \nonumber \\
    &= \int d\mathbf{r} \, \sum_j \hat{n}^j(\mathbf{r}) \Breve{W}_{\mathrm{ext}}^j(\mathbf{r}, t),
\end{align}
where $j\in\{0,+,-,z\}$ and the breve accent is introduced to reverse the ordering of $+$ and $-$ components $\big(\Breve{W}^{j}\big) = \big( V, W^{-}, W^{+}, W^{z} \big)$. 
Using the relations \eqref{eq:n+} and \eqref{eq:n-}, one may also write the four-component susceptibility tensor in circular coordinates, where $\chi^{jk}$ is given by the \textit{Kubo formula} of Eq. \eqref{eq:four component susc tens def}. For example, one obtains
\begin{align}
    \chi^{x0} &= \chi^{+0} + \chi^{-0},
    \nonumber \\
    \chi^{xx} &= \chi^{++} + \chi^{+-} + \chi^{-+} + \chi^{--},
    \nonumber \\
    \chi^{xy} &= -i\chi^{++} + i\chi^{+-} - i\chi^{-+} + i\chi^{--},
    \nonumber \\
    \chi^{xz} &= \chi^{+z} + \chi^{-z},
\end{align}
where the spatial and temporal arguments have been suppressed. Rewriting Eq. \eqref{eq:dn_mu} in this manner yields the response relation in circular coordinates:
\begin{equation}
    \delta n^{j}(\mathbf{r}, t) = \sum_{k}\int_{-\infty}^{\infty}dt'\int d\mathbf{r}' \, \chi^{jk}(\mathbf{r}, \mathbf{r}', t-t') \Breve{W}_{\mathrm{ext}}^{k}(\mathbf{r}', t').
    \label{eq:four-component response relation circ. coord.}
\end{equation}
It should be noted that the circular components satisfy $\chi^{-+}(\mathbf{r},\mathbf{r}',-\omega)=\chi^{+-*}(\mathbf{r},\mathbf{r}',\omega)$.

If the system is collinear, such that the total electronic spin projection in the $z$-direction, $S_z$, can be taken as a good quantum number, the products of transition matrix elements
\begin{equation}
    n^{j}_{0\alpha}(\mathbf{r}) n^{k}_{\alpha0}(\mathbf{r}') = \langle0| \hat{n}^j(\mathbf{r}) |\alpha\rangle \langle \alpha| \hat{n}^k(\mathbf{r}') |0\rangle,
\end{equation}
vanish if $\hat{n}^j(\mathbf{r})\hat{n}^k(\mathbf{r}')$ results in a net change of $S_z$. 
Consequently, several of the components vanish from the Lehmann representation \eqref{eq:lehmann_r} for $\chi^{jk}$, and the tensor becomes block diagonal:
\begin{equation}
    \chi^{[0, +, -, z]} = 
    \begin{pmatrix}
        \chi^{00} & 0 & 0 & \chi^{0z} \\
        0 & 0 & \chi^{+-} & 0 \\
        0 & \chi^{-+} & 0 & 0 \\
        \chi^{z0} & 0 & 0 & \chi^{zz}
    \end{pmatrix},
    \label{eq:four-component susc. tensor sparsity by collinearity circ. coord.}
\end{equation}
\begin{equation}
    \chi^{[0, x, y, z]} = 
    \begin{pmatrix}
        \chi^{00} & 0 & 0 & \chi^{0z} \\
        0 & \chi^{+-} + \chi^{-+} & i\chi^{+-} - i\chi^{-+} & 0 \\
        0 & -i\chi^{+-} + i\chi^{-+} & \chi^{+-} + \chi^{-+} & 0 \\
        \chi^{z0} & 0 & 0 & \chi^{zz}
    \end{pmatrix}.
    \label{eq:four-component susc. tensor sparsity by collineariy}
\end{equation}
Thus, in the collinear case, the transverse magnetic response is completely decoupled from the longitudinal magnetic response, given by $\chi^{zz}$, and the longitudinal dielectric response, given by $\chi^{00}$:
\begin{subequations}
    \begin{equation}
        \delta n^+(\mathbf{r}, t) = \int_{-\infty}^{\infty} dt' \int d\mathbf{r}' \chi^{+-}(\mathbf{r}, \mathbf{r}', t-t') W^+_{\mathrm{ext}}(\mathbf{r}', t'),
    \end{equation}
    \begin{equation}
        \delta n^-(\mathbf{r}, t) = \int_{-\infty}^{\infty} dt' \int d\mathbf{r}' \chi^{-+}(\mathbf{r}, \mathbf{r}', t-t') W^-_{\mathrm{ext}}(\mathbf{r}', t').\label{eq:dyson_n-}
    \end{equation}
\end{subequations}

For a spin-paired (non-magnetic) collinear ground state, spin-rotational symmetry implies that $\chi^{xx}=\chi^{yy}=\chi^{zz}$, but also that $\chi^{xy}=\chi^{zx}$, $\chi^{z0}=\chi^{x0}$ and $\chi^{0z}=\chi^{0x}$ where all the latter terms vanish as argued in Eq. \eqref{eq:four-component susc. tensor sparsity by collineariy}.
Thus, the magnetic response is fully characterized by $\chi^{zz}$ for non-magnetic systems\cite{Wysocki2016}.

\subsection{The spectrum of transverse magnetic excitations}
In periodic crystals, the linear response of a material may be characterized by the four-component plane wave susceptibility, which is defined in terms of the lattice Fourier transform (see Eq. \eqref{eq:plane wave dyn susc. def}):
    \begin{align}
        \chi^{\mu\nu}_{\mathbf{G}\mathbf{G}'}(\mathbf{q},\omega) 
        \equiv 
        &\iint \frac{d\mathbf{r} d\mathbf{r}'}{\Omega} e^{-i(\mathbf{G} + \mathbf{q}) \cdot \mathbf{r}} \chi^{\mu\nu}(\mathbf{r},\mathbf{r}',\omega) e^{i(\mathbf{G}' + \mathbf{q}) \cdot \mathbf{r}'}\notag
        \\
        = 
        &\lim_{\eta \rightarrow 0^+} \frac{1}{\Omega}\sum_{\alpha \neq \alpha_0} \bigg[\frac{n^\mu_{0\alpha}(\mathbf{G}+\mathbf{q}) n^\nu_{\alpha0}(-\mathbf{G}'-\mathbf{q})}{\hbar \omega - (E_\alpha-E_0) + i \hbar \eta}\notag\\
        &\qquad-\frac{n^\nu_{0\alpha}(-\mathbf{G}'-\mathbf{q}) n^\mu_{\alpha0}(\mathbf{G}+\mathbf{q})}{\hbar \omega + (E_\alpha-E_0) + i \hbar \eta}\bigg].
        \label{eq:four-comp. plane wave susc. def.}
    \end{align}
%
Here $\Omega$ is the crystal volume, $\mathbf{G}$ is a reciprocal lattice vector and $\mathbf{q}$ is a wave vector within the first Brillouin zone. The reciprocal space pair densities $n^{\mu}_{\alpha\alpha'}(\mathbf{G}+\mathbf{q})$ are Fourier transforms of the spatial pair densities, see Eqs. \eqref{eq:operator fourier basis}-\eqref{eq:periodic transition matrix elements}. The plane wave susceptibility gives the linear order plane wave response $e^{i([\mathbf{G}+\mathbf{q}]\cdot \mathbf{r} - \omega t)}$ in density component $\mu$ to a plane wave perturbation $e^{i([\mathbf{G}'+\mathbf{q}]\cdot \mathbf{r} - \omega t)}$ in external field component $\nu$ (see Eq. \eqref{eq:plane wave real space response I}). The plane wave response is diagonal in reduced wave vector $\mathbf{q}$ due to the periodicity of the crystal (see Eq. \eqref{eq:chi diagonal spatial FT}).

In analogy with the real space response in Eqs. \eqref{eq:spectrum of density excitations}, \eqref{eq:single frequency density spectral functions} and \eqref{eq:energy dissipation density response chi''}, the energy dissipation in periodic crystals is governed by the dissipative part of $\chi^{\mu\nu}_{\mathbf{G}\mathbf{G}'}(\mathbf{q},\omega)$, that is, the plane wave spectrum of induced excitations \eqref{eq:plane wave spectral function of induced transitions}
\begin{subequations}
    \begin{align}
        S^{\mu\nu}_{\mathbf{G}\mathbf{G}'}(\mathbf{q}, \omega) 
        &= - \frac{1}{2 \pi i} \left\{\chi^{\mu\nu}_{\mathbf{G}\mathbf{G}'}(\mathbf{q}, \omega)
        - \chi^{\nu\mu}_{-\mathbf{G}'-\mathbf{G}}(-\mathbf{q}, -\omega) \right\}
        \\
        &= A^{\mu\nu}_{\mathbf{G}\mathbf{G}'}(\mathbf{q}, \omega) - A^{\nu\mu}_{-\mathbf{G}'-\mathbf{G}}(-\mathbf{q}, -\omega),
    \end{align}
    \label{eq:four-comp. plane wave spectral function}
\end{subequations}
where
\begin{align}
    A^{\mu\nu}_{\mathbf{G}\mathbf{G}'}(\mathbf{q}, \omega) 
    &\equiv \frac{1}{\Omega}\sum_{\alpha \neq \alpha_0} n^\mu_{0\alpha}(\mathbf{G}+\mathbf{q}) n^\nu_{\alpha0}(-\mathbf{G}'-\mathbf{q})
    \nonumber \\
    &\hspace{45pt}\times \delta\big(\hbar \omega - (E_\alpha - E_0)\big). 
    \label{eq:four-component A spectral function}
\end{align}
%
For the reciprocal space pair densities $n^\mu_{0\alpha}(\mathbf{G}+\mathbf{q})$ to be non-zero, it is necessary that $\mathbf{q}_{\alpha 0} = \mathbf{q}$ (see Eq. \eqref{eq:bloch wave transition matrix elements}). Thus, only excited states with a difference in crystal momentum $\hbar \mathbf{q}$ with respect to the ground state have finite weight in the spectral function \eqref{eq:four-component A spectral function}.

Eqs. \eqref{eq:four-comp. plane wave susc. def.},  \eqref{eq:four-comp. plane wave spectral function} and \eqref{eq:four-component A spectral function} also apply to the susceptibility tensor in circular coordinates, $\chi^{jk}$. Because the spin-flip densities $\hat{n}^+(\mathbf{r})$ and $\hat{n}^-(\mathbf{r})$ are hermitian conjugates, it follows that $\chi^{+-*}_{\mathbf{G}\mathbf{G}'}(\mathbf{q},\omega)=\chi^{-+}_{-\mathbf{G}-\mathbf{G}'}(-\mathbf{q},-\omega)$ and consequently, the dissipative parts of $\chi^{\pm\mp}$ are also the imaginary parts along the diagonal:
\begin{subequations}
    \begin{align}
        S^{+-}_{\mathbf{G}}(\mathbf{q}, \omega) 
        &\equiv 
        S^{+-}_{\mathbf{G}\mathbf{G}}(\mathbf{q}, \omega) = - \frac{1}{\pi} \mathrm{Im}\left\{ \chi^{+-}_{\mathbf{G}\mathbf{G}}(\mathbf{q}, \omega) \right\} 
        \nonumber \\
        &= A^{-}_{\mathbf{G}}(\mathbf{q}, \omega) - A^{+}_{-\mathbf{G}}(-\mathbf{q}, -\omega),
        \\
        S^{-+}_{\mathbf{G}}(\mathbf{q}, \omega) 
        &= 
        A^{+}_{\mathbf{G}}(\mathbf{q}, \omega) - A^{-}_{-\mathbf{G}}(-\mathbf{q}, -\omega),
    \end{align}
    \label{eq:transverse fluctuation spectrum def.}
\end{subequations}
where the short-hand notation $A^{\mp}_{\mathbf{G}}(\mathbf{q}, \omega) \equiv A^{\pm\mp}_{\mathbf{G}\mathbf{G}}(\mathbf{q}, \omega)$ has been introduced. 
From Eq. \eqref{eq:four-component A spectral function} it is clear, that $A^+_{\mathbf{G}}(\mathbf{q}, \omega)$ and $A^-_{\mathbf{G}}(\mathbf{q}, \omega)$ are the spectral functions for spin-raising and spin-lowering magnetic excitations respectively. These excitations may be associated with quasi-particles of energy $\hbar \omega$, crystal momentum $\hbar \mathbf{q}$ and spin projections $\pm \hbar$. 
Depending on the character of the excitations, the quasi-particles are either identified as collective magnon quasi-particles, as single-particle electron-hole (Stoner) pairs or something in between. 
Thus, for a ferromagnetic material assumed magnetized along the $z$-direction, one may read off the full spectrum of magnon excitations from the spectral function $S^{+-}_{\mathbf{G}}(\mathbf{q}, \omega)$, with majority-to-minority magnons at positive frequencies and minority-to-majority magnons at negative frequencies.

Finally, the transverse magnetic excitation spectrum does not depend on the reduced wave vector $\mathbf{q}$ only, but also on the reciprocal lattice vector $\mathbf{G}$. The spin-flip pair densities in Eq. \eqref{eq:four-component A spectral function} represent the local field components of the change in spin-orientation from the ground state to the excited state in question. Therefore, different excited states may be visible for different choices of $\mathbf{G}$.
As an example, the macroscopic (unit-cell averaged) $\mathbf{G}=\mathbf{0}$ component represents a dynamic change to the magnetization, where the spin-orientation at different magnetic atomic sites is precessing according to a long-range phase factor of $e^{i\mathbf{q}\cdot\mathbf{r}}$. This corresponds to an acoustic magnon mode, which will dominate the spectrum at small $\mathbf{q}$ and $\omega$. Excited states where different magnetic atoms inside the unit cell precess with opposite phases will not be present in the macroscopic transverse magnetic excitation spectrum $S^{+-}(\mathbf{q}, \omega)$, but in the local field components $\mathbf{G}\neq\mathbf{0}$ that match the spin structure of the given excited state.

\subsection{Linear response time-dependent density functional theory}\label{sec:lr-tddft}
%
It is, in general, a prohibitively demanding task to diagonalize the many-body Hamiltonian $\hat{H}_0$ in order to find the eigenstates entering the susceptibility. However, within the framework of time-dependent density functional theory (TDDFT), it is possible to compute $\chi^{\mu\nu}$ without accessing the many-body eigenstates. 
In particular, it follows from the Runge-Gross theorem\cite{Runge1984} that the time-dependent spin-density can be represented by an auxiliary non-interacting Kohn-Sham system defined by the Hamiltonian
\begin{equation}\label{eq:H_KS}
        \hat{H}_{\mathrm{KS}}(t) = \hat T + \hat V_\mathrm{nuc} + \hat V_\mathrm{Hxc}[n^\mu](t)+\hat H_\mathrm{ext}(t),
\end{equation}
where $\hat H_\mathrm{ext}(t)$ is given by Eq. \eqref{eq:dH} and 
\begin{equation}
        \hat{V}_{\mathrm{Hxc}}[n^\mu](t) = \sum_\mu\int d\mathbf{r}\, \hat n^\mu(\mathbf{r})W^\mu_\mathrm{Hxc}[n^\mu](\mathbf{r},t).
\end{equation}
Here $W^{\mu}_{\mathrm{Hxc}}(\mathbf{r}, t)$ is the four-component time-dependent Hartree-exchange-correlation potential required to reproduce the time-dependent density of the interacting system. It is a functional of the four-component time-dependent density and is typically treated in the adiabatic approximation, where it is evaluated from a given approximation to the static exchange-correlation potential of the electron density at time $t$.

In the Kohn-Sham system, the induced density resulting from a small external perturbation $\delta W^\mu_\mathrm{ext}(\mathbf{r},t)$ can be written as
\begin{equation}
    \delta n^{\mu}(\mathbf{r}, t) = \sum_{\nu}\int_{-\infty}^{\infty}dt'\int d\mathbf{r}' \, \chi_{\mathrm{KS}}^{\mu\nu}(\mathbf{r}, \mathbf{r}', t-t') \delta W_{\mathrm{s}}^{\nu}(\mathbf{r}', t'),
    \label{eq:kohn-sham four-component response relation}
\end{equation}
where $\chi_\mathrm{KS}^{\mu\nu}$
is the non-interacting Kohn-Sham susceptibility and $\delta W_{\mathrm{s}}^\mu=\delta W^\mu_\mathrm{ext} + \delta W^\mu_\mathrm{Hxc}$. Comparing with the response relation \eqref{eq:dn_mu} and using that the induced change in Hartree-exchange-correlation potential $\delta W^\mu_\mathrm{Hxc}$ is a functional of the induced density, one may derive the Dyson equation\cite{Gross1985}:
\begin{align}
    \chi^{\mu\nu}(\mathbf{r}, \mathbf{r}', \omega) = \chi^{\mu\nu}_{\mathrm{KS}}(\mathbf{r}, \mathbf{r}', \omega) + \sum_{\tau_1, \tau_2} &\iint d\mathbf{r}_1 d\mathbf{r}_2 \, 
    \nonumber \\
    \times \chi^{\mu\tau_1}_{\mathrm{KS}}(\mathbf{r}, \mathbf{r}_1, \omega) K_{\mathrm{Hxc}}^{\tau_1\tau_2}(\mathbf{r}_1, \mathbf{r}_2, \omega) &\chi^{\tau_2\nu}(\mathbf{r}_2, \mathbf{r}', \omega),
    \label{eq:four-comp. Dyson real space}
\end{align}
where
\begin{align}
    K_{\mathrm{Hxc}}^{\tau_1\tau_2}(\mathbf{r}_1, \mathbf{r}_2, t_1 - t_2) = \frac{\delta W^{\tau_1}_{\mathrm{Hxc}}(\mathbf{r}_1, t_1)}{\delta n^{\tau_2}(\mathbf{r}_2, t_2)}.
    \label{eq:Hxc kernel def.}
\end{align}
By inverting the Dyson equation \eqref{eq:four-comp. Dyson real space}, the full four-component susceptibility tensor may be computed from the Kohn-Sham susceptibility, which may be obtained directly from  quantities that can be extracted from a routine ground state DFT calculation\cite{Kohn1965,Barth1972, HohenbergP.1973,Rajagopal1973}. The main difficulty then resigns in finding a good approximation for the Hartree-exchange-correlation kernel \eqref{eq:Hxc kernel def.}. Below, the functional form for the transverse components of $K_{\mathrm{Hxc}}^{\tau_1\tau_2}$ is provided within the adiabatic local density approximation for collinear systems. 

\subsection{The Kohn-Sham four-component susceptibility tensor}

In the absence of an external time-dependent electromagnetic field, the (four-component) ground state density can be obtained from the auxiliary Kohn-Sham system, whereupon the Kohn-Sham Hamiltonian \eqref{eq:H_KS} may be diagonalized. With access to the Kohn-Sham eigenstates, the Kohn-Sham susceptibility may be easily evaluated using the Lehmann representation \eqref{eq:lehmann_r}. 
For periodic crystals, the Kohn-Sham eigenstates are Slater determinants composed of Bloch wave spinors 
$\psi_{n\mathbf{k}}(\mathbf{r}) 
= \big(\psi_{n\mathbf{k}\uparrow}(\mathbf{r}), \psi_{n\mathbf{k}\downarrow}(\mathbf{r})\big) / \sqrt{N_k}
$ where $n$ and $\mathbf{k}$ denotes the band index and $k$-point, while the Kohn-Sham orbitals have been normalized to the unit cell by dividing with the square root of the number of $k$-points $N_k$ (number of unit cells in the crystal). By expanding the field operators in Eq. \eqref{eq:n_mu} in terms of the Bloch wave spinors, the four-component density operator may be written in terms of the Kohn-Sham orbitals:
%
%
\begin{equation}
    \hat{n}^{\mu}(\mathbf{r}) = \sum_{s, s'} \sigma^{\mu}_{ss'} \frac{1}{N_k}\sum_{n \mathbf{k}} \sum_{m \mathbf{k}'} \psi_{n\mathbf{k}s}^*(\mathbf{r}) \psi_{m\mathbf{k}'s'}(\mathbf{r}) \hat{c}^{\dagger}_{n \mathbf{k}} \hat{c}_{m \mathbf{k}'}.
\end{equation}
Thus, in the Kohn-Sham system, the four-component density operator simply moves an electron from one spinorial orbital to another. As a consequence, the Kohn-Sham susceptibility is easily evaluated in the Lehmann representation \eqref{eq:lehmann_r}, which only involves states where a single electron from an occupied orbital has been moved to an unoccupied one. Denoting the Kohn-Sham single-particle energies $\epsilon_{n\mathbf{k}}$ and ground state occupancies $f_{n\mathbf{k}}$, one may write the Kohn-Sham four-component susceptibility tensor as
\begin{align}
    \chi^{\mu\nu}_{\mathrm{KS}}(\mathbf{r}, \mathbf{r}', \omega) = \lim_{\eta \rightarrow 0^+} &\frac{1}{N_k^2}\sum_{n\mathbf{k}} \sum_{m \mathbf{k}'} (f_{n\mathbf{k}} - f_{m\mathbf{k}'})
    \nonumber \\
    &\times \frac{n^{\mu}_{n\mathbf{k}, m\mathbf{k}'}(\mathbf{r}) \,  n^{\nu}_{m\mathbf{k}', n\mathbf{k}}(\mathbf{r}')}{\hbar \omega - (\epsilon_{m\mathbf{k}'}-\epsilon_{n\mathbf{k}}) + i \hbar \eta},
    \label{eq:four component KS lehmann}
\end{align}
where the Kohn-Sham four-component pair densities are given by
\begin{align}
    n^{\mu}_{n\mathbf{k}, m\mathbf{k}'}(\mathbf{r}) 
    &=\sum_{s, s'} \sigma^{\mu}_{ss'} \psi_{n\mathbf{k}s}^*(\mathbf{r}) \psi_{m\mathbf{k}'s'}(\mathbf{r}).
    \label{eq:KS four-comp pair dens. def.}
\end{align}

Since $\chi^{\mu\nu}(\mathbf{r}, \mathbf{r}', \omega)$, $\chi^{\mu\nu}_{\mathrm{KS}}(\mathbf{r}, \mathbf{r}', \omega)$ and $K_{\mathrm{Hxc}}^{\mu\nu}(\mathbf{r}, \mathbf{r}', \omega)$ are periodic functions under simultaneous translations of $\mathbf{r}$ and $\mathbf{r}'$ (see Eq. \eqref{eq:periodic susceptibilities}), the Dyson equation \eqref{eq:four-comp. Dyson real space} can be Fourier transformed to reciprocal space, yielding a matrix equation which is diagonal in crystal momentum $\hbar\mathbf{q}$ as well as in energy $\hbar\omega$:
\begin{align}
    &\chi^{\mu\nu}_{\mathbf{G}\mathbf{G}'}(\mathbf{q}, \omega) 
    = \chi^{\mu\nu}_{\mathrm{KS},\mathbf{G}\mathbf{G}'}(\mathbf{q}, \omega) + \sum_{\tau_1, \tau_2} \sum_{\mathbf{G}_1, \mathbf{G}_2}
    \nonumber \\
    &\hspace{16pt} \times \chi^{\mu\tau_1}_{\mathrm{KS}, \mathbf{G}\mathbf{G}_1}\hspace{-1pt}(\mathbf{q}, \omega) K_{\mathrm{Hxc}, \mathbf{G}_1\hspace{-1pt}\mathbf{G}_2}^{\tau_1\tau_2}\hspace{-1pt}(\mathbf{q}, \omega) \chi^{\tau_2\nu}_{\mathbf{G}_2\mathbf{G}'}(\mathbf{q}, \omega).
    \label{eq:dyson eq. reciprocal space}
\end{align}
As a matrix equation, Eq. \eqref{eq:dyson eq. reciprocal space} is straight-forward to invert in order to obtain the many-body susceptibility, $\chi^{\mu\nu}_{\mathbf{G}\mathbf{G}'}(\mathbf{q}, \omega)$, from its Kohn-Sham analogue. From Eq. \eqref{eq:four component KS lehmann}, the Kohn-Sham susceptibility is lattice Fourier transformed, yielding
%
\begin{align}
    \chi^{\mu\nu}_{\mathrm{KS}, \mathbf{G}\mathbf{G}'}(\mathbf{q}, \omega) = &\lim_{\eta \rightarrow 0^+} \frac{1}{\Omega} \sum_{\mathbf{k}} \sum_{n, m} (f_{n\mathbf{k}} - f_{m\mathbf{k}+\mathbf{q}})
    \nonumber \\
    \times &\frac{n^\mu_{n\mathbf{k}, m\mathbf{k}+\mathbf{q}}(\mathbf{G} + \mathbf{q}) \,  n^\nu_{m\mathbf{k}+\mathbf{q}, n\mathbf{k}}(-\mathbf{G}'-\mathbf{q})}{\hbar \omega - (\epsilon_{m\mathbf{k}+\mathbf{q}}-\epsilon_{n\mathbf{k}}) + i \hbar \eta},
    \label{eq:KS plane wave susc. tensor non-collinear}
\end{align}
where
%
\begin{align}
    n^\mu_{n\mathbf{k}, m\mathbf{k}+\mathbf{q}}(\mathbf{G} + \mathbf{q}) =  \int_{\Omega_{\mathrm{cell}}} &d\mathbf{r} \, e^{-i (\mathbf{G} + \mathbf{q}) \cdot \mathbf{r}}\, n^\mu_{n\mathbf{k}, m\mathbf{k}+\mathbf{q}}(\mathbf{r})
    \label{eq:KS four-component reciprocal space pair density}
\end{align}
gives the plane wave coefficients of the Kohn-Sham four-component pair density and $\Omega_\mathrm{cell}$ is the unit cell volume. In the above, $\epsilon_{m\mathbf{k}+\mathbf{q}}$, $f_{m\mathbf{k}+\mathbf{q}}$ and $\psi_{m\mathbf{k}+\mathbf{q}}(\mathbf{r})$ are used to denote the eigenvalue, occupancy and single-particle spinorial wave functions corresponding to the Kohn-Sham orbital with a wave vector $\mathbf{k}'$ within the first Brillouin Zone, satisfying $\mathbf{k}' = \mathbf{k} + \mathbf{q}$ up to a reciprocal lattice vector. The plane wave Hartree-exchange-correlation kernel is simply computed by lattice Fourier transforming Eq. \eqref{eq:Hxc kernel def.}.


For collinear systems, Eqs. \eqref{eq:four-component susc. tensor sparsity by collinearity circ. coord.} and \eqref{eq:four-component susc. tensor sparsity by collineariy} also apply to the Kohn-Sham susceptibility tensor. Furthermore, the spinorial orbitals can all be chosen to have one non-zero component, such that the spin-polarization may be included in the band index $n\rightarrow(n s)$. This lead to a simplification of the Kohn-Sham plane wave susceptibility:
\begin{align}
    \chi^{\mu\nu}_{\mathrm{KS}, \mathbf{G}\mathbf{G}'}(\mathbf{q}, \omega) = &\lim_{\eta \rightarrow 0^+} \frac{1}{\Omega} \sum_{n \mathbf{k}s} \sum_{m s'} (f_{n\mathbf{k}s} - f_{m\mathbf{k}+\mathbf{q}s'}) \sigma^\mu_{ss'} \sigma^\nu_{s's}
    \nonumber \\
    &\hspace{-29pt}\times  \frac{n_{n\mathbf{k}s, m\mathbf{k}+\mathbf{q}s'}(\mathbf{G}+\mathbf{q}) \,  n_{m\mathbf{k}+\mathbf{q}s', n\mathbf{k}s}(-\mathbf{G}'-\mathbf{q})}{\hbar \omega - (\epsilon_{m\mathbf{k}+\mathbf{q}s'}-\epsilon_{n\mathbf{k}s}) + i \hbar \eta},
    \label{eq:dyn. plane wave KS susc. tensor lehmann}
\end{align}
where
\begin{align}
    n_{n\mathbf{k}s, m\mathbf{k}+\mathbf{q}s'}(\mathbf{G} + \mathbf{q}) =  \int_{\Omega_{\mathrm{cell}}} &d\mathbf{r} \, e^{-i (\mathbf{G} + \mathbf{q}) \cdot \mathbf{r}} \nonumber \\
    \times &\psi_{n\mathbf{k}s}^*(\mathbf{r}) \psi_{m\mathbf{k}+\mathbf{q}s'}(\mathbf{r}).
    \label{eq:KS reciprocal space pair densities}
\end{align}
Writing the product of spin matrix elements $\sigma^{\mu}_{s s'}\sigma^{\nu}_{s' s}$ of Eq. \eqref{eq:dyn. plane wave KS susc. tensor lehmann} in terms of the basic matrices
\begin{subequations}
    \begin{align}
        \sigma^{\uparrow} = 
        \begin{pmatrix}
            1 & 0 \\
            0 & 0
        \end{pmatrix},
        \quad
        \sigma^{\downarrow} = 
        \begin{pmatrix}
            0 & 0 \\
            0 & 1
        \end{pmatrix},
        \\
        \sigma^{+} = 
        \begin{pmatrix}
            0 & 1 \\
            0 & 0
        \end{pmatrix},
        \quad
        \sigma^{-} = 
        \begin{pmatrix}
            0 & 0 \\
            1 & 0
        \end{pmatrix},
    \end{align}
\end{subequations}
it is straightforward to see that the only non-vanishing components are $\chi^{+-}_{\mathrm{KS}}$, $\chi^{-+}_{\mathrm{KS}}$ and
\begin{subequations}
    \begin{align}
        \chi^{00}_{\mathrm{KS}} &= \chi^{zz}_{\mathrm{KS}} = \chi^{\uparrow \uparrow}_{\mathrm{KS}} + \chi^{\downarrow \downarrow}_{\mathrm{KS}},
        \\
        \chi^{0z}_{\mathrm{KS}} &= \chi^{z0}_{\mathrm{KS}} = \chi^{\uparrow \uparrow}_{\mathrm{KS}} - \chi^{\downarrow \downarrow}_{\mathrm{KS}}.
    \end{align}
\end{subequations}
Thus, in the collinear case, one only needs to compute $\chi^{\uparrow \uparrow}_{\mathrm{KS}}$, $\chi^{\downarrow \downarrow}_{\mathrm{KS}}$, $\chi^{+-}_{\mathrm{KS}}$ and $\chi^{-+}_{\mathrm{KS}}$ in order to construct the full Kohn-Sham four-component susceptibility tensor.

In the LR-TDDFT formalism described above, one needs in principle all the excited states of the Kohn-Sham system in order to evaluate the Kohn-Sham susceptibility in Eq. \eqref{eq:dyn. plane wave KS susc. tensor lehmann}. It should be noted that the Kohn-Sham construction allows for the calculation of $\chi^{+-}(\mathbf{r}, \mathbf{r}', t-t')$ without explicit use of the excited states of the Kohn-Sham system. Such approaches include propagating the system in real-time for different "transverse magnetic kicks"\cite{Tancogne-Dejean2020} or using the Sternheimer equation from time-dependent density functional perturbation theory\cite{Savrasov1998,Cao2017}. As we will show below, the Kohn-Sham excited states are generally not a main limiting factor for the LR-TDDFT methodology, and we will use these complementing methods only for comparison.

\subsection{Transverse magnetic susceptibility within the adiabatic local spin-density approximation}\label{sec:trans mag susc in ALSDA}
The Hartree part of the Hartree-exchange-correlation kernel is straightforward to evaluate. In frequency space one obtains $K^{\mu\nu}_\mathrm{Hxc}=v_\mathrm{c}\delta^{0\mu}\delta^{0\nu}+K^{\mu\nu}_\mathrm{xc}$, where $v_\mathrm{c}$ is the Coulomb interaction and $K^{\mu\nu}_\mathrm{xc}$ needs to be approximated. In the adiabatic local spin-density approximation (ALDA), $K^{\mu\nu}_\mathrm{xc}$ is approximated by
\begin{align}
    K_{\mathrm{ALDA}}^{\tau_1 \tau_2}(\mathbf{r}_1, \mathbf{r}_2, t_1 - t_2) = &f_{\mathrm{LDA}}^{\tau_1 \tau_2}[n, \mathbf{m}](\mathbf{r}_1)
    \nonumber \\
    &\times \delta\left(\mathbf{r}_1 - \mathbf{r}_2\right) \delta\left(t_1 - t_2\right),
    \label{eq:ALDA four-component kernel}
\end{align}
where $n(\mathbf{r})$ and $\mathbf{m}(\mathbf{r})$ are the ground state electron density and magnetization, while
\begin{equation}
    f_{\mathrm{LDA}}^{\tau_1 \tau_2}\big[n, \mathbf{m}\big](\mathbf{r}) = \left.\frac{\partial^2\big[ \epsilon_\mathrm{xc}(n, |\mathbf{m}|) n\big]}{\partial n^{\tau_1} \partial n^{\tau_2}}\right|_{n(\mathbf{r}), \mathbf{m}(\mathbf{r})},
\end{equation}
where $\epsilon_{\mathrm{xc}}\left(n, m\right)$ is the exchange-correlation energy per electron of a homogeneous electron gas of density $n$ and magnetization $m=|\mathbf{m}|$. The derivatives are evaluated using
\begin{equation}
    m = \sqrt{\left(n^x\right)^2 + \left(n^y\right)^2 + \left(n^z\right)^2} = \sqrt{4 n^+n^- + \left(n^z\right)^2},
\end{equation}
which yields
\begin{equation}
    \frac{\partial}{\partial n^z} = \frac{n^z}{m} \frac{\partial}{\partial m}, \qquad \frac{\partial}{\partial n^\pm} = \frac{2n^\mp}{m} \frac{\partial}{\partial m}.
\end{equation}

Similar to Eq. \eqref{eq:four-component response relation circ. coord.}, the response relation for the Kohn-Sham susceptibility tensor can be rewritten in circular coordinates:
\begin{equation}
    \delta n^{j}(\mathbf{r}, t) = \sum_{k}\int_{-\infty}^{\infty}dt'\int d\mathbf{r}' \, \chi_{\mathrm{KS}}^{jk}(\mathbf{r}, \mathbf{r}', t-t') \delta \Breve{W}_{\mathrm{s}}^{k}(\mathbf{r}', t').
    \label{eq:kohn-sham four-component response relation circ. coord.}
\end{equation}
This results in the Dyson equation
\begin{align}
    \chi^{jk}(\mathbf{r}, \mathbf{r}', \omega) = \chi^{jk}_{\mathrm{KS}}(\mathbf{r}, \mathbf{r}', \omega) + \sum_{l_1, l_2} &\iint d\mathbf{r}_1 d\mathbf{r}_2 \, 
    \nonumber \\
    \times \chi^{j l_1}_{\mathrm{KS}}(\mathbf{r}, \mathbf{r}_1, \omega) \Breve{K}_{\mathrm{Hxc}}^{l_1 l_2}(\mathbf{r}_1, \mathbf{r}_2, \omega) &\chi^{l_2k}(\mathbf{r}_2, \mathbf{r}', \omega),
    \label{eq:four-comp. Dyson real space circ. coord.}
\end{align}
where
\begin{align}
    \Breve{K}_{\mathrm{Hxc}}^{l_1 l_2}(\mathbf{r}_1, \mathbf{r}_2, t_1 - t_2) = \frac{\delta \Breve{W}^{l_1}_{\mathrm{Hxc}}(\mathbf{r}_1, t_1)}{\delta n^{l_2}(\mathbf{r}_2, t_2)}.
    \label{eq:Hxc kernel def. circ. coord.}
\end{align}

In the case of a collinear ground state, spin-polarized in the $z$-direction, the ALDA Hartree-exchange-correlation kernel becomes block diagonal:
\begin{equation}
    \breve{K}_{\mathrm{Hxc}}^{[0, +, -, z]} = 
    \begin{pmatrix}
        v_\mathrm{c} + K_{\mathrm{ALDA}}^{00} & 0 & 0 & K_{\mathrm{ALDA}}^{0z} \\
        0 & 0 & \breve{K}_{\mathrm{ALDA}}^{+-} & 0 \\
        0 & \breve{K}_{\mathrm{ALDA}}^{-+} & 0 & 0 \\
        K_{\mathrm{ALDA}}^{z0} & 0 & 0 & K_{\mathrm{ALDA}}^{zz}
    \end{pmatrix},
\end{equation}
with $K_{\mathrm{ALDA}}^{0z}=K_{\mathrm{ALDA}}^{z0}$ and $\breve{K}_{\mathrm{ALDA}}^{+-}=\breve{K}_{\mathrm{ALDA}}^{-+}$. 
Since both the many-body susceptibility tensor and the Kohn-Sham analogue are block diagonal as well (see Eq.  \eqref{eq:four-component susc. tensor sparsity by collinearity circ. coord.}), the transverse components decouple from the remaining components:
\begin{align}
    \chi^{+-}(\mathbf{r}, \mathbf{r}', \omega) = \chi^{+-}_{\mathrm{KS}}(\mathbf{r}, \mathbf{r}', \omega) + \iint d\mathbf{r}_1 d\mathbf{r}_2 &\, 
    \nonumber \\
    \times \chi^{+-}_{\mathrm{KS}}(\mathbf{r}, \mathbf{r}_1, \omega) \Breve{K}_{\mathrm{ALDA}}^{-+}(\mathbf{r}_1, \mathbf{r}_2, \omega) &\chi^{+-}(\mathbf{r}_2, \mathbf{r}', \omega),
    \label{eq:dyson_+-}
\end{align}
where $+$ and $-$ can be interchanged to obtain the Dyson equation for $\chi^{-+}(\mathbf{r}, \mathbf{r}', \omega)$. The transverse LDA kernel itself turns out to be particularly simple,
\begin{equation}
    f_{\mathrm{LDA}}^{-+}\big[n, n^z\big](\mathbf{r}) = \frac{2 W_{\mathrm{xc,LDA}}^z\big[n, n^z\big](\mathbf{r})}{n^z(\mathbf{r})},
    \label{eq:transverse LDA kernel}
\end{equation}
and in the plane wave representation, the ALDA kernel is independent of $\mathbf{q}$ as well as $\omega$:
\begin{align}
    \breve{K}_{\mathrm{ALDA}, \mathbf{G}_1\hspace{-1pt}\mathbf{G}_2}^{-+}
    &= \frac{1}{\Omega_\mathrm{cell}}\int_{\Omega_{\mathrm{cell}}}d\mathbf{r}  \, e^{-i\left(\mathbf{G}_1 - \mathbf{G}_2\right) \cdot \mathbf{r}} f^{-+}_{\mathrm{LDA}}(\mathbf{r}) 
    \nonumber \\
    &= \frac{1}{\Omega_{\mathrm{cell}}} f^{-+}_{\mathrm{LDA}}\left(\mathbf{G}_1 - \mathbf{G}_2\right).
    \label{eq:transverse ALDA plane wave kernel}
\end{align}
To summarize, the many-body transverse magnetic susceptibility can be calculated directly from the Kohn-Sham susceptibility \eqref{eq:dyn. plane wave KS susc. tensor lehmann} and the kernel \eqref{eq:transverse LDA kernel}-\eqref{eq:transverse ALDA plane wave kernel}. Due to the separation of components, solving the Dyson equation \eqref{eq:dyson eq. reciprocal space} amounts to a simple matrix inversion:
%
\begin{equation}
    \chi^{+-}_{[\mathbf{G}]}(\mathbf{q}, \omega) =\left(1 - \chi^{+-}_{\mathrm{KS}}(\mathbf{q}, \omega) \Breve{K}^{-+}_{\mathrm{ALDA}} \right)^{-1}_{[\mathbf{G}]} \chi^{+-}_{\mathrm{KS}, [\mathbf{G}]}(\mathbf{q}, \omega).
    \label{eq:dyson trans mag plane wave susc.}
\end{equation}

The structure of the susceptibility tensor for a spin-paired ground state will now be briefly discussed. In this case, it is not sensible to distinguish between transverse magnetic and longitudinal magnetic susceptibilities. It is straightforward to show that the full ALDA kernel becomes diagonal, such that
\begin{equation}
    K_{\mathrm{Hxc}}^{[0, x, y, z]} = 
    \begin{pmatrix}
        v_\mathrm{c} + K_{\mathrm{ALDA}}^{00} & 0 & 0 & 0 \\
        0 & K_{\mathrm{ALDA}}^{zz} & 0 & 0 \\
        0 & 0 & K_{\mathrm{ALDA}}^{zz} & 0 \\
        0 & 0 & 0 & K_{\mathrm{ALDA}}^{zz}
    \end{pmatrix},
\end{equation}
with
\begin{subequations}
    \begin{align}
        f_{\mathrm{LDA}}^{00}\big[n\big](\mathbf{r}) &= \left.\frac{\partial^2 \left[\epsilon_{\mathrm{xc}}(n, m) n \right]}{\partial n^2}\right|_{n(\mathbf{r}), m(\mathbf{r})=0},
        \\
        f_{\mathrm{LDA}}^{zz}\big[n\big](\mathbf{r}) &= \left.\frac{\partial^2 \left[\epsilon_{\mathrm{xc}}(n, m) n\right]}{\partial m^2}\right|_{n(\mathbf{r}), m(\mathbf{r})=0}.
    \end{align}
\end{subequations}
Furthermore, one can easily inspect Eq. \eqref{eq:dyn. plane wave KS susc. tensor lehmann} to conclude that $\chi^{\uparrow \uparrow}_{\mathrm{KS}} = \chi^{\downarrow \downarrow}_{\mathrm{KS}} = \chi^{+-}_{\mathrm{KS}} = \chi^{-+}_{\mathrm{KS}} \equiv \chi_{\mathrm{KS}} / 2$, such that the Kohn-Sham four-component susceptibility tensor simplifies significantly: $\chi_{\mathrm{KS}}^{[0, x, y, z]} = \chi_{\mathrm{KS}}\mathrm{I}_{4\times4}$.
%
%
In addition, from the discussion below Eq. \eqref{eq:dyson_n-}, the many-body susceptibility tensor becomes diagonal as well,
\begin{equation}
    \chi^{[0, x, y, z]} = 
    \begin{pmatrix}
        \chi^{00} & 0 & 0 & 0 \\
        0 & \chi^{zz} & 0 & 0 \\
        0 & 0 & \chi^{zz} & 0 \\
        0 & 0 & 0 & \chi^{zz}
    \end{pmatrix},
\end{equation}
and the full magnetic response is contained in a single Dyson equation:
\begin{align}
    \chi^{zz}(\mathbf{r}, \mathbf{r}', \omega) = \chi_{\mathrm{KS}}(\mathbf{r}, \mathbf{r}', \omega) + \iint d\mathbf{r}_1 d\mathbf{r}_2 &\, 
    \nonumber \\
    \times \chi_{\mathrm{KS}}(\mathbf{r}, \mathbf{r}_1, \omega) K_{\mathrm{ALDA}}^{zz}(\mathbf{r}_1, \mathbf{r}_2) 
    &\chi^{zz}(\mathbf{r}_2, \mathbf{r}', \omega).
\end{align}

\subsection{Spectral enhancement and the Goldstone theorem}
Although $\chi^{+-}$ and $\chi^{+-}_{\mathrm{KS}}$ are directly related by the Dyson equation \eqref{eq:dyson trans mag plane wave susc.}, the transverse magnetic excitations as described by the corresponding spectral functions $S^{+-}_{\mathbf{G}}(\mathbf{q}, \omega)$ and $S^{+-}_{\mathrm{KS},\mathbf{G}}(\mathbf{q}, \omega)$ can be quite different. $S^{+-}_{\mathrm{KS},\mathbf{G}}(\mathbf{q}, \omega)$ gives the spectrum of Kohn-Sham spin-flip excitations, also referred to as the Stoner spectrum. In the collinear case, the non-interacting Stoner pairs are generated by removing an electron from an occupied band and $k$-point $\mathbf{k}$, flipping its spin and placing it in an unoccupied band and $k$-point $\mathbf{k}+\mathbf{q}$. The Stoner pairs form a continuum, which for ferromagnetic materials is gapped by the exchange splitting energy $\Delta_{\mathrm{x}}$ at $\mathbf{q}=\mathbf{0}$. 
Whereas the exchange splitting can have a magnitude of several electron volts, the fully interacting spectrum of transverse magnetic excitations, $S^{+-}_{\mathbf{G}}(\mathbf{q}, \omega)$, exhibits a so-called Goldstone mode with $\omega_{\mathbf{q}=\boldsymbol{0}}=0$ for spin-isotropic systems. Physically, this mode arises when a rigid rotation of the direction of magnetization does not cost any energy and it is a manifestation of the more general Goldstone theorem. 
Due to the binding nature of the interaction in Eq. \eqref{eq:transverse LDA kernel}, the many-body transverse magnetic excitations generally exist at energies below the Stoner continuum. However, in itinerant ferromagnets, the Stoner gap will close for wave vectors $\mathbf{q}$ connecting the majority and minority spin Fermi surfaces\cite{Moriya1985,Niesert2011}. As a magnon branch enters the Stoner continuum, it will be dressed by the single-particle excitations leading to a broadening of the spectral width. The corresponding shortening in quasi-particle lifetime is called Landau damping\cite{Landau1965}.

Often $\chi^{+-}$ is referred to as the enhanced susceptibility because the Dyson equation \eqref{eq:dyson trans mag plane wave susc.} can be understood as the formation of collective magnon excitations out of the single-particle Stoner continuum. As it turns out, this procedure preserves the total spectral weight embedded in the susceptibility. 
For the transverse magnetic susceptibility, the zeroth order sum rule (see \eqref{eq:Kubo nth order sum rule}) relates the spectrum of transverse magnetic excitations to the magnetization density of the ground state:
%
\begin{equation}
    \hbar\int_{-\infty}^{\infty} S^{+-}(\mathbf{r}, \mathbf{r}', \omega) \, d\omega = n^z(\mathbf{r})\, \delta(\mathbf{r}-\mathbf{r}').
\end{equation}
Because the spin-polarization density is the same in both the Kohn-Sham and the fully interacting system by construction, the total spectral weight is preserved between the two. 
%
%
By performing a lattice Fourier transform, a similar expression for the plane wave susceptibility is obtained:
\begin{equation}
    \hbar\int_{-\infty}^{\infty} S^{+-}_{\mathbf{G}\mathbf{G}'}(\mathbf{q}, \omega) \, d\omega = \frac{n^z(\mathbf{G}-\mathbf{G}')}{\Omega_{\mathrm{cell}}},
    \label{eq:trans. mag. plane wave sum rule}
\end{equation}
where $n^z(\mathbf{G}-\mathbf{G}')$ denotes the plane wave coefficients of the spin-polarization density, defined similarly to Eq. \eqref{eq:periodic transition matrix elements}. 
As a consequence, the total spectral weight of transverse magnetic excitations at any $\mathbf{G}$ and $\mathbf{q}$ is simply the average spin-polarization density:
\begin{equation}
    \hbar\int_{-\infty}^{\infty} S^{+-}_{\mathbf{G}}(\mathbf{q}, \omega) \, d\omega 
    = \frac{\sigma_z}{\Omega_{\mathrm{cell}}},
    \label{eq:trans. mag. plane wave diagonal sum rule}
\end{equation}
where $\sigma_z$ denotes $n^z(\mathbf{r})$ integrated over the unit cell.

\section{Computational Implementation}\label{sec:impl}
As described above, the transverse magnetic plane wave susceptibility can be computed within linear response time-dependent density functional theory using only quantities that can be obtained from the auxiliary non-interacting Kohn-Sham system. We have implemented this methodology into the GPAW open-source code\cite{Mortensen2005,Enkovaara2010}, which uses the projected augmented wave method\cite{Blochl1994}. The implementation is based on the existing linear response module for GPAW\cite{Yan2011}, which enables computation of the longitudinal dielectric susceptibility $\chi^{00}$ and related material properties. In this section, we present the implementation and make a rigorous performance assessment of the numerical scheme employed.

\subsection{Projected augmented wave method for plane wave susceptibilities}\label{sec:PAW for pw susc}
The list of Kohn-Sham quantities needed for calculating the transverse magnetic plane wave susceptibility is relatively short. The Kohn-Sham orbital energies and occupancies, $\epsilon_{n\mathbf{k}s}$ and $f_{n\mathbf{k}s}$, are easily extracted from any DFT ground state calculation, the Kohn-Sham pair densities \eqref{eq:KS reciprocal space pair densities} are calculated from the Kohn-Sham orbitals and the transverse magnetic plane wave kernel \eqref{eq:transverse LDA kernel}-\eqref{eq:transverse ALDA plane wave kernel} is calculated from the ground state density and spin-polarization density. 

In the projected augmented wave method (PAW), the all-electron Kohn-Sham orbitals $\psi_{n\mathbf{k}s}$ are written in terms of smooth pseudo waves $\Tilde{\psi}_{n\mathbf{k}s}$, which are easy to represent numerically.
\begin{subequations}
    \begin{equation}
        |\psi_{n\mathbf{k}s}\rangle = \hat{\mathcal{T}} |\Tilde{\psi}_{n\mathbf{k}s}\rangle,
    \end{equation}
    \begin{equation}
        \hat{\mathcal{T}} = 1 + \sum_{a,i}\left(|\phi^a_i\rangle - |\Tilde{\phi}^a_i\rangle \right) \langle \Tilde{p}^a_i|.
    \end{equation}
    \label{eq:PAW wave mapping}
\end{subequations}
%
Inside the so-called augmentation sphere, a spherical region of space centered at the position of the $a$'th atomic nuclei $\mathbf{R}_a$, smooth partial waves $\Tilde{\phi}^a_i$ and projector functions $\Tilde{p}^a_i$ are constructed to fulfill $\sum_i |\Tilde{\phi}^a_i\rangle \langle \Tilde{p}^a_i| = 1$, so that the linear operator $\hat{\mathcal{T}}$ effectively maps the smooth pseudo waves onto the all-electron partial waves $\phi^a_i$. Outside the augmentation sphere $\Tilde{\phi}^a_i(\mathbf{r}-\mathbf{R}_a)=\phi^a_i(\mathbf{r}-\mathbf{R}_a)$, making the smooth pseudo wave equal to the all-electron Kohn-Sham orbital in the interstitial region between the augmentation spheres. 
Due to the linear mapping in Eq. \eqref{eq:PAW wave mapping}, matrix elements between Kohn-Sham orbitals can be evaluated from the smooth pseudo waves using a pseudo operator, operating on the space of pseudo waves:
\begin{equation}
    \langle \psi_{n\mathbf{k}s} | \hat{A} | \psi_{m\mathbf{k}'s'} \rangle = \langle \Tilde{\psi}_{n\mathbf{k}s} | \Tilde{A} | \Tilde{\psi}_{m\mathbf{k}'s'} \rangle, \quad \Tilde{A} = \hat{\mathcal{T}}^{\dagger} \hat{A} \hat{\mathcal{T}}.
\end{equation}
For any quasilocal operator $\hat{A}$, the effective pseudo operator can be written\cite{Blochl1994}
\begin{equation}
    \Tilde{A} = \hat{A} + \sum_a \sum_{i,i'} |\Tilde{p}^a_i\rangle \left[ \langle \phi^a_i | \hat{A} | \phi^a_{i'} \rangle - \langle \Tilde{\phi}^a_i | \hat{A} | \Tilde{\phi}^a_{i'} \rangle \right] \langle \Tilde{p}^a_{i'} |.
\end{equation}
Thus, the evaluation of the Kohn-Sham pair densities in Eq. \eqref{eq:KS reciprocal space pair densities} amounts to a direct evaluation using the pseudo waves and a PAW correction:
\begin{align}
    n_{n\mathbf{k}s, m\mathbf{k}+\mathbf{q}s'}(\mathbf{G}+\mathbf{q}) 
    &= \Tilde{n}_{n\mathbf{k}s, m\mathbf{k}+\mathbf{q}s'}(\mathbf{G}+\mathbf{q}) 
    \nonumber \\
    &\hspace{10pt}+ \Delta n_{n\mathbf{k}s, m\mathbf{k}+\mathbf{q}s'}(\mathbf{G}+\mathbf{q}),
    \label{eq:KS pair densities PAW}
\end{align}
where
\begin{align}
    \Delta n_{n\mathbf{k}s, m\mathbf{k}+\mathbf{q}s'}(\mathbf{G}+\mathbf{q}) 
    = \sum_a \sum_{i,i'} &Q^a_{i i'}(\mathbf{G}+\mathbf{q})
    \nonumber \\
    \times &\langle \Tilde{\psi}_{n\mathbf{k}s} | \Tilde{p}^a_i \rangle \langle \Tilde{p}^a_{i'} | \Tilde{\psi}_{m\mathbf{k}+\mathbf{q}s'} \rangle,
\end{align}
with
\begin{align}
    &Q^a_{i i'}(\mathbf{G}+\mathbf{q}) = 
    \int_{\Omega_{\mathrm{cell}}} d\mathbf{r} \, e^{-i (\mathbf{G} + \mathbf{q}) \cdot \mathbf{r}} \nonumber \\
    &\times \left[\phi^{a *}_i(\mathbf{r}-\mathbf{R}_a) \phi^a_{i'}(\mathbf{r}-\mathbf{R}_a) - \Tilde{\phi}^{a *}_i(\mathbf{r}-\mathbf{R}_a) \Tilde{\phi}^a_{i'}(\mathbf{r}-\mathbf{R}_a) \right].
    \label{eq:PAW correction tensor}
\end{align}
In a given DFT calculation, the PAW setups for every atomic species is fixed (fixing $\phi^a_i$, $\Tilde{\phi}^a_i$ and $\Tilde{p}^a_i$), so that the PAW correction tensor, $Q^a_{i i'}(\mathbf{G}+\mathbf{q})$, can be evaluated once and reused for all the Kohn-Sham pair densities as a function of $\mathbf{G}$ and $\mathbf{q}$. As a result, the calculation of pair-densities is a fairly cheap procedure in terms of computational power.

Similarly, the ground state spin-densities may be written in terms of a smooth contribution from the pseudo waves $\Tilde{n}_{\sigma}(\mathbf{r})$ and atom-centered PAW corrections localized to the augmentation spheres:
\begin{equation}
    n_{\sigma}(\mathbf{r}) = \Tilde{n}_{\sigma}(\mathbf{r}) + \sum_a \left[ n^a_{\sigma}(\mathbf{r} - \mathbf{R}_a) - \Tilde{n}^a_{\sigma}(\mathbf{r} - \mathbf{R}_a) \right].
\end{equation}
As a result, ALDA plane wave kernels, such as the transverse magnetic kernel in Eq. \eqref{eq:transverse ALDA plane wave kernel}, can be calculated as a contribution from the smooth density and a PAW correction localized to the augmentation spheres:
\begin{equation}
    \Breve{K}^{-+}_{\mathrm{ALDA},\mathbf{G}_1\hspace{-1pt}\mathbf{G}_2} = 
    \frac{1}{\Omega_{\mathrm{cell}}}
    \Tilde{f}^{-+}_{\mathrm{LDA}}(\mathbf{G}_1 - \mathbf{G}_2) 
    + \Delta \Breve{K}^{-+}_{\mathrm{ALDA},\mathbf{G}_1\hspace{-1pt}\mathbf{G}_2},
    \label{eq:ALDA plane wave kernel PAW}
\end{equation}
where
\begin{align}
    \Delta \Breve{K}^{-+}_{\mathrm{ALDA},\mathbf{G}_1\hspace{-1pt}\mathbf{G}_2} = \sum_a \int_{\Omega_{\mathrm{cell}}} &\frac{d\mathbf{r}_1}{\Omega_{\mathrm{cell}}} \, e^{-i\left(\mathbf{G}_1 - \mathbf{G}_2\right) \cdot \mathbf{r}_1} 
    \nonumber \\
    &\times \Delta f^{a,-+}_{\mathrm{LDA}}(\mathbf{r}_1 - \mathbf{R}_a),
    \label{eq:ALDA plane wave kernel PAW correction}
\end{align}
with atom-centered PAW corrections to the LDA kernel
%
\begin{equation}
    \Delta f^{a,-+}_{\mathrm{LDA}}(\mathbf{r}) 
    = f^{-+}_{\mathrm{LDA}}\left[n^a_{\uparrow}, n^a_{\downarrow}\right](\mathbf{r})
    - f^{-+}_{\mathrm{LDA}}\left[\Tilde{n}^a_{\uparrow}, \Tilde{n}^a_{\downarrow}\right](\mathbf{r}).
\end{equation}

In principle, the PAW method does not lead to any loss in generality, and the PAW corrected Kohn-Sham pair densities and ALDA plane wave kernels can be regarded as all-electron quantities. In practice however, generating partial waves with projector functions to match is not a trivial task, and the partial wave expansion will not be complete.

\subsection{Implementation of the PAW method}
In GPAW, the pseudo waves $\Tilde{\psi}_{n\mathbf{k}s}(\mathbf{r})=e^{i\mathbf{k}\cdot\mathbf{r}} \Tilde{u}_{n\mathbf{k}s}(\mathbf{r})$ are represented on a real-space grid using a plane wave basis set for the periodic parts $\Tilde{u}_{n\mathbf{k}s}(\mathbf{r})$. The smooth contributions to the Kohn-Sham pair densities in Eq. \eqref{eq:KS pair densities PAW} and the transverse ALDA plane wave kernel in Eq. \eqref{eq:ALDA plane wave kernel PAW} are then computed by evaluating the integrand on the real-space grid and performing a fast-fourier-transform to reciprocal space:
\begin{equation}
    \Tilde{n}_{n\mathbf{k}s, m\mathbf{k}+\mathbf{q}s'}(\mathbf{G}+\mathbf{q}) = \mathcal{F}_{\mathbf{G}}\left\{ e^{-i\mathbf{q}\cdot\mathbf{r}} \Tilde{\psi}^*_{n\mathbf{k}s}(\mathbf{r}) \Tilde{\psi}_{m\mathbf{k}+\mathbf{q}s'}(\mathbf{r}) \right\},
\end{equation}
\begin{equation}
    \Tilde{f}^{-+}_{\mathrm{LDA}}(\mathbf{G}_1 - \mathbf{G}_2) = \mathcal{F}_{\mathbf{G}_1-\mathbf{G}_2}\left\{ f^{-+}_{\mathrm{LDA}}\big[\Tilde{n}_{\uparrow}, \Tilde{n}_{\downarrow}\big](\mathbf{r}) \right\}.
\end{equation}
Furthermore, the angular part of the atom-centered partial waves are real spherical harmonics:
\begin{equation}
    \phi^a_i(\mathbf{r}) = Y^{m_i}_{l_i}(\hat{\mathbf{r}}) \phi^a_i(r), \quad \Tilde{\phi}^a_i(\mathbf{r}) = Y^{m_i}_{l_i}(\hat{\mathbf{r}}) \Tilde{\phi}^a_i(r).
\end{equation}
Using the plane wave expansion into real spherical harmonics,
\begin{equation}
    e^{-i \mathbf{K} \cdot \mathbf{r}} = 4\pi \sum_l \sum^{l}_{m=-l} (-i)^l j_l\left(\left|\mathbf{K}\right| r\right) Y^{m}_{l}(\hat{\mathbf{K}}) Y^{m}_{l}(\hat{\mathbf{r}}),
\end{equation}
where $j_l\left(\left|\mathbf{K}\right| r\right)$ are spherical Bessel functions, the angular part of the PAW correction tensor integral in Eq. \eqref{eq:PAW correction tensor} can be carried out analytically:
\begin{align}
    Q^a_{i i'}(\mathbf{K}) 
    = &
    4\pi e^{-i\mathbf{K}\cdot \mathbf{R}_a}
    \sum_l \sum^l_{m=-l} (-i)^l Y^{m}_{l}(\hat{\mathbf{K}}) g^{l m}_{l_i m_i, l_{i'} m_{i'}}
    \nonumber \\
    &\hspace{-48pt}\times \int_{0}^{R^a_{\mathrm{c}}} r^2 dr \, j_l\left(\left|\mathbf{K}\right| r\right) \left[\phi^{a}_i(r) \phi^a_{i'}(r) - \Tilde{\phi}^{a}_i(r) \Tilde{\phi}^a_{i'}(r) \right],
    \label{eq:KS pair densities PAW correction GPAW}
\end{align}
with $\mathbf{K}=\mathbf{G}+\mathbf{q}$. Here $R^a_{\mathrm{c}}$ is the radius of the $a$'th augmentation sphere and $g^{l m}_{l_i m_i, l_{i'} m_{i'}}$ are the Gaunt coefficients. The radial part of each partial wave is stored on the same nonlinear radial grid for a given atom. We use this grid to carry out the radial integral in Eq. \eqref{eq:KS pair densities PAW correction GPAW} by point integration.

For the ALDA plane wave kernel, we approach the PAW correction in a similar fashion. We expand the atom-centered PAW corrections to the LDA kernel in real spherical harmonics,
\begin{equation}
    \Delta f^{a,-+}_{\mathrm{LDA}}(\mathbf{r}) = \sum_l \sum^l_{m=-l} Y^{m}_{l}(\hat{\mathbf{r}}) \Delta f^{a,lm,-+}_{\mathrm{LDA}}(r),
    \label{eq:LDA kernel real spherical harm. expansion}
\end{equation}
such that the angular integral in the PAW correction to the ALDA plane wave kernel \eqref{eq:ALDA plane wave kernel PAW correction} can be carried out analytically:
\begin{align}
    \Delta \Breve{K}^{a,-+}_{\mathrm{ALDA}}(\mathbf{K}) 
    = &\frac{4\pi e^{-i\mathbf{K}\cdot \mathbf{R}_a}}{\Omega_{\mathrm{cell}}} \sum_l \sum^l_{m=-l} (-i)^l Y^{m}_{l}(\hat{\mathbf{K}})
    \nonumber \\
    &\hspace{-10pt}\times \int_{0}^{R^a_{\mathrm{c}}} r^2 dr \, j_l\left(\left|\mathbf{K}\right| r\right) \Delta f^{a,lm,-+}_{\mathrm{LDA}}(r),
\end{align}
with $\mathbf{K}=\mathbf{G}_1 - \mathbf{G}_2$. To obtain the expansion into real spherical harmonics in Eq. \eqref{eq:LDA kernel real spherical harm. expansion}, the atom-centered PAW corrections to the LDA kernel are simply evaluated on an angular grid, a Lebedev quadrature of degree 11, for every radii $r$ on the nonlinear radial grid. Through point integration, the expansion coefficients are calculated for each radii $r$:
\begin{equation}
    \Delta f^{a,lm,-+}_{\mathrm{LDA}}(r) = \int d\hat{\mathbf{r}} \, Y^{m}_{l}(\hat{\mathbf{r}}) \Delta f^{a,-+}_{\mathrm{LDA}}(r \hat{\mathbf{r}}). 
\end{equation}
With a Lebedev quadrature of degree 11, polynomials up to order 11 can be point integrated exactly. This implies that the plane wave expansion remains numerically exact up to $l=5$. In practise, we truncate the expansion at $l=4$, which results in a well converged overall expansion for all the materials covered in this study.

\subsection{Numerical details}\label{sec:numerical details}
In our implementation there are a number of key parameters, with respect to which the calculation needs to be converged. 
The $k$-point summation in Eq. \eqref{eq:dyn. plane wave KS susc. tensor lehmann} is evaluated on the Monkhorst-Pack grid\cite{Pack1976} of the ground state calculation:
\begin{align}
    \chi^{+-}_{\mathrm{KS}, \mathbf{G}\mathbf{G}'}&(\mathbf{q}, z) = 
    \frac{1}{\Omega_{\mathrm{cell}}}\frac{1}{N_k} 
    \sum_{\mathbf{k}}
    \sum_{n, m} (f_{n\mathbf{k}\uparrow} - f_{m\mathbf{k}+\mathbf{q}\downarrow})
    \nonumber \\
    &\times \frac{n_{n\mathbf{k}\uparrow, m\mathbf{k}+\mathbf{q}\downarrow}(\mathbf{G} + \mathbf{q}) \,  n_{m\mathbf{k}+\mathbf{q}\downarrow, n\mathbf{k}\uparrow}(-\mathbf{G}'-\mathbf{q})}{\hbar \omega - (\epsilon_{m\mathbf{k}+\mathbf{q}\downarrow}-\epsilon_{n\mathbf{k}\uparrow}) + i \hbar \eta},
    \label{eq:dyn. trans. mag. plane wave KS susc. tensor lehmann}
\end{align}
where $N_k$ in this case denotes the number of grid points. Because a finite grid is used, the continuum of Kohn-Sham states is discretized. To make up for this fact, we do not take the formal limit $\eta\rightarrow0^+$ in Eq. \eqref{eq:dyn. trans. mag. plane wave KS susc. tensor lehmann}, but leave $\eta$ as a finite broadening parameter in order to smear out the transition energies $(\epsilon_{m\mathbf{k}+\mathbf{q}\downarrow}-\epsilon_{n\mathbf{k}\uparrow})$ and form a continuum. For a detailed discussion of this procedure, see section \ref{sec:KS continuum convergence}. Additionally, the band summation in Eq. \eqref{eq:dyn. trans. mag. plane wave KS susc. tensor lehmann} is truncated to include a finite number of excited states and a finite plane wave basis set is used to invert the Dyson equation in Eq. \eqref{eq:dyson trans mag plane wave susc.}. The effect of these parameters are investigated in sections \ref{sec:sum rule}, \ref{sec:gap error convergence} and \ref{sec:dispersion convergence}. Unless otherwise stated, 12 empty shell bands per atom and a plane wave cutoff of 1000 eV are used.

On top of these convergence parameters, the GPAW implementation has two additional simplifications. As mentioned above, the projected augmented wave method is formally exact, but in reality a finite set of partial waves is used in the expansion of the Kohn-Sham orbitals. For a given number of frozen core electrons, GPAW is distributed with a single PAW setup for each atomic species, meaning that the truncation of the expansion is given in advance, it is not a parameter that can be converged. Furthermore, we do not include the frozen core states in the band summation of Eq. \eqref{eq:dyn. trans. mag. plane wave KS susc. tensor lehmann}. For iron, cobalt and nickel this implies that only transitions from the occupied 4$s$ and 3$d$ electronic orbitals are included. GPAW also supplies an alternative setup for nickel, where also the 3$p$ orbitals are taken as valence states as opposed to being frozen core electronic orbitals. We tested the extended PAW setup, but found it much more difficult to converge the plane wave basis in Eq. \eqref{eq:dyson trans mag plane wave susc.}, only to obtain a small difference in the magnon dispersion. We extract a difference in magnon peak position between the PAW setups of $\Delta\omega_{\mathbf{q}} = 5.2$ meV calculated at the wave vector $\text{X}/3$, where $\omega_{\mathbf{q}}=305$ meV corresponding to a relative difference of $1.7\%$. At the wave vector $2\text{X}/3$ and at the the $\text{X}$-point itself, the relative difference is even smaller. Although including the frozen core states should increase the overall accuracy, the computational cost far exceeds what we seem to stand to gain. The minimal PAW setups are used for the results reported throughout the remainder of this paper.

The crystal structures of the transition metals investigated are described using ASE\cite{Larsen2017} with experimental lattice constants $a=2.867\text{ \AA}$ for bcc-Fe, $a=3.524\text{ \AA}$ for fcc-Ni, $a=3.539\text{ \AA}$ for fcc-Co and $a=2.507\text{ \AA}$ for hcp-Co taken from \cite{Buczek2011b,Singh2018} and the references therein. We investigate only reduced wave vectors $\mathbf{q}$ commensurate with the Monkhorst-Pack grid of the ground state calculation.

\subsection{Sum rule check}\label{sec:sum rule}
\begin{figure*}
    \centering
    \includegraphics{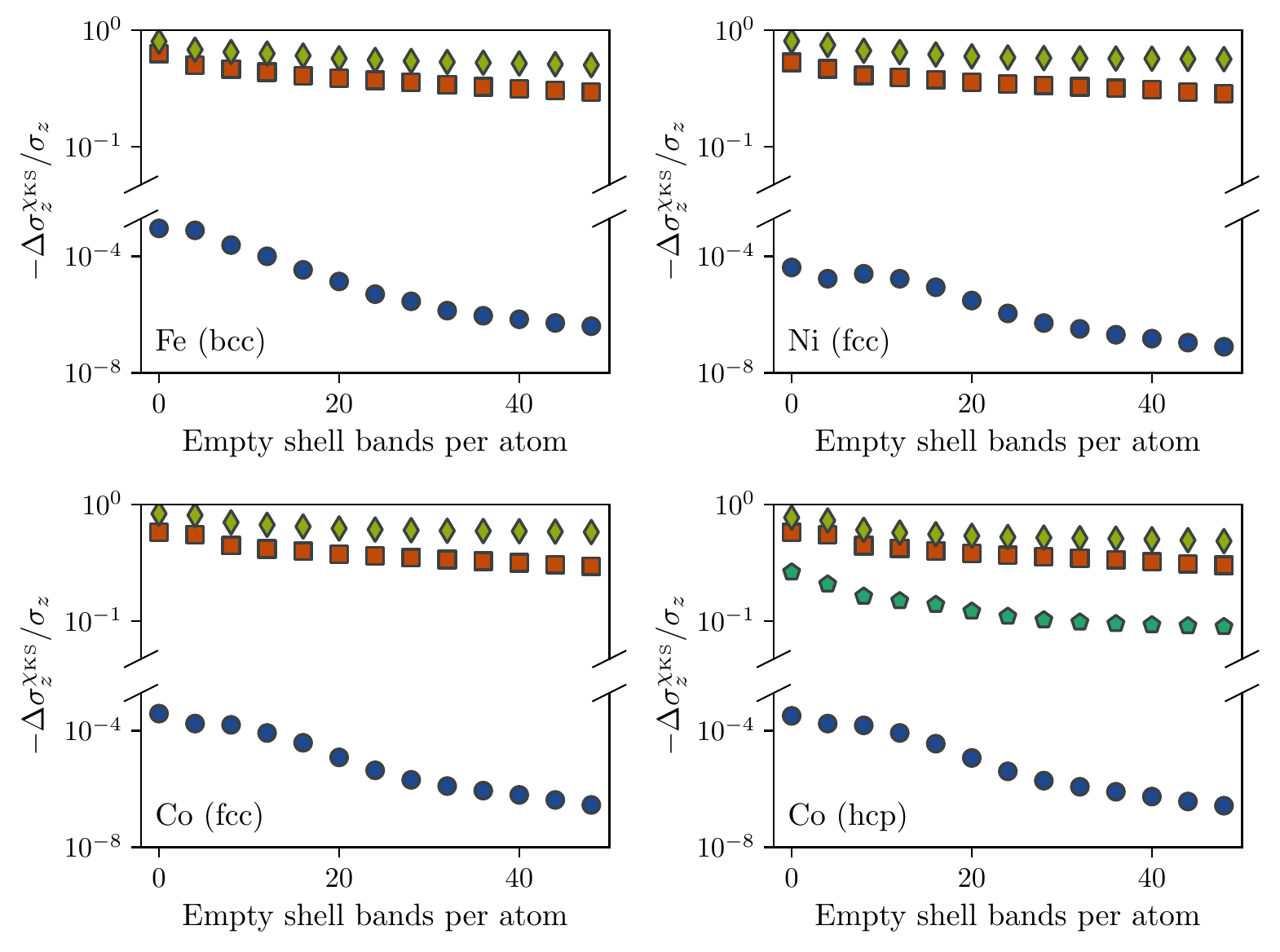}
    \caption{Relative error in the pair spin-polarization of iron, nickel and cobalt calculated from the sum rule \eqref{eq:pair spin-polarization} at $\mathbf{q}=\mathbf{0}$ as a function of empty shell bands per atom. The markers represent different reciprocal lattice vectors $\mathbf{G}$: Blue circles represent $(0, 0, 0)$, red squares $(0, 0, 1)$ for Fe, Ni and fcc-Co and $(0, 0, 2)$ for hcp-Co, green rhombi $(1, 1, -1)$ for Fe, $(1, 0, -1)$ for Ni, fcc-Co and $(1, 0, -2)$ for hcp-Co. The teal pentagons represent the $(0, 0, 1)$ reciprocal lattice vector in hcp-Co.}
    \label{fig:pair_magnetization_unocc_convergence}
\end{figure*}
As a check of our implementation, we have computed the average spin-polarization from the Kohn-Sham transverse magnetic susceptibility. Inserting the diagonal components of Eq. \eqref{eq:dyn. plane wave KS susc. tensor lehmann} into the sum rule \eqref{eq:trans. mag. plane wave diagonal sum rule} and performing the frequency integral analytically,
\begin{align}
    \frac{1}{\Omega} \sum_{\mathbf{k}}\sum_{n,m} &(f_{n\mathbf{k}\uparrow} - f_{m\mathbf{k}+\mathbf{q}\downarrow})
    \nonumber \\
    &\times \left|n_{m\mathbf{k}+\mathbf{q}\downarrow, n\mathbf{k}\uparrow}(-\mathbf{G}-\mathbf{q})\right|^2 = \frac{\sigma_z}{\Omega_{\mathrm{cell}}}.
    \label{eq:pair spin-polarization}
\end{align}
We refer to the average spin-polarization calculated in this manner as the pair spin-polarization, $\sigma_z^{\chi_{\mathrm{KS}}}$. 

We have computed the pair spin-polarization and compared it to the average spin-polarization extracted from the ground state for iron, nickel and cobalt at $\mathbf{q}=\mathbf{0}$ and different reciprocal lattice vectors $\mathbf{G}$. The comparison is presented in Fig. \ref{fig:pair_magnetization_unocc_convergence} as a function of the number of empty shell bands per atom included in the band summation of Eq. \eqref{eq:pair spin-polarization}. The pair spin-polarization is consistently smaller that the average spin-polarization of the ground state, $\Delta\sigma_z^{\chi_{\mathrm{KS}}}=\sigma_z^{\chi_{\mathrm{KS}}} - \sigma_z < 0$, but rapidly converges towards it for $\mathbf{G}=\mathbf{0}$ as the number of empty shell bands is increased. Thus, the PAW implementation seems to provide a good description of the macroscopic spatial variation embedded in the transverse magnetic susceptibility. 

For $\mathbf{G} \neq \mathbf{0}$ the convergence is orders of magnitude slower. The convergence is governed by the pair densities, which are calculated as simple overlap integrals between two Kohn-Sham orbitals and a plane wave (see Eq. \eqref{eq:KS reciprocal space pair densities}). We believe that this slow convergence arises because many Kohn-Sham orbitals are needed to represent a single plane-wave, or conversely, that many plane-waves are needed to represent a single Kohn-Sham orbital. This interpretation is supported by the fact, that the $\mathbf{G}=(0, 0, 1)$ pair spin-polarization in hcp-Co has an improved convergence with respect to more local reciprocal lattice vectors. The $\mathbf{G}=(0, 0, 1)$ plane wave is better represented in terms of Kohn-Sham orbitals as it gives the two atoms in the unit cell exactly opposite phases. To fully converge the pair spin-polarization for all reciprocal lattice vectors, one would also need to include the frozen core states in the band summation. This slow convergence with respect to the number of bands is much less pronounced for the transverse magnetic susceptibility at small frequencies, as we will show in sections \ref{sec:gap error convergence} and \ref{sec:dispersion convergence}, because transitions to highly excited states are suppressed by a factor $\Delta\epsilon^{-1}$ in Eq. \eqref{eq:dyn. trans. mag. plane wave KS susc. tensor lehmann}. Thus, the pair spin-polarization convergence is generally not a necessary requirement for obtaining an accurate description of the magnons.

\subsection{Convergence of the Kohn-Sham continuum}\label{sec:KS continuum convergence}
\begin{figure*}
    \centering
    \includegraphics{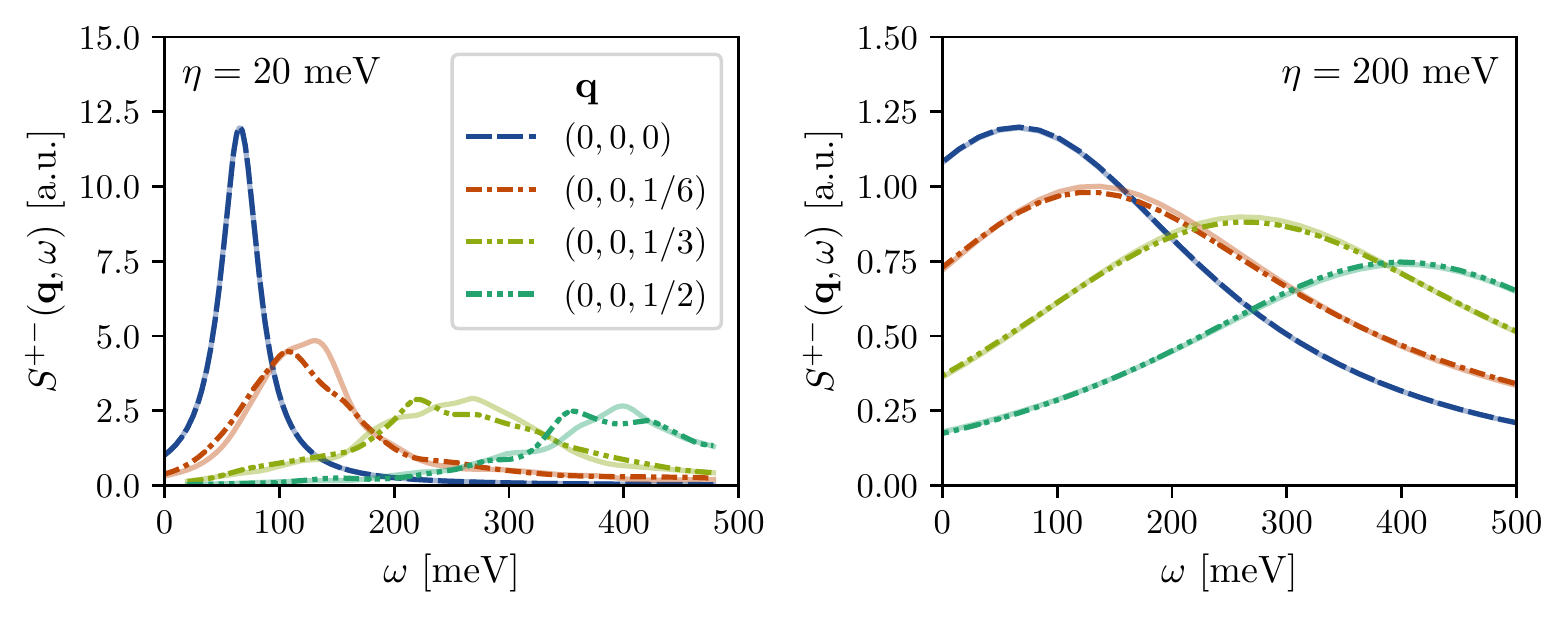}
    \caption{Macroscopic transverse magnetic excitation spectrum of bcc-Fe in the ALDA ($S^{+-}_{\mathbf{G}=\mathbf{0}}(\mathbf{q}, \omega)$, see Eq. \eqref{eq:transverse fluctuation spectrum def.}) calculated at a range of different wave vectors $\mathbf{q}$. The calculations were performed on $(42, 42, 42)$ $k$-point grids, where the dash-dotted lines indicate results from a regular Monkhorst-Pack grid, whereas the translucent lines are results from a $\Gamma$-centered grid. The panels show the spectra calculated at two different broadening parameters $\eta$.}
    \label{fig:trans_mag_susc_eta_example}
\end{figure*}
For the itinerant ferromagnets of this study, the $k$-point grid refinement of Eq. \eqref{eq:dyn. trans. mag. plane wave KS susc. tensor lehmann} is an important numerical parameter to converge. Even though the bands of different spin character are split by exchange, there are metallic Kohn-Sham bands of both majority and minority spin character in all four materials. This means that the Stoner continuum will extend downwards from the exchange splitting energy $\Delta_{\mathrm{x}}$ to $\omega=0$ for reduced wave vectors $\mathbf{q}$ that connect the Fermi surfaces of different spin character. For such $\mathbf{q}$, the collective magnon modes will unavoidably be dressed by these low frequency Stoner excitations and be Landau damped as a result. Thus, to accurately describe the magnon modes, the discretized Stoner continuum obtained from Eq. \eqref{eq:dyn. trans. mag. plane wave KS susc. tensor lehmann} must be broadened into a continuum by leaving $\eta$ as a finite broadening parameter. In the end, one should use a sufficiently dense $k$-point grid such that $\eta$ can be chosen small enough not to have an overall influence on the magnon dispersion, yet large enough to effectively broaden the low frequency spectrum of Stoner excitations into a continuum.

In Fig. \ref{fig:trans_mag_susc_eta_example}, we illustrate the effect of the broadening procedure by plotting the macroscopic transverse magnetic excitation spectrum at a fixed $k$-point density, but with regular and centered Monkhorst-Pack grids and different values for $\eta$. For the spectral peak at $\mathbf{q}=\mathbf{0}$, the two grid alignments yield consistent results with a Lorentzian lineshape of half-width $\eta$, corresponding to a magnon mode free of Landau damping. However, this is not the case for the spectra at finite crystal momentum transfer. With a broadening of $\eta=20$ meV, spurious finite-grid effects dominate the lineshapes, and the magnon peak positions, i.e. the frequencies corresponding to the maximum of the spectral function for a given $\mathbf{q}$, cannot be consistently extracted. At $\eta=200$ meV the discrepancy between the two grid alignments is more or less cured, as the discrete spectrum of low frequency Stoner excitations has been broadened into a continuum. Unfortunately, the effect of Landau damping is now hard to discern and, as will be shown below, the magnon peak positions have been shifted towards higher frequencies.

\begin{figure*}
    \centering
    \includegraphics{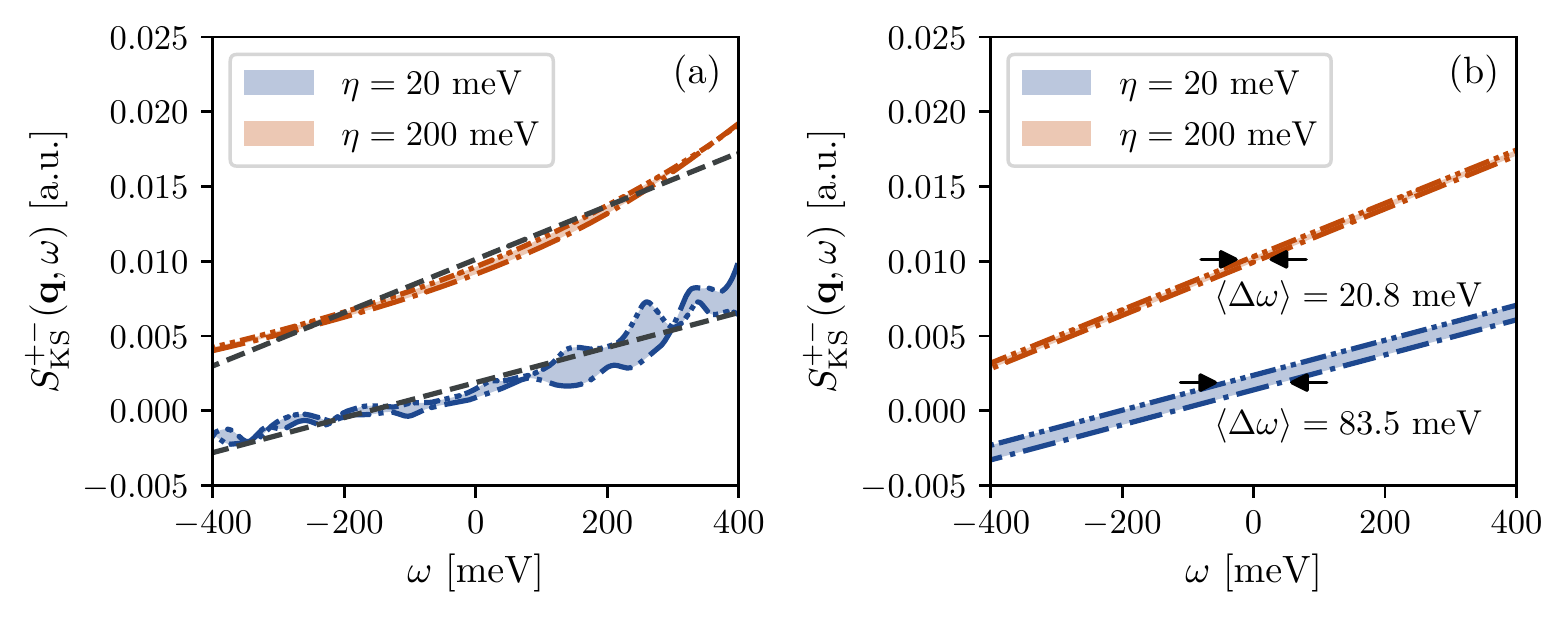}
    \caption{The average frequency displacement, $\langle \Delta \omega \rangle$, calculated from the macroscopic spectral function of Stoner excitations $S^{+-}_{\mathrm{KS},\mathbf{G}=\mathbf{0}}(\mathbf{q}, \omega)$. Panel (a) shows the Kohn-Sham spectral functions of iron calculated using a $(30, 30, 30)$ $k$-point grid and two different values for $\eta$. Dash-dotted lines represents a $\Gamma$-centered grid and the dash-double-dotted lines a regular Monkhorst-Pack grid. A simultaneous fit to the two spectral functions is also shown, which along with the colored area between the functions determines $\langle \Delta \omega \rangle$ as defined in Eq. \eqref{eq:avg freq displ}. Panel (b) shows linear spectral functions with the gradient from the fit in panel (a) and of the same area between the curves. For these parallel spectral functions, $\langle \Delta \omega \rangle$ gives the horizontal frequency displacement of the curves.
    }
    \label{fig:ksspectra_comparison}
\end{figure*}
To study the convergence of the low frequency Stoner continuum further, it is worthwhile to remark, that the macroscopic spectrum of Stoner excitations is much cheaper to compute than the full transverse magnetic excitation spectrum, as no extra plane wave components are needed, when the Dyson equation \eqref{eq:dyson trans mag plane wave susc.} does not have to be inverted. Thus, it would be of great value, if the convergence of the low frequency Stoner continuum could be assessed from the Kohn-Sham spectral function itself. To that end, we introduce the average frequency displacement $\langle \Delta\omega \rangle$. The idea is to consider the Stoner continuum truly converged when different $k$-point grid alignments yield the same Kohn-Sham spectral function. The average frequency displacement is defined as the integrated absolute difference between the Kohn-Sham spectral functions calculated on regular and $\Gamma$-centered Monkhorst-Pack $k$-point grids, normalized by the effective absolute change in spectral function intensity over the integration range:
\begin{align}
    \langle \Delta \omega \rangle_{[\omega]} &\equiv 
    \frac{1}{\left| \Delta S_{\mathrm{KS}}^{+-}(\mathbf{q})\right|_{[\omega]}}
    \nonumber \\
    &\times \int_{[\omega]}\left| S_{\mathrm{KS},\mathrm{r}}^{+-}(\mathbf{q}, \omega) - S_{\mathrm{KS},\mathrm{c}}^{+-}(\mathbf{q}, \omega)\right| d\omega.
    \label{eq:avg freq displ}
\end{align}
Here $S_{\mathrm{KS},\mathrm{r}/\mathrm{c}}^{+-}$ denotes the macroscopic Kohn-Sham spectral function ($\mathbf{G}=0$) calculated using a regular/centered Monkhorst-Pack grids and $[\omega]$ denotes a given choice of frequency integration range. 
The effective absolute change in spectral function intensity, $\left| \Delta S_{\mathrm{KS}}^{+-}(\mathbf{q})\right|_{[\omega]}$, 
is calculated from the gradient of a linear fit to both spectral functions as illustrated in Fig. \ref{fig:ksspectra_comparison}.a. In a similar setup, but where the two spectral functions happened to be straight parallel lines with the same gradient as the linear fit and the same integrated absolute difference between the spectral functions, see Fig. \ref{fig:ksspectra_comparison}.b, this definition exactly corresponds to the horizontal frequency displacement of the two spectral functions, hence the name. 
Now, the idea is to choose a frequency integration range that overlaps with the magnon bandwidth (the actual region of interest for Landau damping) and in which the spectral function is approximately a linear function of frequency. Due to the normalization, the average frequency displacement does not depend on the actual intensity of the low frequency Stoner continuum, which may vary substantially between different materials. As a consequence of the construction illustrated in Fig. \ref{fig:ksspectra_comparison}, $\langle \Delta \omega \rangle$ quantifies the actual frequency displacement of the spectra calculated on differently aligned grids, which should correlate strongly with the discrepancy in magnon peak position between the grids, that is, the quantity we want to converge. This said, the discrepancies between the spectral functions are spurious in nature, and the computed average frequency displacement will vary with the chosen frequency integration range in actual calculations. However, when calculating $\langle \Delta \omega \rangle$ also as an average over different wave vectors $\mathbf{q}$, the spurious effects can be averaged out sufficiently well to make $\langle \Delta \omega \rangle$ stable enough for comparisons of different $k$-point densities and values of $\eta$. The stability towards changes in the frequency integration range is documented in the Supplementary Material. In the main text a frequency integration range of $[-0.4\,\mathrm{eV}, 0.4\,\mathrm{eV}]$ is used for all materials.

\begin{figure*}
    \centering
    \includegraphics{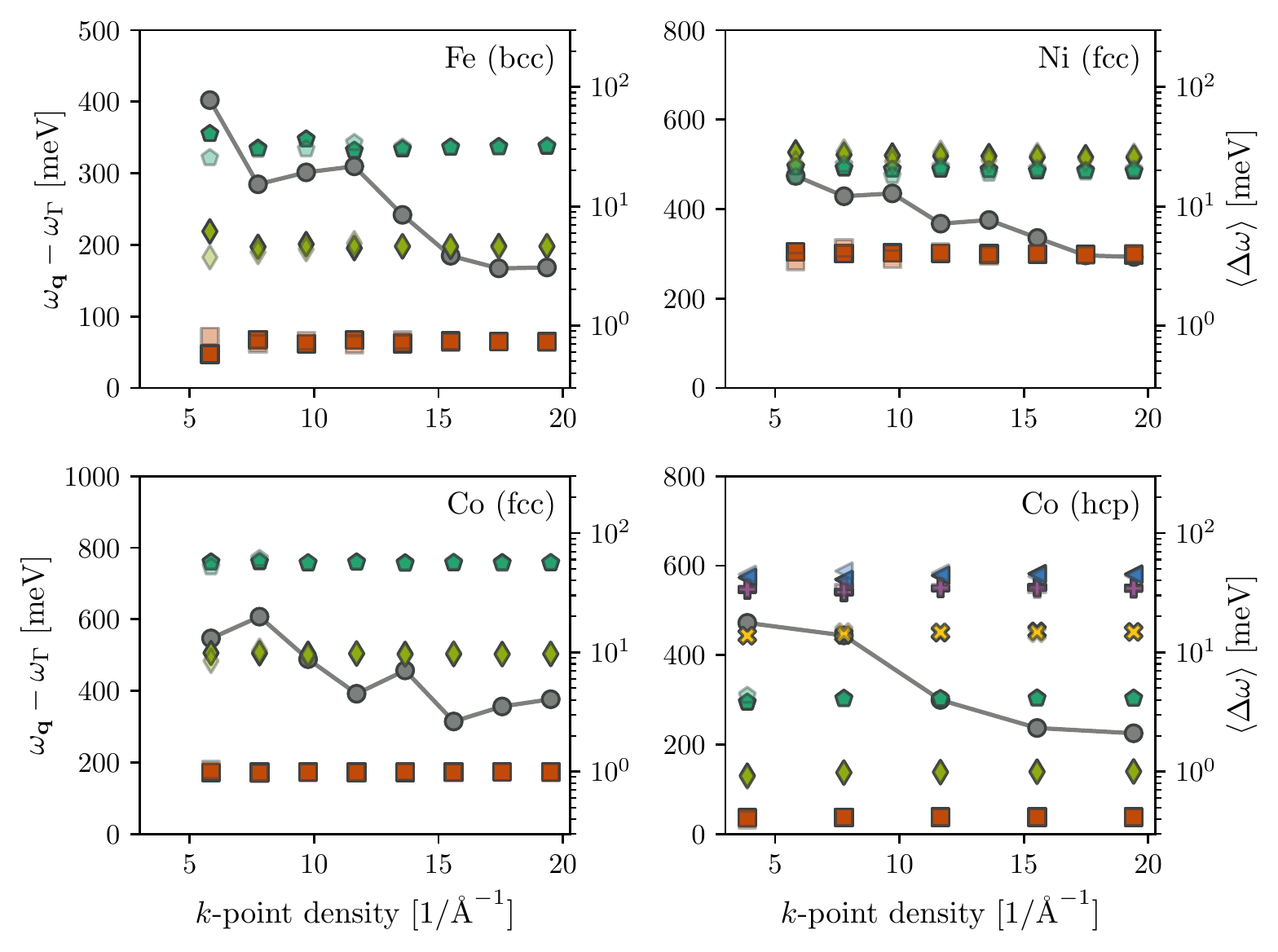}
    \caption{Magnon peak positions relative to the $\Gamma$-peak, in color (left axis), and average displacement frequency, in grey (right axis), as a function of $k$-point density with $\eta=200$ meV. For hcp-Co, the $k$-point density along the $c$-direction is plotted. The colors red, green and teal indicate the magnon peaks at wave vectors $1/3$ of the way, $2/3$ of the way and at the end of the paths $\Gamma\rightarrow \mathrm{N}$, $\Gamma\rightarrow \mathrm{X}$ and $\Gamma\rightarrow \mathrm{A}$ (for bcc, fcc and hcp). The yellow, purple and blue colors indicate similar points on the path $\mathrm{A}\rightarrow\Gamma$ in the second Brillouin Zone of hcp-Co. The opaque and translucent markers represent results calculated using regular and $\Gamma$-centered Monkhorst-Pack grids respectively.
    }
    \label{fig:magdisp_kptdens_convergence}
\end{figure*}
To assess the convergence of the low frequency Kohn-Sham spectrum and the applicability of $\langle \Delta \omega \rangle$ as a method of quantifying the related convergence in magnon peak positions, we have calculated the magnon peak positions at a range of different wave vectors $\mathbf{q}$ in iron, nickel and cobalt at different $k$-point grid densities, using $\eta=200$ meV. The $\mathbf{q}$ wave vectors are all chosen to lie on the same path through the first Brillouin zone, $\Gamma\rightarrow \mathrm{N}$ for bcc-Fe, $\Gamma\rightarrow \mathrm{X}$ for fcc-Ni and fcc-Co and $\Gamma\rightarrow \mathrm{A}$ for hcp-Co. To accurately obtain the magnon peak position, the transverse magnetic excitation spectrum is calculated on a frequency grid with a spacing $\delta \omega \leq \eta / 8$ and the peak position is extracted from a parabolic fit to the spectral function maximum. Along with the magnon peak positions, $\langle \Delta \omega \rangle$ has been calculated averaging over (up to 9) wave vectors on the given path starting $1/3$ of the way to the first Brillouin zone edge, such that the Stoner gap is closed for all the wave vectors in the average. In Fig. \ref{fig:magdisp_kptdens_convergence}, the magnon peak positions calculated on a regular and $\Gamma$-centered Monkhorst-Pack grid are compared as a function of $k$-point density, showing also the calculated values for $\langle \Delta \omega \rangle$. Interestingly, the $k$-point density itself does not seem to influence the overall magnon dispersion. No net change in magnon peak positions is observed as the density is increased, but with increasing grid density, the spurious effects seem to disappear as $\eta=200$ meV becomes sufficient to broaden the low frequency Stoner spectrum into a continuum. Moreover, the disappearance of spurious effects seems strongly correlated with the average frequency displacement. For all materials, the general trend is that the average displacement frequency drops with increasing $k$-point grid density, but not in a monotonic way. We believe that the non-monotonic behaviour reflects the fact that the low frequency Stoner spectrum is highly sensitive to the sampling of Fermi surfaces, which does not only depend on the density of the grid, but also the geometry of the surfaces and how they are situated on the grid. For the same reasons, the spurious displacements of magnon peak positions do not decrease monotonically either and the two trends seem correlated. As an example, both the average frequency displacement and magnon peak position displacement in iron were found to be larger for $k$-point densities of $9.7\, \text{\AA}$ and $11.6\, \text{\AA}$ compared to the grid with density $7.7\, \text{\AA}$. For all materials, $k$-point densities, which result in an average frequency displacement below 8 meV, yield consistent results.

\begin{figure*}
    \centering
    \includegraphics{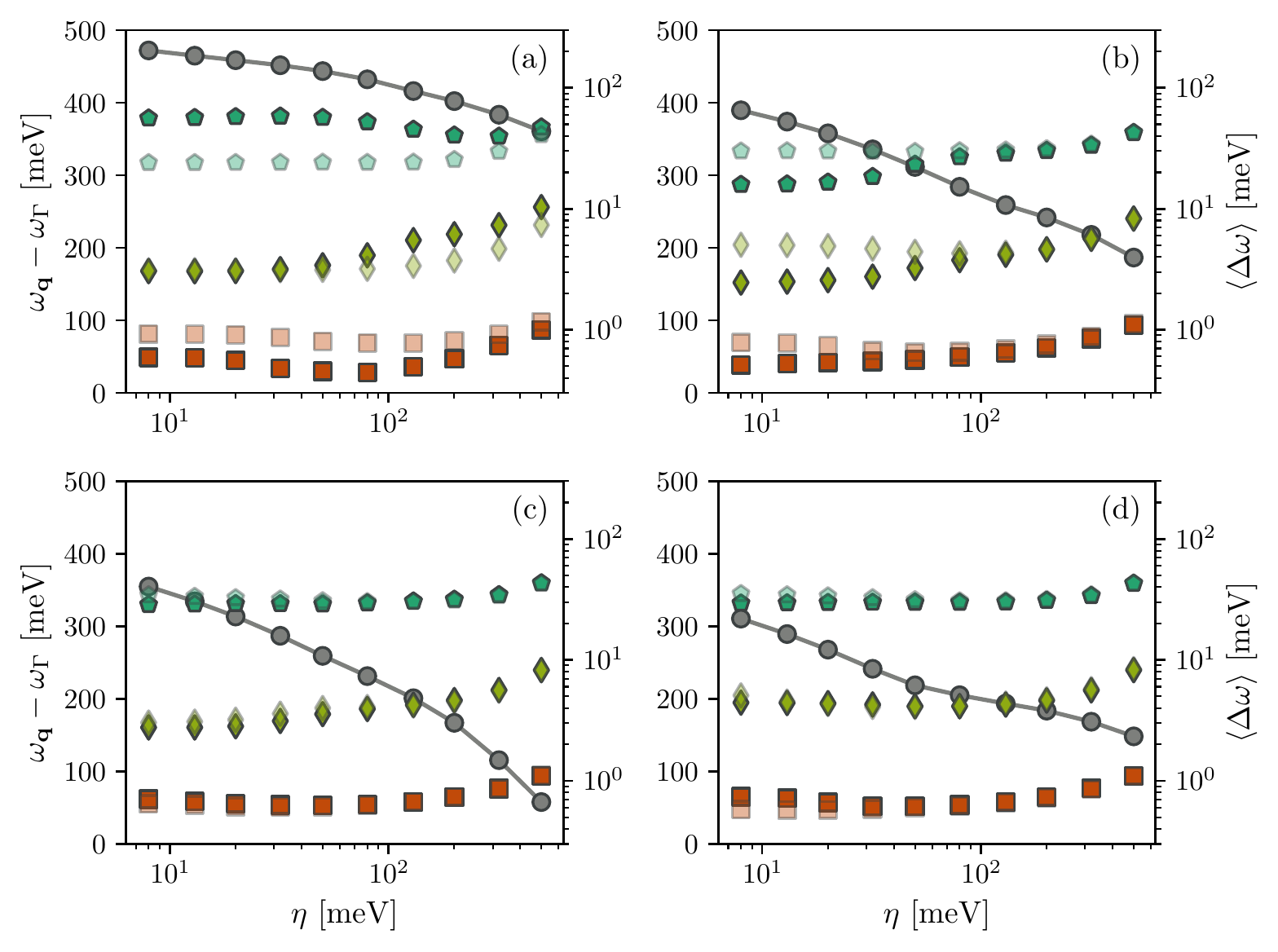}
    \caption{Magnon peak positions of iron relative to the $\Gamma$-peak, in color (left axis), and average displacement frequency, in grey (right axis), as a function of broadening parameter $\eta$. The panels (a), (b), (c) and (d) were calculated using $(18, 18, 18)$, $(42, 42, 42)$, $(54, 54, 54)$ and $(78, 78, 78)$ $k$-point grids respectively. The colors red, green and teal indicate the magnon peaks at wave vectors $1/3$ of the way, $2/3$ of the way and at the end of the path $\Gamma\rightarrow \mathrm{N}$. The opaque and translucent markers represent results calculated using regular and $\Gamma$-centered Monkhorst-Pack grids respectively.}
    \label{fig:magdisp_eta_convergence_bcc-Fe}
\end{figure*}
Now, to further investigate the correlation between the average frequency displacement and the convergence of magnon peak positions, we computed both as a function of broadening parameter $\eta$ for selected $k$-point densities in iron and nickel. In Fig. \ref{fig:magdisp_eta_convergence_bcc-Fe} we show a selection of results for iron, whereas the results for nickel are given in the Supplementary Material. For coarse $k$-point grids, such as in Fig. \ref{fig:magdisp_eta_convergence_bcc-Fe}.a, we never obtain consistency of results between the two different grid alignments, but as the $k$-point density increases, consistency is achieved for a broadening above some threshold $\eta_{\mathrm{t}}$. A similar picture is obtained for nickel, but with magnon frequency discrepancies smaller in magnitude below the threshold $\eta_{\mathrm{t}}$. For both materials, there is consistency of results for all the $k$-point grids and broadening parameters $\eta$ that yield an average frequency displacement $\langle \Delta \omega \rangle \leq 5$ meV. Inductively, this may be used as a criterion to guarantee strictly converged low frequency Stoner spectra. 

To illustrate the use of this criterion, we have computed the average frequency displacement as a function of $\eta$ for a wide selection of $k$-point grids in iron, nickel and cobalt using also different frequency integration ranges. All show a smooth monotonic decrease in $\langle \Delta \omega \rangle$ as a function of $\eta$, similar to the behaviour shown in Fig. \ref{fig:magdisp_eta_convergence_bcc-Fe}. In the Supplementary Material, we supply a table of threshold values $\eta_{\mathrm{t}}$ corresponding to the intersection with $\langle \Delta \omega \rangle = 5$ meV found by linear interpolation. As an example, we find $\eta_{\mathrm{t}} = 126$ meV for iron with the $(54, 54, 54)$ $k$-point grid shown in Fig. \ref{fig:magdisp_eta_convergence_bcc-Fe}.c and $\eta_{\mathrm{t}} = 87$ meV with the $(78, 78, 78)$ $k$-point grid shown in Fig. \ref{fig:magdisp_eta_convergence_bcc-Fe}.d, both using a frequency integration range of $[-0.4\,\mathrm{eV}, 0.4\,\mathrm{eV}]$. Using $[\omega] = [-0.2\,\mathrm{eV}, 0.2\,\mathrm{eV}]$ and $[\omega] = [-0.6\,\mathrm{eV}, 0.6\,\mathrm{eV}]$ instead, threshold values of $\eta_{\mathrm{t}} = 150$ meV, $\eta_{\mathrm{t}} = 127$ meV and $\eta_{\mathrm{t}} = 92$ meV, $\eta_{\mathrm{t}} = 105$ meV are obtained for the two different $k$-point densities. The variations with frequency integration range are small enough to make the general approach applicable as a computationally cheap rule of thumb, but 
in the general case one should mostly use it as a starting point for a more careful analysis. Depending on the desired accuracy, a more relaxed criterion of $\langle \Delta \omega \rangle \leq 8$ meV should yield converged magnon peak positions, except for a few cases, and if only the general trends are important, not the actual peak positions themselves, an even larger threshold could be applied to achieve spectra similar to the one shown in Fig. \ref{fig:trans_mag_susc_eta_example}.a. In the context at present, we want to eliminate spurious effects in the magnon peak positions all together to enable the conduction of a convergence study in other numerical parameters and to obtain magnon dispersions that are suitable for benchmarking against literature. To achieve this, we apply the strict $\langle \Delta \omega \rangle \leq 5$ meV criterion.

\begin{figure*}
    \centering
    \includegraphics{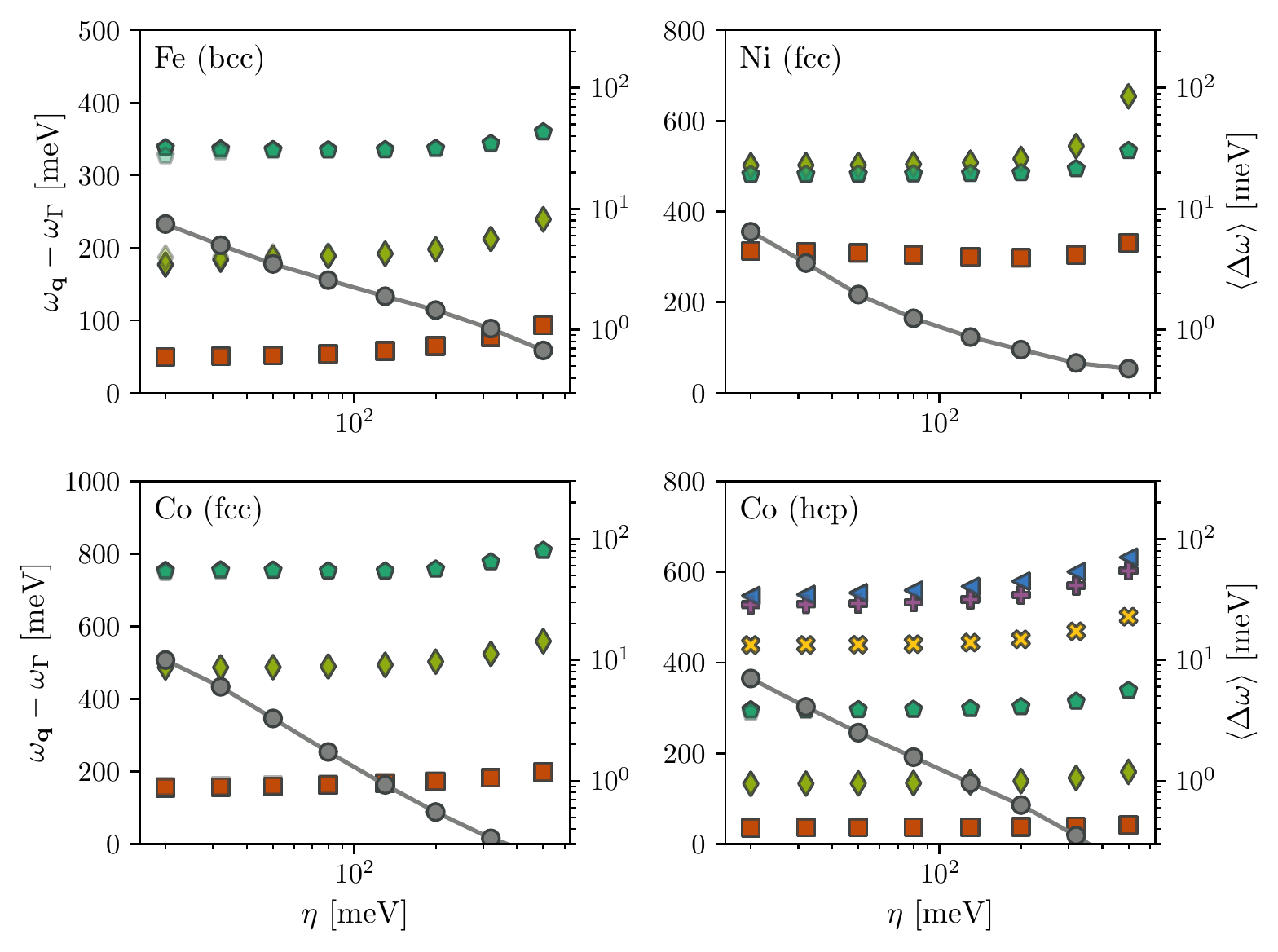}
    \caption{Magnon peak positions relative to the $\Gamma$-peak, in color (left axis), and average displacement frequency, in grey (right axis), as a function of broadening parameter $\eta$. Calculations were performed on a $(102, 102, 102)$ $k$-point grid for bcc-Fe, fcc-Ni and fcc-Co, and a $(84, 84, 48)$ grid for hcp-Co. The colors red, green and teal indicate the magnon peaks at wave vectors $1/3$ of the way, $2/3$ of the way and at the end of the paths $\Gamma\rightarrow \mathrm{N}$, $\Gamma\rightarrow \mathrm{X}$ and $\Gamma\rightarrow \mathrm{A}$ (for bcc, fcc and hcp). The yellow, purple and blue colors indicate similar points on the path $\mathrm{A}\rightarrow\Gamma$ in the second Brillouin Zone of hcp-Co. The opaque and translucent markers represent results calculated using regular and $\Gamma$-centered Monkhorst-Pack grids respectively.}
    \label{fig:magdisp_eta_convergence}
\end{figure*}
So far, we only discussed the effect of $\eta$ on the grid alignment consistency, but clearly $\eta$ also has an effect on the overall magnon dispersion, as seen in Fig. \ref{fig:magdisp_eta_convergence_bcc-Fe}. Even though we use a finite $\eta>0$ to broaden the Stoner spectrum into a continuum, we need also to choose it small enough that $\eta$ itself does not influence the overall dispersion. To find out how small an $\eta$ that is, we computed the magnon peak positions as a function of $\eta$ for all four materials on dense $k$-point grids, where an $\langle \Delta \omega \rangle \leq 5$ meV criterion leads to $\eta_{\mathrm{t}} = 32$ meV, $\eta_{\mathrm{t}} = 26$ meV, $\eta_{\mathrm{t}} = 39$ meV and $\eta_{\mathrm{t}} = 28$ meV for bcc-Fe, fcc-Ni, fcc-Co and hcp-Co respectively. These results are presented in Fig. \ref{fig:magdisp_eta_convergence}. It seems to be a general trend, that the magnon peak positions shift to higher energies as $\eta$ is increased. In fact, a broadening parameter less than $120$ meV is needed in order to achieve a good convergence, except for a few points that require a value as low as $\eta=50$ meV. Together with the spurious discretization effetcs, this requires us to use quite dense $k$-point grids. In order to use $\eta=50$ meV within the $\langle \Delta \omega \rangle \leq 5$ meV criterion, a $(90, 90, 90)$ $k$-point grid is needed for bcc-Fe, a $(84, 84, 84)$ grid for fcc-Ni and fcc-Co and a $(60, 60, 30)$ grid for hcp-Co. For the materials investigated here, performing such dense $k$-point samplings does not itself pose any computational problem, as there are at most 2 atoms in the unit cell. For larger systems however, grids that dense will quickly be prohibitive. To circumvent this problem, one can either apply analytic continuation to an alignment consistent calculation performed with a large broadening parameter $\eta$, or refine the $k$-point summation in Eq. \eqref{eq:dyn. trans. mag. plane wave KS susc. tensor lehmann} using methods such as linear tetrahedron interpolation in order to improve the continuum description of the Stoner spectrum.

\subsection{Gap error convergence}\label{sec:gap error convergence}

\begin{figure}[ht]
    \centering
    \includegraphics{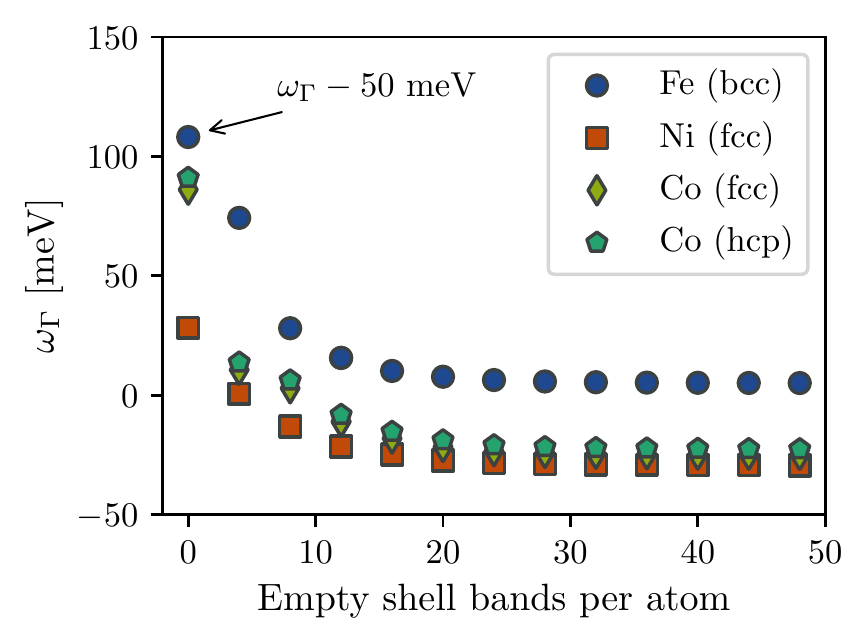}
    \caption{Magnon peak position at the $\Gamma$-point in iron, nickel and cobalt as a function of the number of empty shell bands per atom included in the band summation of Eq. \eqref{eq:dyn. trans. mag. plane wave KS susc. tensor lehmann}. Calculations were performed on a $(54, 54, 54)$ regular Monkhorst-pack grid for bcc and fcc structures and a $(48, 48, 30)$ grid for hcp-Co. For all materials, a broadening parameter of $\eta=200$ meV was used. For iron, $\omega_{\Gamma} - 50$ meV is plotted in order for all the points to be visible on a single axis.}
    \label{fig:gaperr_unocc_convergence}
\end{figure}
As our treatment of the transverse magnetic susceptibility is collinear, all the itinerant ferromagnetic materials of this study should have a so-called Goldstone mode with a macroscopic magnon peak at $\omega_{\mathbf{q}=\mathbf{0}}=0$. In reality though, this is not necessarily guaranteed numerically for linear response TDDFT calculations, and transverse magnetic excitation spectra, such as the one shown in Fig. \ref{fig:trans_mag_susc_eta_example}, display finite gap errors $\omega_{\Gamma} \neq 0$. In literature\cite{Buczek2011b,Lounis2011,Rousseau2012}, the gap error is usually attributed to numerical approximations as well as inconsistencies between the Kohn-Sham susceptibility and the exchange-correlation kernel. Regarding the latter, one needs to use an exchange-correlation kernel that in the static limit gives the same ground state spin-densities as the ground state DFT calculation, on the basis of which the Kohn-Sham susceptibility is computed. Otherwise, $\big(\delta W^{\mu}_{\mathrm{s}}(\mathbf{r}, t)\big)$ cannot be considered a perturbative quantity, so that the linear response relation \eqref{eq:kohn-sham four-component response relation} and consequently also the Dyson equation \eqref{eq:four-comp. Dyson real space} no longer holds. As an example, using an ALDA kernel on top of a GGA ground state calculation will result in a gap error, why we are restricted to the use of LDA for the ground state at present. In many-body perturbation theory similar considerations have to be made\cite{Muller2016}.

\begin{figure}[ht]
    \centering
    \includegraphics{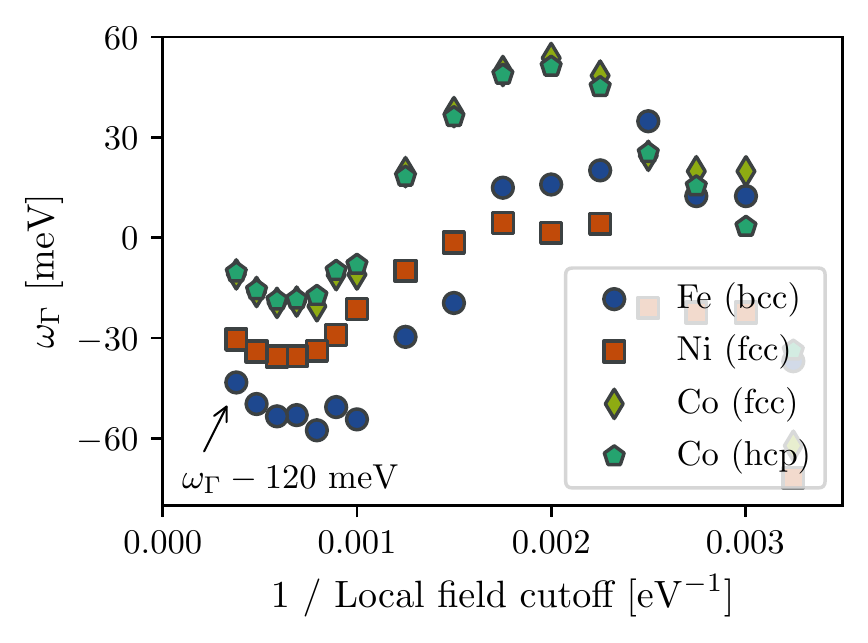}
    \caption{Magnon peak position at the $\Gamma$-point in iron, nickel and cobalt as a function of the inverse local field cutoff. Calculations were performed on a $(54, 54, 54)$ regular Monkhorst-pack grid for bcc and fcc structures and a $(48, 48, 30)$ grid for hcp-Co, for all materials using a broadening parameter of $\eta=200$ meV. For iron, $\omega_{\Gamma} - 120$ meV is plotted in order for all the points to be visible on a single axis.}
    \label{fig:gaperr_ecut_convergence}
\end{figure}
For our calculations, we have identified two main numerical parameters that need to be converged in order to minimize the gap error, namely the truncation in band summation and plane wave representation of the Kohn-Sham susceptibility. Neither the $k$-point density nor the broadening parameter, $\eta$, investigated above had any significant influence on $\omega_{\Gamma}$ due to the Stoner gap. In Fig. \ref{fig:gaperr_unocc_convergence}, we show the gap error as a function of the number of empty shell bands per atom. For all four materials the convergence follows a similar pattern in which the gap error falls off as the number of bands is increased and beyond 20 empty shell bands per atom or so, the gap error can be considered to be converged. However, it does not vanish, which in part is due to the plane wave cutoff of $1000$ eV used in these calculations. In Fig. \ref{fig:gaperr_ecut_convergence} we present the gap error dependence on the plane wave representation. Unfortunately, the gap error does not converge even at cutoffs as high as $3600$ eV. Extrapolating the trend at high cutoffs, it seems that one in principle would need an infinite cutoff to converge the gap error, and even so, the gap error does not seem to vanish completely, especially in the case of iron. Using the extended PAW setup for nickel, where also the 3$p$ electronic orbitals are included as valence states in the band summation of Eq. \eqref{eq:dyn. trans. mag. plane wave KS susc. tensor lehmann}, slows down the gap error convergence even more, but yields a smaller gap error for a plane wave cutoff extrapolated to infinity. Based on these results, it would seem that in order to eliminate the gap error altogether, one would need to drop the frozen core approximation, use an infinite plane wave cutoff and possibly also improve the all-electron partial wave completeness of the PAW datasets. This is bad news, of course, but there are several practical ways to circumvent these limitations. As an example, one can invert the Dyson equation \eqref{eq:dyson trans mag plane wave susc.} in another basis set than plane waves, a strategy previously shown to yield smaller gap errors than the ones reported here\cite{Buczek2011b}. Additionally, different strategies have been developed to enforce Goldstone's theorem by introducing information about the exchange-correlation kernel into the Kohn-Sham susceptibility\cite{Rousseau2012} or vice-versa\cite{Buczek2011b,Lounis2011} and in that way achieve the consistency needed to guarantee a Goldstone mode.

\subsection{Magnon dispersion convergence}\label{sec:dispersion convergence}

\begin{figure*}
    \centering
    \includegraphics{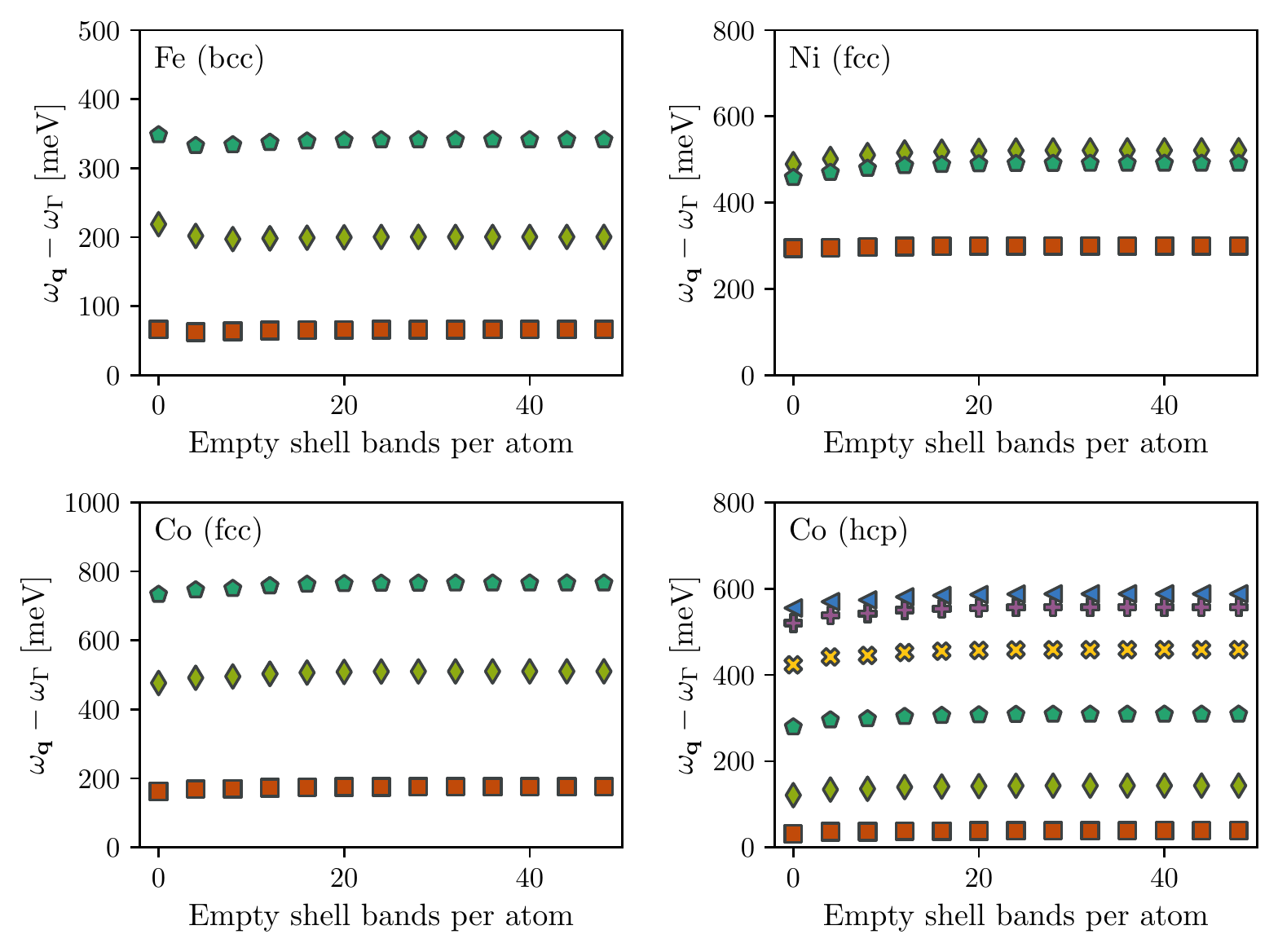}
    \caption{Magnon peak positions relative to the $\Gamma$-peak as a function of the number of empty shell bands per atom included in the band summation of Eq. \eqref{eq:dyn. trans. mag. plane wave KS susc. tensor lehmann}. 
    The colors red, green and teal indicate the magnon peaks at wave vectors $1/3$ of the way, $2/3$ of the way and at the end of the paths $\Gamma\rightarrow \mathrm{N}$, $\Gamma\rightarrow \mathrm{X}$ and $\Gamma\rightarrow \mathrm{A}$ (for bcc, fcc and hcp). The yellow, purple and blue colors indicate similar points on the path $\mathrm{A}\rightarrow\Gamma$ in the second Brillouin Zone of hcp-Co.}
    \label{fig:magdisp_unocc_convergence}
\end{figure*}
As shown above, we are able to converge the gap error $\omega_{\Gamma}$ within a finite band summation, but not within a finite plane wave representation. A natural question arises: Can we converge the magnon dispersion itself? To investigate this, we have computed the magnon peak positions for a set of wave vectors in all four materials and as a function of empty shell bands per atom and plane wave cutoff. Generally, the gap error itself should not strongly influence the magnon dispersion. However, it is important for the Landau damping that the transverse magnetic excitation spectrum and the Stoner continuum is appropriately aligned as a function of frequency. For the magnon dispersion convergence, we have used a broadening parameter of $\eta=200$ meV and applied a $(54, 54, 54)$ regular Monkhorst-Pack grid for the bcc and fcc structures, while a $(48, 48, 30)$ grid has been used for hcp-Co. With these grids, we satisfy the $\langle \Delta \omega \rangle \leq 5$ meV criterion. Even though $\eta$ itself is not converged, the effect of the broadening parameter seen in Fig. \ref{fig:magdisp_eta_convergence} is sufficiently smooth, that we believe the results to be transferable to lower broadening. After extracting the magnon peak positions from the transverse magnetic excitation spectrum, we shift the peak positions by $\omega_{\Gamma}$ to minimize the effect of the gap error convergence on the convergence of the full dispersion.

In Fig. \ref{fig:magdisp_unocc_convergence}, the magnon dispersion convergence as a function of empty shell bands per atom is presented. Clearly, the magnon dispersion only weakly depends on inclusion of excited states above the $3d$-shell and above approximately 12 empty shell bands per atom, the magnon dispersion can be considered well converged. Even without empty shell bands, a good description of the overall magnon dispersion is achieved. This is reassuring for the scalability to larger systems and shows that the band summation in excited states is not a practical limitation in linear response TDDFT for magnon spectroscopy.

\begin{figure*}
    \centering
    \includegraphics{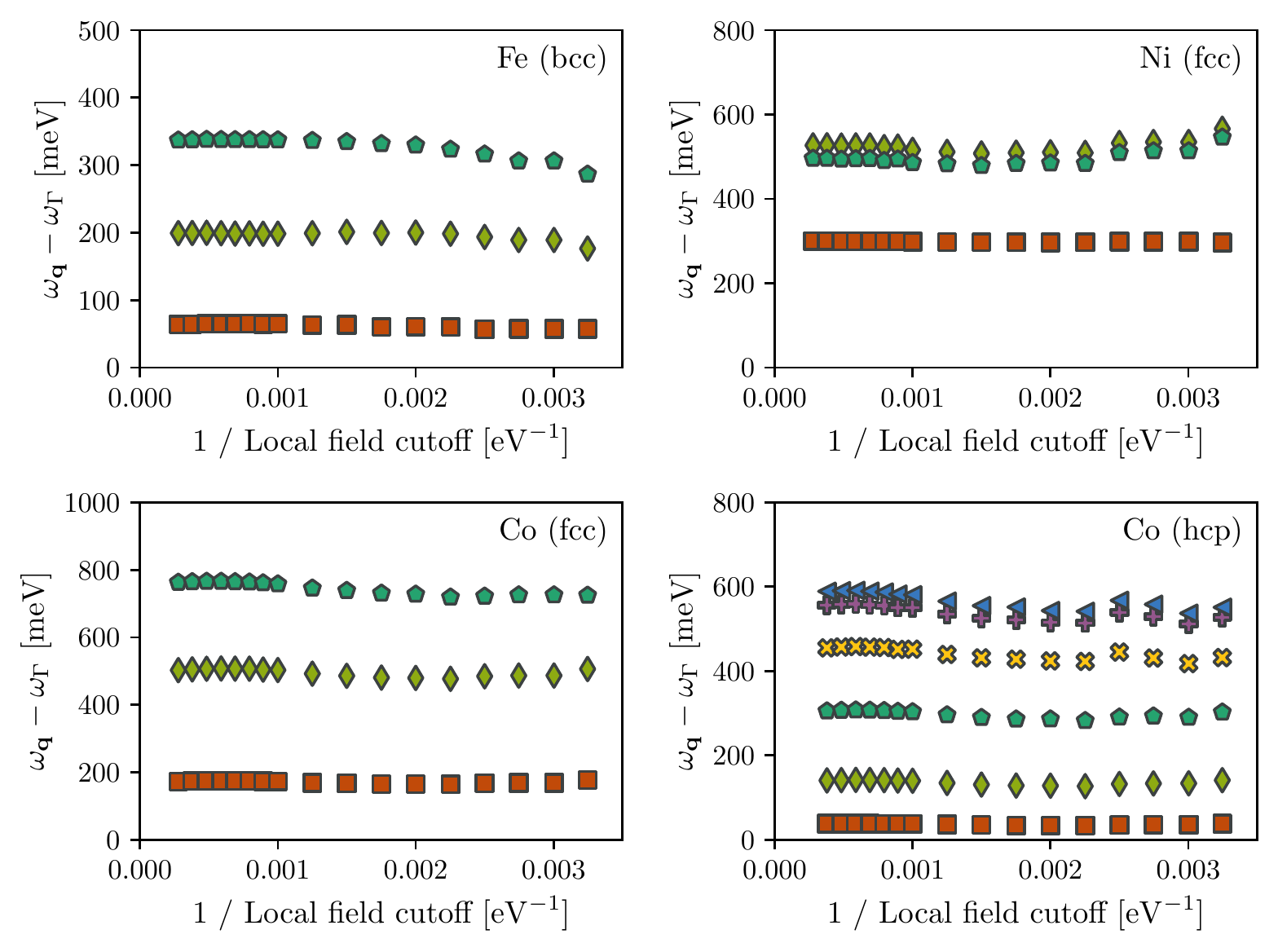}
    \caption{Magnon peak positions relative to the $\Gamma$-peak as a function of inverse local field cutoff.
    The colors red, green and teal indicate the magnon peaks at wave vectors $1/3$ of the way, $2/3$ of the way and at the end of the paths $\Gamma\rightarrow \mathrm{N}$, $\Gamma\rightarrow \mathrm{X}$ and $\Gamma\rightarrow \mathrm{A}$ (for bcc, fcc and hcp). The yellow, purple and blue colors indicate similar points on the path $\mathrm{A}\rightarrow\Gamma$ in the second Brillouin Zone of hcp-Co.}
    \label{fig:magdisp_ecut_convergence}
\end{figure*}
The magnon dispersion convergence in plane wave representation is presented in Fig. \ref{fig:magdisp_ecut_convergence}. In comparison to the gap error, it is much easier to converge the magnon dispersion in terms of the plane wave cutoff as variations in the relative magnon peak positions become insignificant above a 1000 eV cutoff. For nickel with the extended PAW setups, we need a cutoff of $1450$ eV to converge the magnon peak positions, yielding only small differences $\leq1.7\%$ from the minimal setup, as previously discussed. This illustrates the usefulness of a gap error correction scheme. For a given cutoff, the spectra can be shifted such that the Goldstone condition of $\omega_\mathbf{q}=0$ is satisfied and one is then not limited by the slow gap error convergence. This implies that the numerical scheme can be considered exact up to the limitations in PAW projectors and frozen core states discussed above. However, the convergence study also illustrates an important disadvantage of the present implementation. Even though  we are able to circumvent the slow convergence of the gap error, a plane wave cutoff of 1000 eV becomes prohibitive for larger structures. The Dyson equation \eqref{eq:dyson trans mag plane wave susc.} is expressed in matrices that scale in size with the number of plane wave coefficients squared and as a result, the memory requirements quickly become a computational bottleneck. Nevertheless, the results in Fig. \ref{fig:magdisp_ecut_convergence} illustrate that less accurate, yet qualitatively correct magnon dispersions can be extracted at significantly smaller plane wave cutoffs - especially when using minimal PAW setups. Once again, a different representation of the spatial coordinates in the Dyson equation may help to overcome this problem, but even within the limitations of the plane wave representation and present computational resources, the ALDA transverse magnetic susceptibility can be calculated for a wide range of collinear materials.

\section{Results}\label{sec:results}

On the basis of the convergence study above, we have computed the transverse magnetic excitation spectrum of bcc-Fe, fcc-Ni, fcc-Co and hcp-Co within the ALDA. For these calculations, 12 empty shell bands per atom were used in the band summation of Eq. \eqref{eq:dyn. trans. mag. plane wave KS susc. tensor lehmann} and a 1000 eV plane wave cutoff was used in the plane wave representation of the Dyson equation \eqref{eq:dyson trans mag plane wave susc.}. Furthermore, a constant frequency shift was applied in order to fulfill the Goldstone condition. To converge the magnon dispersion for reduced wave vectors $\mathbf{q}$ inside the low frequency Stoner continuum, a broadening parameter of $\eta=50$ meV was used as well as $(90, 90, 90)$, $(84, 84, 84)$ and $(60, 60, 30)$ $\Gamma$-centered Monkhorst-Pack $k$-point grids for the bcc, fcc and hcp structures respectively. Below the Stoner continuum, where the acoustic magnon mode is free of Landau damping, the magnon peak positions do not depend on the broadening, and the the limit $\eta \rightarrow 0^+$ should be taken. To resolve the full magnon spectrum in a single figure, we do this in an approximate fashion by letting $\eta$ be $\mathbf{q}$-dependent. We increase $\eta$ quadratically as a function of $|\mathbf{q}|$ from $\eta=5$ meV at $\mathbf{q}=\mathbf{0}$ to $\eta=50$ meV at a threshold $q_{\mathrm{t}}$. For wave vectors $|\mathbf{q}| > q_{\mathrm{t}}$, $\eta$ is held constant. We use a threshold $q_{\mathrm{t}}$ of $|\mathbf{q}_{\mathrm{N}}| / 3$, $|\mathbf{q}_{\mathrm{L}}| / 3$ and $|\mathbf{q}_{\mathrm{M}}| / 3$ for the bcc, fcc and hcp structures respectively. 

\subsection{Fe (bcc)}\label{sec:bcc-Fe}

\begin{figure*}
    \centering
    \includegraphics{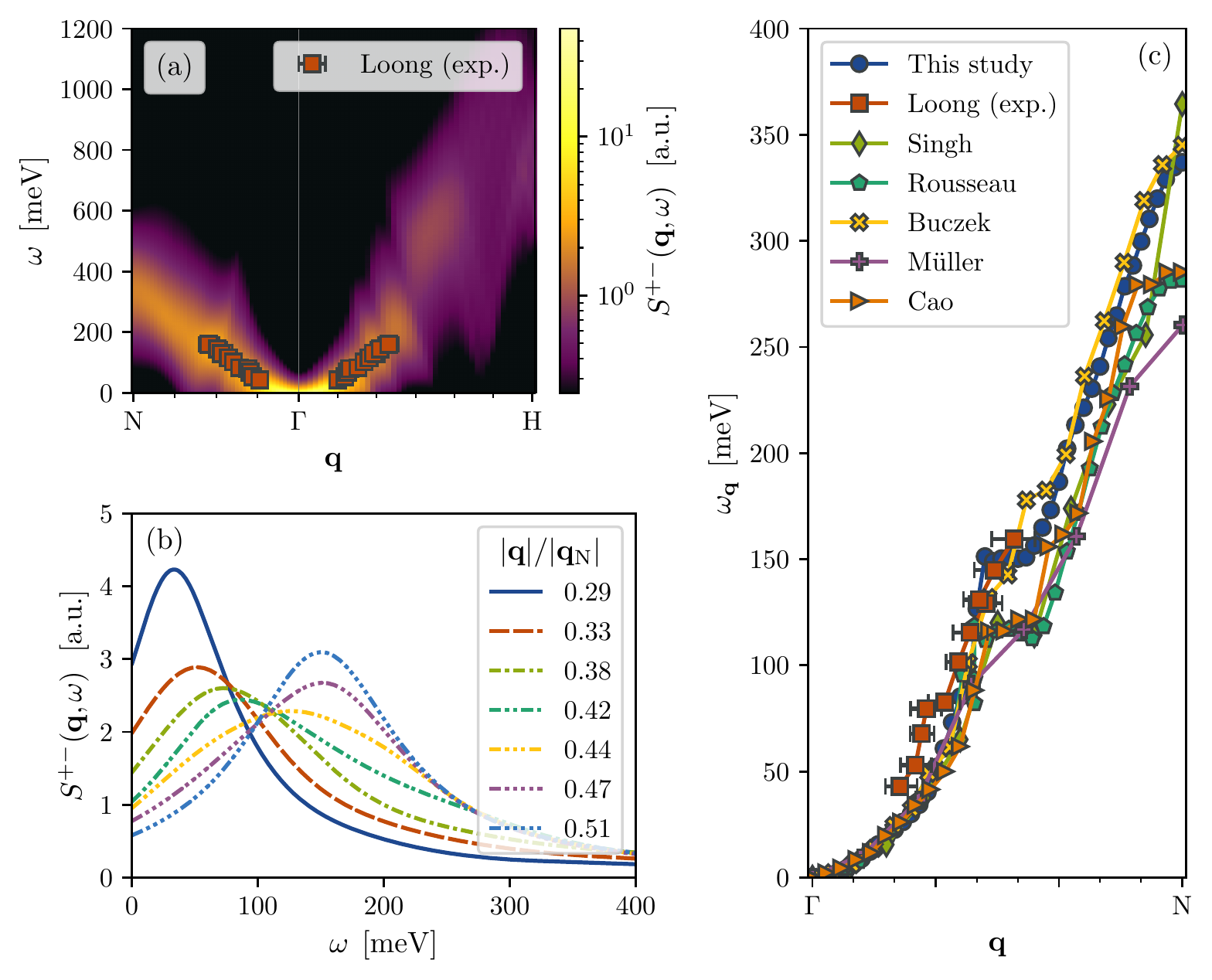}
    \caption{Transverse magnetic excitation spectrum of bcc-Fe computed within the ALDA. In (a) the macroscopic (unit-cell averaged) $\mathbf{G}=0$ component of the spectrum is shown as a heat map and function of wave vector $\mathbf{q}$ and frequency $\omega$. The spectrum was computed on a N-$\Gamma$-H band path and is compared to inelastic neutron scattering data\cite{Loong1984}. In (b) the spectral intensity is shown as a function of frequency for a range of fixed values for $\mathbf{q}$ along the $\Gamma$-N path. In (c) the magnon peak positions extracted along the $\Gamma$-N path are shown and compared to experimental\cite{Loong1984} as well as \textit{ab initio} references\cite{Singh2018,Rousseau2012,Buczek2011b,Muller2016,Cao2017}.}
    \label{fig:bcc-Fe_magnon_spectrum-and-dispersion}
\end{figure*}
For bcc-Fe, applying the LDA and using the experimental lattice constant of $a=2.867\text{ \AA}$, we obtain a ferromagnetic ground state with a spatially averaged spin-polarization of $2.20$ $\mu_{\mathrm{B}}$ per iron atom. In Fig. \ref{fig:bcc-Fe_magnon_spectrum-and-dispersion}.a we present the calculated macroscopic transverse magnetic excitation spectrum as a function of wave vector $\mathbf{q}$ and compare it to inelastic neutron scattering (INS) data gathered in the $[1\bar{1}0]$ scattering plane\cite{Loong1984}. The transverse magnetic excitation spectrum has been corrected for a gap error of $\omega_{\Gamma} = 65.6$ meV. The experimental comparison is made to the same dataset in both the $\Gamma\rightarrow\mathrm{N}$ and $\Gamma\rightarrow\mathrm{H}$ directions, as the experimentally observed magnon dispersion is isotropic for frequencies up to at least 120 meV\cite{Mook1973}. For wave vectors shorter than $0.5\,\text{\AA}^{-1}$, the magnon dispersion in our transverse magnetic excitation spectrum is completely isotropic. At $0.5\,\text{\AA}^{-1}$ the dispersion in magnon peak positions flattens out in the $\Gamma \rightarrow \mathrm{H}$ direction, before making a jump to a plateau around 140 meV, where the magnon dispersion takes a negative slope. The first jump is shortly followed by a second jump to a new plateau, again with a decreasing magnon frequency from 215 meV at $0.88\,\text{\AA}^{-1}$ to 180 meV at $1.07\,\text{\AA}^{-1}$. At this point, the dispersion makes a third jump to 500 meV and the lineshape gets severely broadened. There continues to be a well-defined peak position up to $q\sim 1.5\,\text{\AA}^{-1}$, where the magnon frequency is 600 meV, but beyond this point the spectrum becomes dominated by the low frequency Stoner excitations and it is not possible to discern a collective magnon mode. 
This is in contrast to the $\Gamma \rightarrow \mathrm{N}$ direction, in which the magnon mode remains well-defined throughout the entire first Brillouin Zone with a single plateau around 150 meV and a total bandwidth of 337 meV. The observed jumps in magnon dispersion as well as the disappearance of the magnon mode in the $\Gamma \rightarrow \mathrm{H}$ direction agree well with previous theoretical results\cite{Friedrich2014,Friedrich2020,Buczek2011b,Rousseau2012,Cao2017,Singh2018}. The magnon frequency jumps arise because the magnon mode crosses stripe-like features in the Kohn-Sham spectrum corresponding to well-defined Stoner excitations residing below the main Stoner continuum. The appearance of stripe-like features is an itinerant electrons effect and is further discussed in the context of fcc-Ni in the following section as well as in the work of Friedrich and coworkers\cite{Friedrich2020}. Experimentally, a significant intensity drop has been reported for wave vectors longer than $0.6\,\text{\AA}^{-1}$\cite{Mook1973}, but a full experimental picture is not available as the present data is restricted to frequencies below 160 meV. In the frequency range available, the ALDA transverse magnetic excitation spectrum seems to match the experimentally extracted magnon dispersion well.

In Fig. \ref{fig:bcc-Fe_magnon_spectrum-and-dispersion}.c, the extracted dispersion in magnon peak positions along the $\Gamma \rightarrow \textrm{N}$ direction is compared with experimental as well as \textit{ab initio} references. Singh\cite{Singh2018}, Rousseau\cite{Rousseau2012}, Buczek\cite{Buczek2011b}, and coworkers use different implementations of the LR-TDDFT methodology in the ALDA, removing the gap error by applying a constant frequency shift, adding a corrective contribution to $\chi^{+-}_{\mathrm{KS}}$ and forcing the smallest $\mathbf{q}=\mathbf{0}$ energy eigenvalue of $\chi^{+-}$ to zero, respectively. M{\"{u}}ller and coworkers\cite{Muller2016} apply MBPT in the LDA, but with an \textit{ad hoc} adjustment of the exchange splitting to remove the gap error. Cao and coworkers\cite{Cao2017} apply the LDA to TD-DFPT, which does not suffer from any gap error. At short wave vectors, all theoretical dispersion relations agree nicely, but for wave vectors longer than $|\mathbf{q}|=0.3\,|\mathbf{q}_{\mathrm{N}}|$, the Stoner continuum starts to skew the magnon lineshape and discrepancies between results start to form. Similar to the magnon dispersion presented here, Singh, Rousseau and Cao all report a plateau midway between the $\Gamma$ and $\textrm{N}$ points, but at lower energies than the plateau we find. The upper plateau frequency seems to match better the experimental dispersion, however it is unclear from the experimental evidence, whether there should be a plateau or not. Buczek and collaborators report an overall magnon dispersion that agrees very well with our results, except that is does not display a frequency plateau. 
Finally, a wide range of values are reported for the bandwidth among the different theoretical methods.

Most likely, the discrepancies between theoretical (A)LDA results arise from details in the representation of the Stoner continuum. In Fig. \ref{fig:bcc-Fe_magnon_spectrum-and-dispersion}.b, we present the transverse magnetic excitation spectrum for wave vectors below the plateau and around the onset of the plateau. Just below the plateau, the magnon peak intensity is attenuated as the lineshape attains a long tail towards higher frequencies, resembling the magnon lineshapes of wave vectors on the plateau itself. On the plateau, the magnon lineshape more closely resembles a Lorentzian with a less pronounced Landau damping. In this way, the plateau shape is intimately related to the low frequency Stoner continuum, which is sensitive to both broadening procedure and \textit{k}-point sampling, as shown above, as well as details in the DFT ground state calculation. Hopefully, the rigorous convergence analysis presented here can be a step towards resolving some of the discrepancies between different implementations in regards of the former. Concerning the DFT ground states, there are discrepancies already in the ground state magnetization reported. Singh, Rousseau, M{\"{u}}ller, Cao and collaborators reports values for the LDA average spin-polarization of $2.00$ $\mu_{\mathrm{B}}$, $2.11$ $\mu_{\mathrm{B}}$, $2.20$ $\mu_{\mathrm{B}}$ and $2.16$ $\mu_{\mathrm{B}}$ within their respective ground state methodologies. This 
implies quantitatively different Fermi surfaces, which will influence the low frequency Stoner continuum and the magnon modes embedded in it. Furthermore, the gap error correction procedure can affect the frequency alignment of magnon mode and Stoner continuum, which may also influence the magnon dispersion. 

\subsection{Ni (fcc)}
\begin{figure*}
    \centering
    \includegraphics{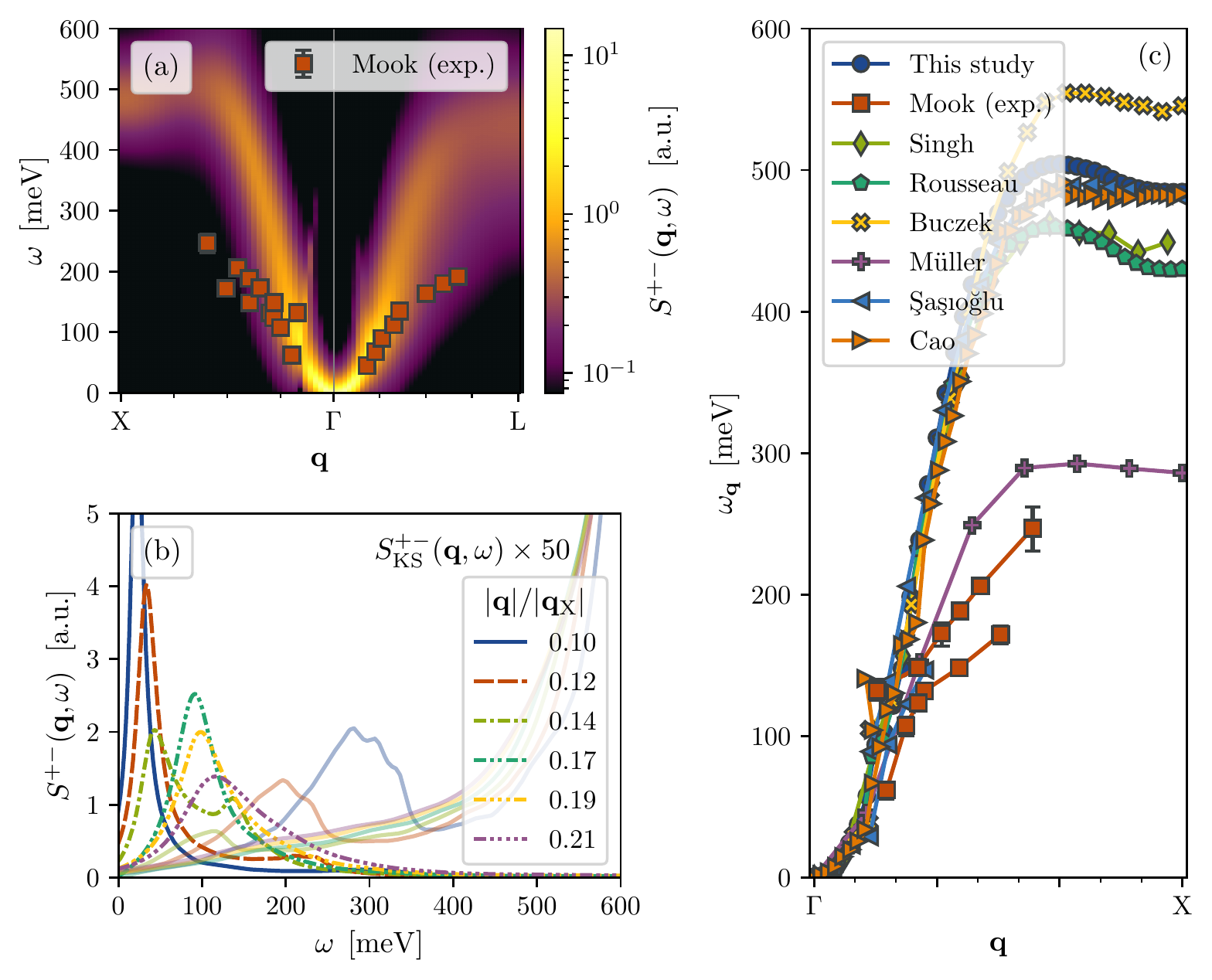}
    \caption{Transverse magnetic excitation spectrum of fcc-Ni computed within the ALDA. In (a) the macroscopic (unit-cell averaged) $\mathbf{G}=0$ component of the spectrum is shown as a heat map and function of wave vector $\mathbf{q}$ and frequency $\omega$. The spectrum was computed on a X-$\Gamma$-L band path and is compared to inelastic neutron scattering data\cite{Mook1985}. In (b) the spectral intensity is shown as a function of frequency for a range of fixed values for $\mathbf{q}$ along the $\Gamma$-X path. The corresponding Kohn-Sham spectrum of Stoner excitations (multiplied in intensity by a factor of $25$) is shown as translucent lines. In (c) the magnon peak positions extracted along the $\Gamma$-X path are shown and compared to experimental\cite{Mook1985} as well as \textit{ab initio} references\cite{Singh2018,Rousseau2012,Buczek2011b,Muller2016,SasIoglu2010,Cao2017}.
    }
    \label{fig:fcc-Ni-setup10_magnon_spectrum-and-dispersion}
\end{figure*}
In Fig. \ref{fig:fcc-Ni-setup10_magnon_spectrum-and-dispersion}.a, we present the transverse magnetic excitation spectrum of ferromagnetic fcc-Ni. The spectrum is based on a LDA ground state calculation with lattice constant $a=3.524\,\text{\AA}$, resulting in an average spin-polarization per nickel atom of $0.627\,\mu_{\textrm{B}}$. The spectrum is presented as a function of wave vector $\mathbf{q}$ along the X-$\Gamma$-L path and is compared to the magnon dispersion as measured by inelastic neutron scattering\cite{Mook1985}. A gap error of $\omega_{\Gamma}=-21.5\,\textrm{meV}$ was accounted for. The magnon dispersion extracted from the transverse magnetic excitation spectrum is isotropic for small wave vectors, but at $\mathbf{q}=0.17\,\mathbf{q}_{\mathrm{X}}$ ($q=0.3\,\text{\AA}^{-1}$) there is a sudden increase in the magnon frequency, which is not present in the $\Gamma\rightarrow \mathrm{L}$ direction. For wave vectors longer than $q=0.3\,\text{\AA}^{-1}$, the magnon dispersion remains slightly anisotropic. 
The acoustic magnon mode remains well-defined in both directions all the way to the first Brillouin Zone edge, although the spectral width of the mode is more severely broadened due to Landau damping along the $\Gamma\rightarrow\mathrm{L}$ direction for long wave vectors. Along the $\Gamma\rightarrow \mathrm{X}$ path, the magnon dispersion attains a maximum frequency of 504 meV at $q=1.19\,\text{\AA}^{-1}$ before decreasing to a value of 484 meV at the BZ edge. Along the $\Gamma \rightarrow \mathrm{L}$ direction, the magnon frequency is maximal at the BZ edge itself resulting in a bandwidth of 441 meV.

Except for short wave vectors along the $\Gamma\rightarrow\mathrm{L}$ direction, the computed magnon excitation spectrum fails to reproduce the experimentally observed magnon dispersion. The ALDA treatment results in a significantly more dispersive magnon mode compared to experiment, and where two coexisting modes are observed experimentally along the $\Gamma\rightarrow\mathrm{X}$ direction, we observe mostly just one. In accordance with previous (A)LDA studies\cite{Buczek2011b,SasIoglu2010,Cao2017,Friedrich2020}, a double-peak lineshape is observed around $\mathbf{q} \sim 0.15\,\mathbf{q}_{\mathrm{X}}$, that is, at the point where there is a jump in the magnon frequency, but the coexistence only happens in a very narrow range of wave vectors $\mathbf{q}$. In Fig. \ref{fig:fcc-Ni-setup10_magnon_spectrum-and-dispersion}.b we present the spectral lineshapes around this value, both for the spectrum of transverse magnetic excitations as well as the single-particle Stoner excitations encoded in $S_{\mathrm{KS}}^{+-}(\mathbf{q}, \omega)$. For the wave vectors shorter than $0.17\,|\mathbf{q}_{\mathrm{X}}|$, the lineshape of $S^{+-}(\mathbf{q}, \omega)$ has a shoulder above the main magnon peak, clearly originating from a well-defined single-particle Stoner peak sitting below the main Stoner continuum in $S_{\mathrm{KS}}^{+-}(\mathbf{q}, \omega)$. As the magnon mode and Stoner peak become close in frequency, a new collective peak is developed above the Stoner peak, coexisting with the Goldstone mode only at $\mathbf{q}=0.14\,\mathbf{q}_{X}$, where the Stoner peak is wedged in between the two collective peaks. At $\mathbf{q}=0.17\,\mathbf{q}_{X}$, the Stoner peak disappears to negative frequencies and the upper collective magnon mode acquires the entire spectral weight. 
A comprehensive discussion of this phenomena stemming from a stripe-like feature in the single-particle Stoner spectrum can be found in previous literature\cite{Friedrich2020,Buczek2011b,SasIoglu2010,Karlsson2000}. 

\begin{figure*}
    \centering
    \includegraphics{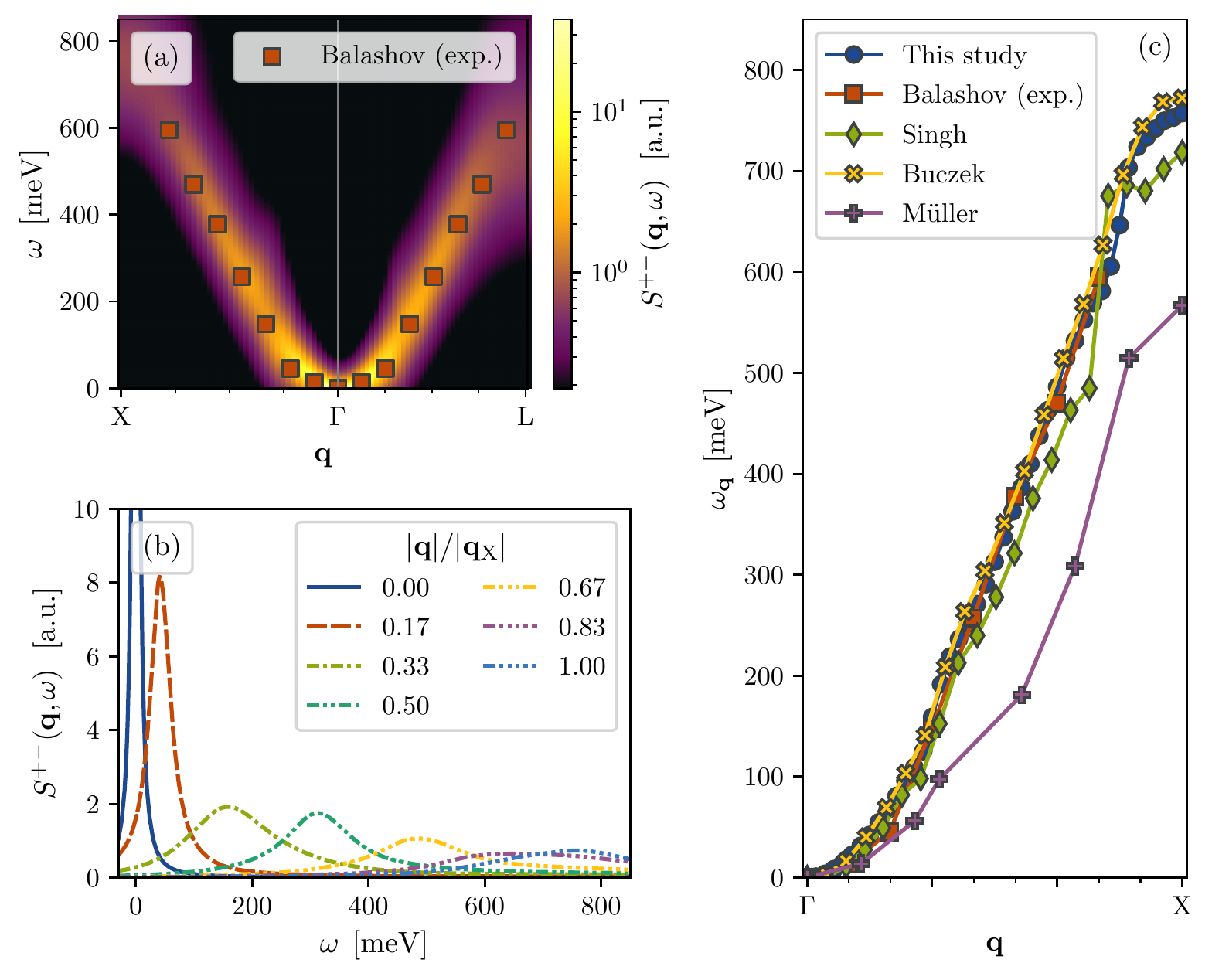}
    \caption{Transverse magnetic excitation spectrum of fcc-Co computed within the ALDA. In (a) the macroscopic (unit-cell averaged) $\mathbf{G}=0$ component of the spectrum is shown as a heat map and function of wave vector $\mathbf{q}$ and frequency $\omega$. The spectrum was computed on a X-$\Gamma$-L band path and is compared to inelastic scanning tunneling spectroscopy data\cite{Balashov2009}. In (b) the spectral intensity is shown as a function of frequency for a range of fixed values for $\mathbf{q}$ along the $\Gamma$-X path. In (c) the magnon peak positions extracted along the $\Gamma$-X path are shown and compared to experimental\cite{Balashov2009} as well as \textit{ab initio} references\cite{Singh2018,Buczek2011b,Muller2016}.
    }
    \label{fig:fcc-Co_magnon_spectrum-and-dispersion}
\end{figure*}
In Fig. \ref{fig:fcc-Ni-setup10_magnon_spectrum-and-dispersion}.c, we compare the extracted magnon dispersion along the $\Gamma\rightarrow \mathrm{X}$ direction with theoretical literature values as well as the experimental data. \c{S}a\c{s}\i o\u{g}lu and coworkers\cite{SasIoglu2010} treat the problem within MBPT, employing the LDA and scaling the screened Coulomb potential in order to remove the gap error. The other theoretical references are described in section \ref{sec:bcc-Fe}. Between different methodologies, there seems to be a good agreement for the (A)LDA magnon dispersion of wave vectors up to $\mathbf{q}\sim 4/9\,\mathbf{q}_{\mathrm{X}}$. Beyond this point there are significant differences in the extracted magnon frequency, resulting once again in a broad range of different values for the bandwidth. However, there seems to be a good agreement about the position of the magnon dispersion maxima. As argued in section \ref{sec:bcc-Fe}, at least some of the quantitative discrepancies in the magnon dispersion inside the Stoner continuum can be attributed to differences in the underlying DFT ground states and to improve consistency of ALDA results in the future, one would need to investigate why the different ground state DFT methodologies result in different Stoner spectra. To actually match the experimental dispersion, one would need to go beyond the (A)LDA. The poor performance of (A)LDA in the case of fcc-Ni is known to originate from the exchange splitting being overestimated by roughly a factor of two\cite{SasIoglu2010}. As seen in Fig. \ref{fig:fcc-Ni-setup10_magnon_spectrum-and-dispersion}.c, M{\"{u}}ller and coworkers obtain an improved description of the magnon dispersion, which is due to their adjustment of the exchange splitting in connection with the removal of the gap error. To get an improved \textit{ab initio} description of fcc-Ni within LR-TDDFT, one would need an exchange-correlation functional that improves the exchange splitting in its own right. Furthermore, one can also expect inclusion of non-local effects in the exchange-part of the kernel to decrease the magnon (spin-wave) stiffness\cite{Eich2018}.

\subsection{Co (fcc)}
Similar to the treatment of bcc-Fe and fcc-Ni presented above, we have computed the transverse magnetic excitation spectrum for fcc-Co and compared the extracted magnon peak positions with experimental as well as theoretical references. These results are presented in Fig. \ref{fig:fcc-Co_magnon_spectrum-and-dispersion}. The spectrum was computed on the basis of a LDA ground state with average spin-polarization per Co atom of $1.62\,\mu_{\textrm{B}}$, using $a=3.539\,\text{\AA}$ for the lattice constant. The original gap error was $\omega_{\Gamma}=-10.8\,\textrm{meV}$. The computed magnon spectrum in \ref{fig:fcc-Co_magnon_spectrum-and-dispersion}.a is fairly isotropic even at long wave vectors. For wave vectors longer than $q=0.44\,\text{\AA}^{-1}$, local differences in the dispersion between directions start occurring, but only beyond $q=1.35\,\text{\AA}^{-1}$ do the branches start to split. At this point,  ($q=0.88\,|\mathbf{q}_{\mathrm{L}}|$), the magnon mode approaches the BZ edge in the $\Gamma \rightarrow \mathrm{L}$ direction and starts to flatten out, whereas the mode continues to disperse towards higher frequencies in the $\Gamma \rightarrow \mathrm{X}$ direction. As such, we end up with bandwidths of 555 meV and 757 meV in the two directions respectively. 

\begin{figure*}
    \centering
    \includegraphics{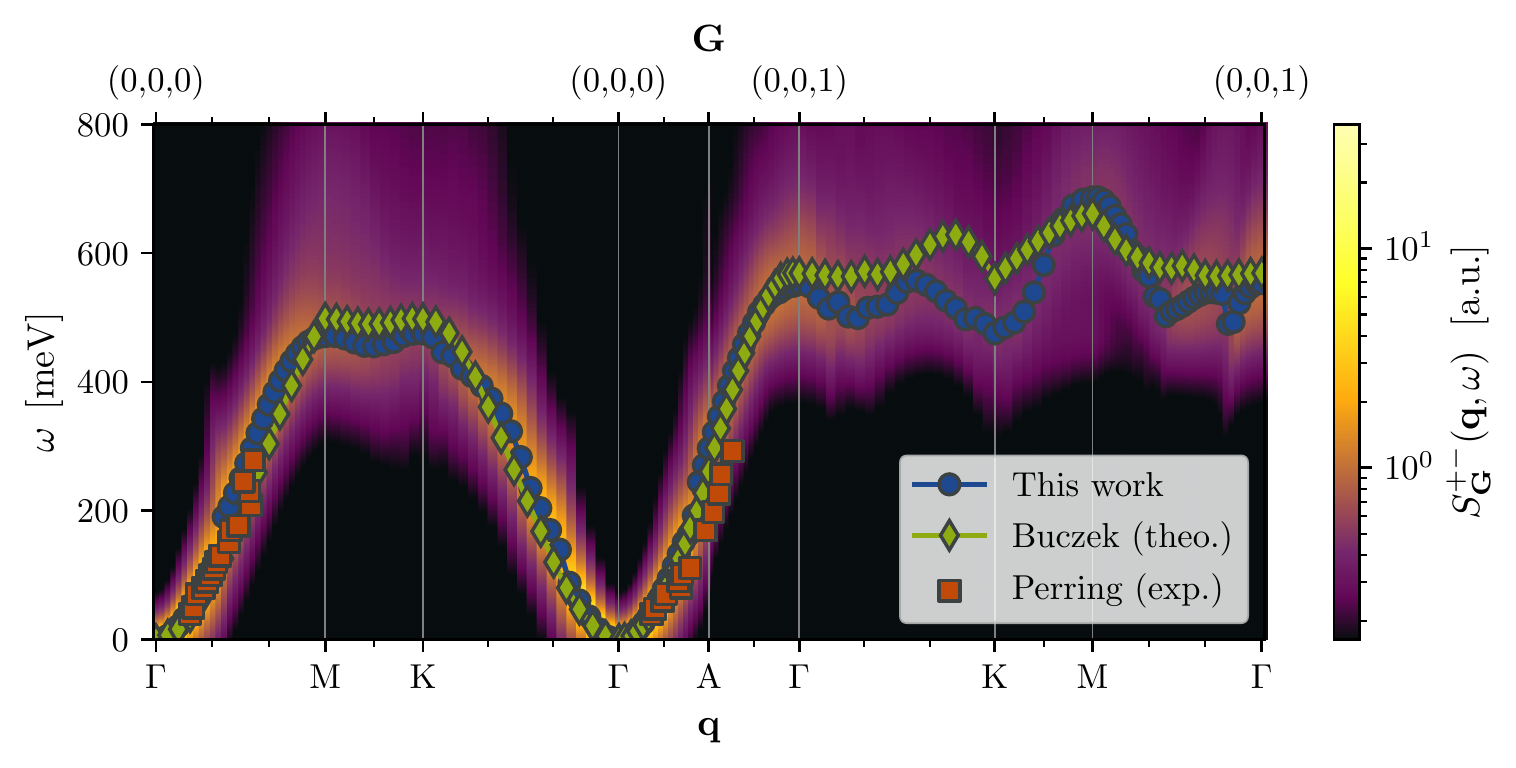}
    \caption{Transverse magnetic excitation spectrum of hcp-Co computed within the ALDA. The spectrum is shown as a heat map and function of wave vector $\mathbf{G}+\mathbf{q}$ and frequency $\omega$. The spectrum was computed on a $\Gamma$-M-K-$\Gamma$-A band path for the reduced wave vector $\mathbf{q}$ (lower axis) and is shown in the first and second Brillouin Zones (upper axis). The magnon peak positions are plotted on top of the heat map and compared to inelastic neutron scattering data\cite{Perring1995} as well as ALDA results from the literature\cite{Buczek2011b}.
    }
    \label{fig:hcp-Co_magnon_spectrum}
\end{figure*}
As evident from Figs. \ref{fig:fcc-Co_magnon_spectrum-and-dispersion}.a and \ref{fig:fcc-Co_magnon_spectrum-and-dispersion}.c, the computed magnon dispersion compares very well to the reference experimental dispersion, which itself was inferred from inelastic scanning tunneling spectroscopy data measured on a 9 monolayer Co/Cu(100) film\cite{Balashov2009}. We compare with the same data set in both directions, as most of the data points lie within the isotropic dispersion range. In addition to the experimental comparison, there is also a good agreement between the entire dispersion computed within ALDA using different implementations of LR-TDDFT. This may be a result of the excitation spectrum having a more trivial dependence of the lineshape as a function of $\mathbf{q}$ compared to the cases of bcc-Fe and fcc-Ni. In Fig. \ref{fig:fcc-Co_magnon_spectrum-and-dispersion}.b, the spectral lineshapes are shown for wave vectors evenly distributed along the $\Gamma \rightarrow \mathrm{X}$ path. Most of the lineshapes are well approximated by Lorentzians of increasing width, meaning that the Stoner continuum mainly broadens the collective magnon mode without altering its shape. If the low frequency Stoner continuum does not strongly influence the magnon peak positions, this implies that the theoretical magnon dispersion is less susceptible to subtle differences in the DFT ground state calculation on which it is based.

\subsection{Co (hcp)}
As the last material investigated in this study, we present the transverse magnetic excitation spectrum of hcp-Co in Fig. \ref{fig:hcp-Co_magnon_spectrum}. Using $a=2.507\,\text{\AA}$ for the lattice constant, we obtain a LDA ground state with an average spin-polarization of $1.59\,\mu_{\textrm{B}}$, very close to the value in fcc-Co. The spectrum has been corrected for a gap error of $\omega_{\Gamma}=-8.1\,\textrm{meV}$. Because hcp-Co has two magnetic atoms in the unit cell, the magnon spectrum include an optical mode as well as the acoustic (Goldstone) mode. The spectral function of transverse magnetic excitations, $S^{+-}_{\mathbf{G}}(\mathbf{q}, \omega)$, record excited states where the spin-orientation is precessing with a wave vector $\mathbf{G} + \mathbf{q}$ with respect to the ground state. Accordingly, the optical mode manifests itself for wave vectors with which the spin-orientation of the two magnetic atoms in the same unit-cell are precessing out of phase. This is the case for the second Brillouin Zone in hcp-Co, and in Fig. \ref{fig:hcp-Co_magnon_spectrum} we show the spectral function in the second BZ as well as the first. We present also the extracted magnon peak positions and compare them to experimental INS data\cite{Perring1995} as well as reference ALDA values from a literature LR-TDDFT calculation\cite{Buczek2011b}.

We obtain well-defined magnon modes for all investigated wave vectors $\mathbf{G}+\mathbf{q}$, although the optical mode is substantially attenuated by Landau damping. The magnon dispersion is isotropic along all three directions up to  $q=0.48\,\text{\AA}^{-1}$. Beyond this point, the magnon dispersion is generally steepest in the $\Gamma \rightarrow \mathrm{A}$ direction, and at the second BZ center, $1.545\,\text{\AA}^{-1}$ from the reciprocal space origo, the magnon dispersion attains a maximum with a frequency of 553 meV. The magnon frequencies at the first BZ edge is very similar at the M and K points, with 471 meV and 475 meV respectively. Because the M-point ($q=1.447\,\text{\AA}^{-1}$) lies closer to the $\Gamma$-point compared to the K-point ($q=1.671\,\text{\AA}^{-1}$), the upper part of the magnon dispersion is generally slightly steeper along the $\Gamma \rightarrow \mathrm{M}$ path compared to the $\Gamma \rightarrow \mathrm{K}$ path. 

Overall, hcp-Co has a relatively isotropic magnon dispersion, as is the case of fcc-Co. The extracted magnon peak positions match quite well with experiment along the $\Gamma \rightarrow \mathrm{M}$ direction, whereas the upper part of the dispersion towards the second BZ center is somewhat overestimated. These conclusions are consistent with previous ALDA results (also plotted). However, we see some discrepancies for the magnon dispersion of the optical branch between the LR-TDDFT implementations. The entire $\Gamma$-K-M-$\Gamma$ optical magnon branch lies in close proximity to a dense region of the Stoner continuum. As such, the magnon dispersion is strongly influenced by local variations in $S_{\mathrm{KS},\mathbf{G}}^{+-}(\mathbf{q}, \omega)$ and at least some of the discrepancies can be attributed to subtle differences in the respective DFT ground states. Meanwhile, the small bumps in the $\Gamma$-K-M-$\Gamma$ optical magnon dispersion might also indicate that the Stoner continuum was not appropriately converged with respect to the $k$-point density and broadening parameter $\eta$. In the convergence analysis underlying the present choice of parameters, the average frequency displacement, $\langle \Delta\omega \rangle$, was analysed for $S_{\mathrm{KS},\mathbf{G}}^{+-}(\mathbf{q}, \omega)$ within the first BZ only.

\section{Summary and outlook}\label{sec:summary}
We have applied the Kubo formalism to time-dependent spin-density functional theory and shown how to compute the four-component plane wave susceptibility from first principles. Although the theory is already well-known, we have provided a self-contained compilation suitable for plane wave treatments within LR-TDDFT.  
The methodology has been implented in the GPAW electronic structure package, enabling accurate computations of the transverse magnetic susceptibility. Within the limitations of the frozen core approximation and a finite set of PAW projector functions, the implemented methodology is formally exact, given that proper convergence in computational parameters is achieved. Thus, all approximations are due to the collinear spin-density functional theory framework and the chosen exchange-correlation functional/kernel.
    
A detailed convergence analysis was performed regarding spectral broadening, $k$-point sampling, plane wave representation and truncation of the unoccupied bands. In particular, it was shown that in order to obtain an appropriate description of the low frequency Stoner continuum, the $k$-point density and broadening parameter $\eta$ need to be converged in parallel. To this end, we have introduced the average displacement frequency $\langle \Delta \omega \rangle$, which provides reliable guidance for choosing values of $\eta$ that result in converged magnon dispersion relations. $\langle \Delta \omega \rangle$ is calculated from the single-particle Stoner spectrum only, which itself is fast to compute. We have assessed the gap error convergence and found that it is not possible to converge $\omega_{\Gamma}$ within a finite plane wave basis. However, the gap error can be effectively accounted for by applying a constant shift to the spectrum of transverse magnetic excitations, such that the Goldstone condition is fulfilled. As a result, it is possible to attain convergence of the magnon dispersion relation itself within a finite basis set and a modest number of unoccupied bands. 
    
Using the implemented methodology and converged numerical parameters, the transverse magnetic excitation spectrum was computed for 3$d$ transition metals iron, nickel and cobalt. For bcc-Fe, fcc-Co and hcp-Co, the ALDA was shown to reproduce experimental magnon dispersions in a satisfactory manner, whereas the magnon dispersion in fcc-Ni is overestimated due to the well-known overestimation of the ground state exchange splitting energy $\Delta_{\mathrm{x}}$ with LDA. All results match previous (A)LDA literature well for short wave vectors $\mathbf{q}$, but inside the low frequency Stoner continuum, literature values for the magnon peak positions vary substantially. These discrepancies were discussed in detail and mostly attributed subtle differences in the underlying DFT ground states.

First principles calculations of magnons are rather scarce in the literature and most studies have focused on iron, nickel or cobalt. This is likely due to the conspicuous role of these materials when discussing magnetic solids and partly due to the fact that these materials can be described within small unit cells, rendering otherwise prohibitively demanding TDDFT computations feasible. There is, however, a vast experimental literature on transverse magnetic excitations in a wide range of solids and it is our hope that first principles calculation of the transverse magnetic susceptibility can be carried out routinely in the future. The convergence study of this work implies that the treatment of complex magnetic materials requires additional method development in order to lower the demands on the computational power, but several well-known magnetic materials with small unit cells should be within reach using the present framework\cite{Buczek2009,Odashima2013}. To this end, itinerant magnets seem to be the most challenging, as the Stoner spectrum is gapped for insulators and the magnons less sensitive towards $k$-point sampling and broadening. In addition, the Heisenberg model often provides a rather accurate description for insulators, with parameters that can be obtained directly from ground state DFT calculations\cite{Xiang2013,Torelli2019,Torelli2020}. Still, it would be of fundamental interest to compare the dispersion relations obtained from a first principles Heisenberg model with a direct computation from TDDFT. Such a comparison could yield valuable insight into the limitations and virtues of both methods.

In this work, we have applied a collinear description of the 3$d$ transition metals, as spin-orbit interactions are nearly negligible for iron, cobalt and nickel. However, materials with strong spin-orbit coupling may exhibit a wealth of interesting effects. Specifically, spin-orbit effects provide a coupling between the transverse and longitudinal magnetic excitations as well as to the density response. This implies, for example, that magnons can be accessed by perturbing electric fields and that magnons may couple to plasmons and excitons in metals and insulators respectively. Moreover, spin-orbit coupling may induce topological gaps between magnon branches\cite{Mook2014, Costa2020b}, which implies the existence 
of topological robust surface magnons, or induce nonreciprocity in the magnon dispersion relation\cite{Costa2020a}. We believe that first principle calculations could help unravel such exotic phenomena in the future.

\appendix

\section{Linear response theory}\label{sec:theory}

\subsection{Dynamic susceptibilities and spectral functions}\label{sec:susceptibility}
For experimental as well as theoretical spectroscopy, the central object of interest is the susceptibility of the system. In the framework of linear response theory, the retarded susceptibility $\chi_{BA}(t-t')$ gives the change in a system coordinate $\hat{B}=\hat{B}^{\dagger}$ at time $t$ to a weak external perturbation in the system coordinate $\hat{A}=\hat{A}^{\dagger}$ at time $t'$, to linear order:
\begin{equation}
    \hat{H}(t) = \hat{H}_0 + \hat{H}_{\mathrm{ext}}(t), \quad \hat{H}_{\mathrm{ext}}(t) = \hat{A} f(t),
    \label{eq:linear response H}
\end{equation}
\begin{equation}
    \langle \delta\hat{B}(t) \rangle = \langle \hat{B}(t) \rangle - \langle \hat{B} \rangle_0 = \int_{-\infty}^{\infty}dt'\, \chi_{BA}(t-t') f(t').
    \label{eq:linear response def.}
\end{equation}
Here, $\hat{H}_0$ is the time-independent system Hamiltonian, $f(t)$ is a coordinate external to the system and $\langle \hat{B} \rangle_0$ is the expectation value of the coordinate $\hat{B}$ in the absence of the external perturbation $\hat{H}_{\mathrm{ext}}(t)$.

The retarded susceptibility can be computed from the \textit{Kubo formula}\cite{Kubo1957}
\begin{equation}
    \chi_{BA}(t-t') = - \frac{i}{\hbar} \theta(t - t') \langle\, [\hat{B}_0(t), \hat{A}_0(t')] \,\rangle_0,
    \label{eq:Kubo formula}
\end{equation}
where $\theta(t-t')$ is the step function, making the susceptibility retarded, while $\hat{A}_0(t')=e^{i\hat{H}_0t'/\hbar} \hat{A}\, e^{-i\hat{H}_0t'/\hbar}$ and $\hat{B}_0(t)$ carry the time-dependence in the interaction picture. 
Due to the step function in Eq. \eqref{eq:Kubo formula}, the retarded susceptibility is analytic in the upper half of the complex frequency plane. Inserting a complete set of energy eigenstates to the system Hamiltonian $\hat{H}_0$ and carrying out the Fourier-Laplace transform (see Appendix \ref{sec:app temp FT definition} for definitions), one obtains the dynamic susceptibility in the Lehmann representation:
\begin{equation}
        \chi_{BA}(z=\omega+i\eta) = \sum_{\alpha, \alpha'}\frac{\langle \alpha| \hat{B} |\alpha'\rangle  \langle \alpha'| \hat{A} |\alpha\rangle}{\hbar \omega - (E_{\alpha'}-E_{\alpha}) + i \hbar \eta} (n_{\alpha} - n_{\alpha'}),
    \label{eq:lehmann}
\end{equation}
where $z$ is the complex frequency and $\omega$ and $\eta$ are real with $\eta > 0$. $|\alpha\rangle$ denotes an energy eigenstate of $\hat{H}_0$ with energy $E_{\alpha}$ and population factor $n_{\alpha}$ (when the system in the absence of the perturbation is in thermal equilibrium with a bath of temperature $T$). 

If $\hat{H}_0$ is known and can be diagonalized, the Lehmann representation \eqref{eq:lehmann} can be used to evaluate the dynamic susceptibility. Conversely, \eqref{eq:lehmann} can be used to interpret a measured or computed susceptibility in terms of the fundamental excitations of the system. In particular, it is useful to split up the dynamic susceptibility in its reactive and dissipative parts, $\chi_{BA}'(z)$ and $\chi_{BA}''(z)$\cite{Jensen1991}:
\begin{subequations}
    \begin{equation}
        \chi_{BA}(z) = \chi_{BA}'(z) + i \chi_{BA}''(z),
        \label{eq:reac diss split}
    \end{equation}
    \begin{equation}
        \chi_{BA}'(z) = \chi_{AB}'(-z^*) = \frac{1}{2}\left\{\chi_{BA}(z) + \chi_{AB}(-z^*)\right\},
        \label{eq:reactive part}
    \end{equation}
    \begin{equation}
        \chi_{BA}''(z) = - \chi_{AB}''(-z^*) = \frac{1}{2i}\left\{\chi_{BA}(z) - \chi_{AB}(-z^*)\right\}.
        \label{eq:dissipative part}
    \end{equation}
    \label{eq:reactive and dissipative parts}
\end{subequations}
This operation has the effect of splitting the simple poles in the Lehmann representation \eqref{eq:lehmann} into its real and imaginary parts:
\begin{subequations}
    \begin{align}
        \chi_{BA}'(\omega+i\eta) = \sum_{\alpha, \alpha'} &\langle \alpha| \hat{B} |\alpha'\rangle  \langle \alpha'| \hat{A} |\alpha\rangle (n_{\alpha} - n_{\alpha'}) \nonumber \\
        & \hspace{-13pt} \times \mathrm{Re}\left\{\frac{1}{\hbar \omega - (E_{\alpha'}-E_{\alpha}) + i \hbar \eta}\right\}
        \label{eq:lehmann reactive part}
    \end{align}
    \begin{align}
        \chi_{BA}''(\omega+i\eta) = \sum_{\alpha, \alpha'} &\langle \alpha| \hat{B} |\alpha'\rangle  \langle \alpha'| \hat{A} |\alpha\rangle (n_{\alpha} - n_{\alpha'}) \nonumber \\
        & \hspace{-13pt} \times \mathrm{Im}\left\{\frac{1}{\hbar \omega - (E_{\alpha'}-E_{\alpha}) + i \hbar \eta}\right\},
        \label{eq:lehmann dissipative part}
    \end{align}
    \label{eq:lehmann reactive and dissipative parts}
\end{subequations}
of which the imaginary part of the simple poles are Lorentzians of width $2\hbar\eta$ and amplitude $-\pi$. 
In the limit $\eta \rightarrow 0^{+}$, for which the notation $\chi_{BA}(\omega) = \chi_{BA}(\omega + i0^+)$ is used, the Lorentzians become $\delta$-functions. 

For a system $\hat{H}_0$ with a non-degenerate ground state $|\alpha_0\rangle$ and ground state energy $E_0$, the Lehmann representation \eqref{eq:lehmann} reduces to a single sum over excited states in the zero temperature limit:
\begin{align}
    \chi_{BA}(\omega + i\eta) = \sum_{\alpha \neq \alpha_0} \Bigg( &\frac{\langle \alpha_0| \hat{B} |\alpha\rangle  \langle \alpha| \hat{A} |\alpha_0\rangle}{\hbar \omega - (E_{\alpha}-E_{0}) + i \hbar \eta}
    \nonumber \\
    -&\frac{\langle \alpha_0| \hat{A} |\alpha\rangle  \langle \alpha| \hat{B} |\alpha_0\rangle}{\hbar \omega + (E_{\alpha}-E_{0}) + i \hbar \eta} \Bigg).
    \label{eq:lehmann T=0}
\end{align}
Moreover, the dissipative part of the dynamic susceptibility may be expressed as a spectral function for the induced excitations:
\begin{align}
    S_{BA}(\omega) 
    &\equiv 
    - \frac{\chi_{BA}''(\omega)}{\pi} 
    = A_{BA}(\omega) - A_{AB}(- \omega),
    \label{eq:connection to spectral functions}
\end{align}
%
%
\begin{equation}
    A_{BA}(\omega) = \sum_{\alpha \neq \alpha_0} \langle \alpha_0| \hat{B} |\alpha\rangle  \langle \alpha| \hat{A} |\alpha_0\rangle \, \delta \big(\hbar \omega - (E_{\alpha} - E_{0})\big).
    \label{eq:spectral function def.}
\end{equation}
%
Thus, the dissipative part of the dynamic susceptibility contains both the spectrum of excited states generated by $\hat{A}$, reversed by $\hat{B}$, at positive frequencies, and the spectrum generated by $\hat{B}$, reversed by $\hat{A}$, at negative frequencies. In this way, the susceptibility is not only a quantity characterizing the system response to external perturbations, but it also contains valuable information about the eigenstates of the underlying quantum system.

The intimate relation between the underlying quantum system and the dynamic susceptibility is further illustrated by the spectral moments of its dissipative part. The moments generate a range of expectation values of the quantum system, valid also at finite temperatures\cite{Jensen1991,Kubo1966}:
\begin{equation}
    \int_{-\infty}^{\infty} (\hbar \omega)^n 
    S_{BA}(\omega) \,
    d \hbar\omega = (-\hbar)^n \left\langle \left[\hat{\mathcal{L}}_0^n \hat{B}, \hat{A}\right] \right\rangle_0,
    \label{eq:Kubo nth order sum rule}
\end{equation}
where $\hat{\mathcal{L}}_0$ is the Liouville operator of the system,
\begin{equation}
    \hat{\mathcal{L}}_0 \hat{B} = \frac{1}{\hbar} \left[\hat{H}_0, \hat{B}\right].
    \label{eq:liouville operator definition}
\end{equation}
Eq. \eqref{eq:Kubo nth order sum rule} is commonly refered to as the $n$'th order sum rule.

\subsection{Linear response theory and spectroscopy}\label{sec:spectroscopy}
In the context of a spectroscopic experiment, the dissipative part of the dynamic susceptibility, $\chi_{BA}''(\omega)$, gives the spectrum of transitions between energy eigenstates induced by the perturbation in question \eqref{eq:lehmann dissipative part}. By the virtue of the \textit{fluctuation-dissipation theorem}\cite{Nyquist1928,Callen1951,Kubo1957,Kubo1966}, this spectrum is directly related to the fundamental fluctuations of the system as well as the energy dissipation. 

More specifically, one may consider the response to a harmonic perturbation\cite{Jensen1991}
%
%
\begin{equation}
    f(t) = f_0 \cos(\omega_0 t) = \frac{f_0}{2}\left(e^{-i\omega_0 t} + cc.\right). 
    \label{eq:simple harmonic perturbation}
\end{equation}
Insertion into the response relation \eqref{eq:linear response def.} and application of the convolution theorem yields
\begin{equation}
    \langle \delta \hat{B}(t) \rangle = \frac{f_0}{2} \left[ \chi_{BA}(\omega_0) e^{-i\omega_0 t} + \chi_{BA}(-\omega_0) e^{i \omega_0 t} \right].
    \label{eq:simple harmonic perturbation response}
\end{equation}
Now, using the Lehmann representation \eqref{eq:lehmann}, it is straightforward to show that any retarded susceptibility as defined by the \textit{Kubo formula} \eqref{eq:Kubo formula} satisfy
\begin{equation}
    \chi_{B^{\dagger}A^{\dagger}}(-z^*) = \chi_{BA}^*(z).
    \label{eq:relation daggers cc}
\end{equation}
Insertion into Eq. \eqref{eq:simple harmonic perturbation response} reveals that the real and imaginary parts of the dynamic susceptibility gives the response in- and out-of-phase of the harmonic perturbation respectively:
\begin{align}
    \langle \delta \hat{B}(t) \rangle = f_0 \big[ &\mathrm{Re}\left\{\chi_{BA}(\omega_0)\right\} \cos(\omega_0 t) 
    \nonumber \\
    + &\mathrm{Im}\left\{\chi_{BA}(\omega_0)\right\} \sin(\omega_0 t) \big].
     \label{eq:in and out of phase response}    
\end{align}
%
%
Finally, only the out-of-phase response contribute to energy dissipation on average. Consequently, the mean rate of energy absorption in the system ($Q=d\langle\hat{H}\rangle/dt =\langle\hat{A}(t)\rangle df/dt$) is proportional to $\mathrm{Im}\left\{\chi_{AA}(\omega_0)\right\}$: 
\begin{equation}
    \bar{Q} = - \frac{1}{2} f_0^2 \omega_0 \mathrm{Im}\left\{\chi_{AA}(\omega_0)\right\} = - \frac{1}{2} f_0^2 \omega_0 \chi_{AA}''(\omega_0).
    \label{eq:mean energy dissipation}
\end{equation}
In the last equality, it was used that Eq. \eqref{eq:relation daggers cc} implies $\chi_{A A^{\dagger}}(-z^*) = \chi_{A^{\dagger} A}^*(z)$, meaning that
\begin{subequations}
    \begin{equation}
        \chi_{A^{\dagger}A}'(z) 
        = \mathrm{Re}\left\{ \chi_{A^{\dagger}A}(z) \right\},
        \label{eq:reactive part = real part}
    \end{equation}
    \begin{equation}
        \chi_{A^{\dagger}A}''(z) 
        = \mathrm{Im}\left\{ \chi_{A^{\dagger}A}(z) \right\}.
        \label{eq:dissipative part = imaginary part}
    \end{equation}
\end{subequations}
With this in hand, various spectroscopic techniques can directly probe $\chi_{AA}''(\omega)$ by tracking the energy dissipated from the source of the perturbation. Finally, 
the energy dissipation is related directly to the transitions between system eigenstates through Eq. \eqref{eq:lehmann dissipative part} or specific ground state excitations through Eqs. \eqref{eq:connection to spectral functions} and \eqref{eq:spectral function def.}.
%

\subsection{Dynamic susceptibilities of periodic crystals}\label{sec:susceptibility periodic crystals}
As discussed, the dynamic susceptibility is a fundamental property of any quantum system $\hat{H}_0$. In particular, it gives the system response to a weak external perturbation and characterizes the spectrum of system excitations that the perturbation generates to linear order.  
In the case of real life materials, one has to consider a perturbation which varies in both time and space. If the Born-Oppenheimer approximation is employed, such that $\hat{H}_0$ only needs to describe the electronic degrees of freedom in the material, such a perturbation may be written
\begin{equation}
    \hat{H}_\mathrm{ext}(t) = \int d\mathbf{r}\, \hat{A}(\mathbf{r}) f(\mathbf{r}, t),
    \label{eq:linear response Hext periodic crystals}
\end{equation}
where $\hat{A}(\mathbf{r})=\hat{A}^{\dagger}(\mathbf{r})$ is taken to be an electronic one-body operator. The Kubo formalism itself is not restricted to the consideration of one-body operators and what follows can be easily generalized if needed. Now the retarded susceptibility gives the electronic system response in some coordinate $\hat{B}$ (also taken to be a one-body operator) at position $\mathbf{r}$ and time $t$ to a weak perturbation of the system coordinate $\hat{A}$ at position $\mathbf{r}'$ and time $t'$:
\begin{equation}
    \langle \delta\hat{B}(\mathbf{r}, t) \rangle = \int_{-\infty}^{\infty} dt' \int d\mathbf{r}'\, \chi_{BA}(\mathbf{r}, \mathbf{r}', t-t') f(\mathbf{r}', t').
    \label{eq:linear response periodic crystals}
\end{equation}
The Kubo theory described above can be easily applied to $\chi_{BA}(\mathbf{r}, \mathbf{r}', t-t')$, simply by letting $\hat{A}=\hat{A}(\mathbf{r})$ and $\hat{B}=\hat{B}(\mathbf{r})=\hat{B}^{\dagger}(\mathbf{r})$. In particular, the transition matrix elements entering the Lehmann representation \eqref{eq:lehmann} will now depend on position:
\begin{equation}
    A_{\alpha'\alpha}(\mathbf{r}) \equiv \langle \alpha'| \hat{A}(\mathbf{r}) |\alpha\rangle.
    \label{eq:transition matrix element def}
\end{equation}

In the study of periodic crystals, the material in question is represented as a quantum system which is invariant under lattice translations, $[\hat{T}_{\mathbf{R}}, \hat{H}_0] = 0$. Here, $\hat{T}_{\mathbf{R}}$ denotes the unitary generator of translations $\mathbf{r}\rightarrow\mathbf{r} - \mathbf{R}$, where $\mathbf{R}$ is any lattice vector connecting two points on the Bravais lattice of the crystal. The commutation relation implies that the eigenstates of $\hat{T}_{\mathbf{R}}$ diagonalize $\hat{H}_0$ and, according to Bloch's theorem, the eigenvalues may be written in terms of real wave vectors $\mathbf{k}_{\alpha}$:
\begin{equation}
    \hat{T}_{\mathbf{R}} |\alpha\rangle = e^{i \mathbf{k}_{\alpha} \cdot \mathbf{R}} |\alpha\rangle.
\end{equation}
This has important consequences for the dynamic susceptibility of the system. It implies that all transition matrix elements transform as Bloch waves under lattice translations:
\begin{align}
    A_{\alpha'\alpha}(\mathbf{r} + \mathbf{R}) 
    &= \langle \alpha'| \hat{A}(\mathbf{r} + \mathbf{R}) |\alpha\rangle 
    \nonumber \\
    &= \langle \alpha'| \hat{T}^{\dagger}_{\mathbf{R}} \hat{A}(\mathbf{r}) \hat{T}_{\mathbf{R}} |\alpha\rangle 
    \nonumber \\
    &= e^{-i \mathbf{q}_{\alpha'\alpha} \cdot \mathbf{R}} A_{\alpha'\alpha}(\mathbf{r}).
    \label{eq:trans. mat. ele. under lattice translations}
\end{align}
The reduced wave vector $\mathbf{q}_{\alpha'\alpha} \equiv (\mathbf{k}_{\alpha'} - \mathbf{k}_{\alpha}) - \mathbf{G}_{\alpha'\alpha}$, represents the difference in crystal momentum between the two states $|\alpha'\rangle$ and $|\alpha\rangle$, where $\mathbf{G}_{\alpha'\alpha}$ is a reciprocal lattice vector chosen such that $\mathbf{q}_{\alpha'\alpha}$ lies within the first Brillouin zone. Following Eq. \eqref{eq:trans. mat. ele. under lattice translations}, the transition matrix elements can be written on a Bloch wave form, with periodic parts $a_{\alpha'\alpha}(\mathbf{r}+\mathbf{R}) = a_{\alpha'\alpha}(\mathbf{r})$:
\begin{equation}
    A_{\alpha'\alpha}(\mathbf{r}) = \frac{\Omega_{\mathrm{cell}}}{\Omega} e^{-i \mathbf{q}_{\alpha'\alpha} \cdot \mathbf{r}} a_{\alpha'\alpha}(\mathbf{r}).
    \label{eq:trans mat ele Bloch wave}
\end{equation}
Here, the periodic parts have been normalized by the crystal volume $\Omega$ and the unit cell volume $\Omega_{\mathrm{cell}}$, so as to make $a_{\alpha'\alpha}(\mathbf{r})$ size intensive, that is, independent of the crystal volume. As a consequence of Eq. \eqref{eq:trans mat ele Bloch wave},
\begin{equation}
    B_{\alpha\alpha'}(\mathbf{r} + \mathbf{R}) A_{\alpha'\alpha}(\mathbf{r}' + \mathbf{R}) = B_{\alpha\alpha'}(\mathbf{r}) A_{\alpha'\alpha}(\mathbf{r}'),
\end{equation}
and from Eq. \eqref{eq:lehmann}, it is concluded that also the dynamic susceptibility is a periodic function:
\begin{equation}
    \chi_{BA}(\mathbf{r} + \mathbf{R}, \mathbf{r}' + \mathbf{R}, z) = \chi_{BA}(\mathbf{r}, \mathbf{r}', z).
    \label{eq:periodic susceptibilities}
\end{equation}

The retarded susceptibility as defined by Eq. \eqref{eq:linear response periodic crystals} describes the system response on all time and length scales simultaneously. $\chi_{BA}(\mathbf{r}, \mathbf{r}_0, t-t_0)$ gives the response at a specific position $\mathbf{r}$ and time $t$ to a perturbation which is completely local in space and time: $f(\mathbf{r}, t) \propto \delta(\mathbf{r} - \mathbf{r}_0) \delta(t - t_0)$. Thus, it describes the microscopic details and mechanisms that can be activated by an external source, but it does not directly describe the macroscopic properties of the material. To investigate the macroscopic properties embedded in the susceptibility, the response to a plane wave perturbation is considered:
\begin{equation}
    \hat{H}_{\mathrm{ext}}(t) = \int d\mathbf{r}\, \hat{A}(\mathbf{r})\, \frac{f_0}{2} \left[e^{i(\mathbf{k}_0 \cdot \mathbf{r} - \omega_0 t)} + c.c. \right].
    \label{eq:sinusoidal plane wave perturbation}
\end{equation}
Then one can ask: To linear order in $f_0 \in \mathbb{R}$, what is the strength of induced plane wave fluctuations in the system coordinate $\hat{B}$,
\begin{equation}
    \langle \delta\hat{B}(\mathbf{k}, \omega) \rangle = \int_{-\infty}^{\infty} dt \int 
    d\mathbf{r}\, 
    e^{-i(\mathbf{k} \cdot \mathbf{r} - \omega t)} \langle \delta \hat{B}(\mathbf{r}, t) \rangle.
    \label{eq:FT of periodic response}
\end{equation}
To answer this question, the plane wave susceptibility is introduced as the lattice Fourier transform of the dynamic susceptibility:
\begin{equation}
    \chi_{BA}^{\mathbf{G} \mathbf{G}'}(\mathbf{q}, z) \equiv \iint \frac{d\mathbf{r} d\mathbf{r}'}{\Omega} e^{-i(\mathbf{G} + \mathbf{q}) \cdot \mathbf{r}} \chi_{BA}(\mathbf{r}, \mathbf{r}', z) e^{i(\mathbf{G}' + \mathbf{q}) \cdot \mathbf{r}'}
    \label{eq:plane wave dyn susc. def}
\end{equation}
where $\mathbf{G}$ and $\mathbf{G}'$ are reciprocal lattice vectors, while $\mathbf{q}$ is a wave vector within the first Brillouin zone.
Now, due to the periodicity of the dynamic susceptibility \eqref{eq:periodic susceptibilities}, its spatial Fourier transform is diagonal in wave vectors $\mathbf{q}$ and $\mathbf{q}'$ (see Appendix \ref{sec:app double spatial FT}):
\begin{equation}
    \chi_{BA}(\mathbf{G} + \mathbf{q}, \mathbf{G}' + \mathbf{q}', z) = \frac{(2\pi)^D}{\Omega} \chi_{BA}^{\mathbf{G} \mathbf{G}'}(\mathbf{q}, z) \delta(\mathbf{q} - \mathbf{q}'),
    \label{eq:chi diagonal spatial FT}
\end{equation}
where $D$ is the dimensionality of the problem. With this, Eqs. \eqref{eq:linear response periodic crystals} and \eqref{eq:sinusoidal plane wave perturbation} are inserted into Eq. \eqref{eq:FT of periodic response} and the convolution theorem is used to obtain
%
\begin{align}
    \frac{\langle \delta\hat{B}(\mathbf{k}, \omega) \rangle}
    {(2\pi)^{D+1}} 
    = \frac{f_0}{2} \Big[ \chi_{BA}^{\mathbf{G} \mathbf{G}_0}(\mathbf{q}, \omega) &\delta(\mathbf{q} - \mathbf{q}_0) \delta(\omega - \omega_0) 
    \nonumber \\
    + \chi_{BA}^{\mathbf{G} -\mathbf{G}_0}(\mathbf{q}, \omega) &\delta(\mathbf{q} + \mathbf{q}_0) \delta(\omega + \omega_0) \Big],
\end{align}
where $\mathbf{k}=\mathbf{G} + \mathbf{q}$ and $\mathbf{k}_0=\mathbf{G}_0 + \mathbf{q}_0$. Inverting the Fourier transforms of Eq. \eqref{eq:FT of periodic response},
\begin{align}
    \langle \delta \hat{B}(\mathbf{r}, t) \rangle 
    = & \frac{f_0}{2} \sum_{\mathbf{G}} \Big[
    \chi_{BA}^{\mathbf{G} \mathbf{G}_0}(\mathbf{q}_0, \omega_0) e^{i([\mathbf{G}+\mathbf{q}_0]\cdot\mathbf{r} - \omega_0 t)}
    \nonumber \\ 
    +
    &\chi_{BA}^{-\mathbf{G} -\mathbf{G}_0}(-\mathbf{q}_0, -\omega_0) e^{-i([\mathbf{G}+\mathbf{q}_0]\cdot\mathbf{r} - \omega_0 t)}
    \Big].
    \label{eq:plane wave real space response I}
\end{align}
From this, the physical interpretation of the plane wave susceptibility, $\chi_{BA}^{\mathbf{G} \mathbf{G}'}(\mathbf{q}, \omega)$, is clear: It is a fundamental material property, giving the plane wave coefficients of the material response in coordinate $\hat{B}$ to a plane wave perturbation in coordinate $\hat{A}$ with wave vector $\mathbf{G}'+\mathbf{q}$ and frequency $\omega$ per source strength $f_0$, a response which is diagonal in both $\omega$ and $\mathbf{q}$.

\subsection{Lehmann representation of the plane wave susceptibility}\label{sec:reac and diss parts of pw susceptibility}
So far, the retarded susceptibility was introduced and defined in terms of the linear response in system coordinates assumed to be hermitian $\hat{A}^{\dagger}=\hat{A}$ and $\hat{B}^{\dagger}=\hat{B}$, see Eqs. \eqref{eq:linear response H} and \eqref{eq:linear response def.}. More generally, operators that are not necessarily hermitian may be considered taking the \textit{Kubo formula} \eqref{eq:Kubo formula} itself as the definition of a retarded susceptibility. Starting from the Kubo formula, the Lehmann representation \eqref{eq:lehmann} and the separation into reactive and dissipative parts \eqref{eq:lehmann reactive and dissipative parts} still hold, meaning that also dynamic susceptibilities of non-hermitian operators are made up out of spectra of excited states in the system.

For the plane wave susceptibility, one may use the field operators in the Fourier basis,
\begin{equation}
    \hat{A}(\mathbf{Q}) \equiv
    \int d\mathbf{r} \, e^{-i \mathbf{Q} \cdot \mathbf{r}} \hat{A}(\mathbf{r}),
    \label{eq:operator fourier basis}
\end{equation}
with which the susceptibility can be written on a form consistent with Kubo theory:
\begin{subequations}
    \begin{equation}
        \chi_{BA}^{\mathbf{G} \mathbf{G}'}(\mathbf{q}, z) = 
        \frac{1}{\Omega}  
        \, \chi_{\beta \alpha}(z),
    \end{equation}
    \begin{equation}
        \hat{\beta} = \hat{B}(\mathbf{G} + \mathbf{q}), 
        \quad
        \hat{\alpha} = \hat{A}(-\mathbf{G}' - \mathbf{q}).
    \end{equation}
    \label{eq:plane wave on kubo form}
\end{subequations}
%
%
%
Resultantly, also the plane wave susceptibility can be split up in reactive and dissipative parts by applying Eq. \eqref{eq:reactive and dissipative parts} to $\chi_{\beta \alpha}(z)$. In order to do this, it is used that the transition matrix elements are Bloch waves \eqref{eq:trans mat ele Bloch wave}:
%
%
\begin{equation}
    \langle \alpha' | \hat{A}(-\mathbf{G} - \mathbf{q}) | \alpha \rangle = 
    a_{\alpha'\alpha}(-\mathbf{G}) \, \delta_{\mathbf{q},\mathbf{q}_{\alpha'\alpha}},
    \label{eq:bloch wave transition matrix elements}
\end{equation}
where $\delta_{\mathbf{q},\mathbf{q}_{\alpha'\alpha}}$ is a Kroenecker-$\delta$ counting pairs of energy eigenstates $\alpha', \alpha$ with $\mathbf{q}_{\alpha'\alpha} = \mathbf{q}$ and $a_{\alpha'\alpha}(\mathbf{G})$ is the plane wave coefficient of the periodic part of the transition matrix elements
\begin{equation}
    a_{\alpha'\alpha}(\mathbf{G}) = 
    \int_{\Omega_{\mathrm{cell}}} d\mathbf{r} \, e^{-i\mathbf{G} \cdot \mathbf{r}} a_{\alpha'\alpha}(\mathbf{r}).
    \label{eq:periodic transition matrix elements}
\end{equation}
%
%
%
With this, the Lehmann representation of the plane wave susceptibility may be written up directly from Eq. \eqref{eq:lehmann},
\begin{align}
    \chi_{BA}^{\mathbf{G} \mathbf{G}'}(\mathbf{q}, z) = \frac{1}{\Omega} \sum_{\alpha,\alpha'} &\frac{b_{\alpha\alpha'}(\mathbf{G}) a_{\alpha'\alpha}(-\mathbf{G}')}{\hbar \omega - (E_{\alpha'}-E_{\alpha}) + i \hbar \eta}
    \nonumber \\
    &\times (n_{\alpha} - n_{\alpha'})\delta_{\mathbf{q},\mathbf{q}_{\alpha'\alpha}},
\end{align}
which may be seen as a generalization of the results for the dielectric function from Alder\cite{Adler1962} and Wiser\cite{Wiser1963}. 
Likewise, the reactive and dissipative parts of the plane wave susceptibility can be written up directly from Eq. \eqref{eq:lehmann reactive and dissipative parts}:
\begin{subequations}
    \begin{align}
        \chi_{\beta\alpha}'(\omega+i\eta) 
        &=
        \sum_{\alpha, \alpha'} b_{\alpha\alpha'}(\mathbf{G}) a_{\alpha'\alpha}(-\mathbf{G}') (n_{\alpha} - n_{\alpha'}) \delta_{\mathbf{q},\mathbf{q}_{\alpha'\alpha}}
        \nonumber \\
        &\times  \mathrm{Re}\left\{\frac{1}{\hbar \omega - (E_{\alpha'}-E_{\alpha}) + i \hbar \eta}\right\},
        \label{eq:plane wave reactive part}
    \end{align}
    \begin{align}
        \chi_{\beta\alpha}''(\omega+i\eta) 
        &= 
        \sum_{\alpha, \alpha'} b_{\alpha\alpha'}(\mathbf{G}) a_{\alpha'\alpha}(-\mathbf{G}') (n_{\alpha} - n_{\alpha'}) \delta_{\mathbf{q},\mathbf{q}_{\alpha'\alpha}}
        \nonumber \\
        &\times \mathrm{Im}\left\{\frac{1}{\hbar \omega - (E_{\alpha'}-E_{\alpha}) + i \hbar \eta}\right\}.
        \label{eq:plane wave dissipative part}
    \end{align}
    \label{eq:plane wave reac. and diss. parts}
\end{subequations}
Thus, in the case of the plane wave susceptibility, $\chi_{BA}^{\mathbf{G} \mathbf{G}'}(\mathbf{q}, z)$, the dissipative part is a spectral function for transitions between energy eigenstates with a difference in crystal momentum $\hbar \mathbf{q}$ and energy $\hbar \omega$, transitions which can be induced by $\hat{A}(-\mathbf{G}'-\mathbf{q})$, reversed by $\hat{B}(\mathbf{G}+\mathbf{q})$ and vice-versa. 

\subsection{Energy dissipation in periodic crystals}\label{sec:energy abs. in periodic crystals}
%
In section \ref{sec:susceptibility periodic crystals}, it was shown that the plane wave susceptibility, $\chi_{BA}^{\mathbf{G} \mathbf{G}'}(\mathbf{q}, z)$, gives the plane wave coefficients of the linear response in system coordinate $\hat{B}$ to a plane wave perturbation in coordinate $\hat{A}$. As in section \ref{sec:susceptibility}, the out-of-phase response to the full sinusoidal perturbation \eqref{eq:sinusoidal plane wave perturbation} is needed to compute the energy dissipation. Once again, with $\hat{A}^{\dagger}(\mathbf{r})=\hat{A}(\mathbf{r})$ and $\hat{B}^{\dagger}(\mathbf{r})=\hat{B}(\mathbf{r})$, Eq. \eqref{eq:relation daggers cc} may be used to rewrite
\begin{align}
    \chi_{BA}^{-\mathbf{G} -\mathbf{G}'}(-\mathbf{q}, -z^*) 
    &= 
    \frac{1}{\Omega}  
    \chi_{B(-\mathbf{G}-\mathbf{q})\,A(\mathbf{G}' + \mathbf{q})}(-z^*)
    \nonumber \\
    &= 
    \frac{1}{\Omega}  
    \chi_{B(\mathbf{G}+\mathbf{q})^{\dagger}\,A(-\mathbf{G}' - \mathbf{q})^{\dagger}}(-z^*)
    \nonumber \\
    &= 
    \frac{1}{\Omega}  
    \chi_{B(\mathbf{G}+\mathbf{q})\,A(-\mathbf{G}' - \mathbf{q})}^*(z)
    \nonumber \\
    &= \chi_{BA}^{\mathbf{G} \mathbf{G}' \,*}(\mathbf{q}, z).
\end{align}
With this result, insertion into Eq. \eqref{eq:plane wave real space response I} yields
\begin{align}
    \langle \delta \hat{B}(\mathbf{r}, t) \rangle = f_0
    \sum_{\mathbf{G}} \Big[ \mathrm{Re}\Big\{ \chi_{BA}^{\mathbf{G} \mathbf{G}_0}&(\mathbf{q}_0, \omega_0) \Big\}
    \nonumber \\
    \times &\cos([\mathbf{G}+\mathbf{q}_0]\cdot\mathbf{r} - \omega_0 t)
    \nonumber \\
    - \mathrm{Im}\left\{ \chi_{BA}^{\mathbf{G} \mathbf{G}_0}(\mathbf{q}_0, \omega_0) \right\}
    \sin([\mathbf{G}&+\mathbf{q}_0]\cdot\mathbf{r} - \omega_0 t)
    \Big].
    \label{eq:plane wave real space response II}
\end{align}
In full analogy with Eq. \eqref{eq:in and out of phase response}, the real and imaginary parts of the plane wave susceptibility gives the response in- and out-of-phase of the harmonic perturbation respectively.
Using Eqs. \eqref{eq:sinusoidal plane wave perturbation} and \eqref{eq:plane wave real space response II}, the mean rate of energy absorption may be computed:
\begin{equation}
    \bar{Q} = - \frac{1}{2} f_0^2 \omega_0 \, \Omega \, \mathrm{Im}\left\{ \chi_{AA}^{\mathbf{G}_0 \mathbf{G}_0}(\mathbf{q}_0, \omega_0) \right\}.
    \label{eq:mean energy absorption plane wave}
\end{equation}
Compared to Eq. \eqref{eq:mean energy dissipation}, the plane wave susceptibility has simply been normalized by the crystal volume, such as to make it a size intensive material property, whereas the dynamic susceptibility in Eq. \eqref{eq:mean energy dissipation} is a property of the quantum system $\hat{H}_0$. 

For a more general perturbation than that of a single plane wave component in Eq. \eqref{eq:sinusoidal plane wave perturbation}, the energy dissipation will be governed by the full plane wave spectrum of induced transitions
\begin{align}
    S_{BA}^{\mathbf{G}\mathbf{G}'}(\mathbf{q}, \omega)
    &\equiv 
    - \frac{1}{\Omega}\frac{\chi_{\beta\alpha}''(\omega)}{\pi}
    \label{eq:plane wave spectral function of induced transitions}
    \\
    &= - \frac{1}{2\pi i} \left\{\chi_{BA}^{\mathbf{G}\mathbf{G}'}(\mathbf{q}, \omega) - \chi_{AB}^{-\mathbf{G}'-\mathbf{G}}(-\mathbf{q}, -\omega) \right\}.
    \nonumber
\end{align}
From Eq. \eqref{eq:dissipative part = imaginary part} it follows that the imaginary part and the dissipative part of the plane wave susceptibility are the same along the diagonal. 
Then, using Eq. \eqref{eq:plane wave dissipative part}, a clear connection from the Kubo theory to the quasi-particle picture can be made:
\begin{align}
    S_{AA}^{\mathbf{G}\mathbf{G}}(\mathbf{q}, \omega) &=
    -\frac{\mathrm{Im}\left\{ \chi_{AA}^{\mathbf{G} \mathbf{G}}(\mathbf{q}, \omega) \right\}}{\pi} 
    \nonumber \\
    &= 
    \frac{1}{\Omega}  
    \sum_{\alpha, \alpha'} \left| a_{\alpha'\alpha}(-\mathbf{G})\right|^2 (n_{\alpha} - n_{\alpha'})
    \nonumber \\
    &\hspace{30pt}\times
    \delta_{\mathbf{q},\mathbf{q}_{\alpha'\alpha}} \delta\big(\hbar \omega - (E_{\alpha'} - E_{\alpha})\big).
    \label{eq:plane wave spectrum}
\end{align}
When various spectroscopic experiments are carried out, energy dissipation is a direct manifestation of transitions between the energy eigenstates of the system. Through Eqs. \eqref{eq:mean energy absorption plane wave} and \eqref{eq:plane wave spectrum}, the rate of energy absorption in a material at momentum transfer $\hbar \mathbf{q}$ and transition energy $\hbar \omega$ is proportional to the spectral density of eigenstate transitions associated with quasi-particles of crystal momentum $\hbar \mathbf{q}_{\alpha'\alpha}=\hbar \mathbf{q}$ and energy $E_{\alpha'} - E_{\alpha} = \hbar \omega$. The spectrum is weighted by the periodic part of the transition matrix elements associated to the spectroscopic technique in question.

\section{Fourier transforms}
\subsection{Temporal Fourier transform}\label{sec:app temp FT definition}
We use the following definition for the temporal Fourier-Laplace transform to complex frequencies:
\begin{equation}
    \chi_{BA}(z) = \int_{-\infty}^{\infty} dt \, \chi_{BA}(t) e^{i z t}.
\end{equation}
For retarded susceptibilities, $\chi_{BA}(z)$ is analytic in the upper half complex plane and has the inverse transform
\begin{equation}
    \chi_{BA}(t) = \lim_{\eta \rightarrow 0^+} \int_{-\infty}^{\infty} \frac{d\omega}{2\pi} \, \chi_{BA}(\omega + i\eta) e^{- i \omega t}.
\end{equation}

\subsection{Spatial Fourier transform}\label{sec:app double spatial FT}
For the spatial Fourier transform, the following definition is used:
\begin{equation}
    f(\mathbf{Q}) = 
    \int d\mathbf{r} \, f(\mathbf{r}) e^{-i\mathbf{Q}\cdot \mathbf{r}}. 
\end{equation}
%
For two-point functions, $f(\mathbf{r}, \mathbf{r}')$, the spatial Fourier transform is generalized as
\begin{equation}
    f(\mathbf{Q}, \mathbf{Q}') = \frac{1}{\Omega} \iint d\mathbf{r} d\mathbf{r}' \, e^{-i\mathbf{Q}\cdot \mathbf{r}} f(\mathbf{r}, \mathbf{r}') e^{i\mathbf{Q}'\cdot \mathbf{r}'},  
\end{equation}
where $\Omega$ is the crystal volume.

Considering a crystal with Bravais lattice points $\mathbf{R}$ and unit cell volume $\Omega_{\mathrm{cell}}$, we may change the integration variables:
\begin{equation}
    \iint d\mathbf{r} d\mathbf{r}' \, g(\mathbf{r}, \mathbf{r}') = \sum_{\mathbf{R}}\int_{\Omega_{\mathrm{cell}}} d\mathbf{r} \int d\mathbf{r}' \, g(\mathbf{r} + \mathbf{R}, \mathbf{r}' + \mathbf{R}).
\end{equation}
Now, take $\mathbf{G}$, $\mathbf{G}'$ to be reciprocal lattice vectors and $\mathbf{q}$, $\mathbf{q}'$ to be a wave vectors within the first Brillouin zone, then the Fourier transform of a periodic two-point function $f(\mathbf{r} + \mathbf{R}, \mathbf{r}' + \mathbf{R}) = f(\mathbf{r}, \mathbf{r}')$ reduces:
\begin{widetext}
\begin{align}
    f(\mathbf{G} + \mathbf{q}, \mathbf{G}' + \mathbf{q}) 
    &= \frac{1}{\Omega} \sum_{\mathbf{R}} \int_{\Omega_{\mathrm{cell}}} d\mathbf{r} \int d\mathbf{r}' \, 
    e^{-i(\mathbf{G} + \mathbf{q}) \cdot \mathbf{r}} e^{-i\mathbf{q}\cdot \mathbf{R}} f(\mathbf{r} + \mathbf{R}, \mathbf{r}' + \mathbf{R}) e^{i(\mathbf{G}' + \mathbf{q}) \cdot \mathbf{r}'} e^{i\mathbf{q}\cdot \mathbf{R}}
    \nonumber \\
    &= \frac{1}{\Omega_{\mathrm{cell}}} \int_{\Omega_{\mathrm{cell}}} d\mathbf{r} \int d\mathbf{r}' \, 
    e^{-i(\mathbf{G} + \mathbf{q}) \cdot \mathbf{r}} f(\mathbf{r}, \mathbf{r}') e^{i(\mathbf{G}' + \mathbf{q}) \cdot \mathbf{r}'},
\end{align}
\begin{align}
    f(\mathbf{G} + \mathbf{q}, \mathbf{G}' + \mathbf{q}')
    &= \frac{1}{\Omega} \sum_{\mathbf{R}} \int_{\Omega_{\mathrm{cell}}} d\mathbf{r} \int d\mathbf{r}' \, 
    e^{-i(\mathbf{G} + \mathbf{q}) \cdot \mathbf{r}} f(\mathbf{r} + \mathbf{R}, \mathbf{r}' + \mathbf{R}) e^{i(\mathbf{G}' + \mathbf{q}') \cdot \mathbf{r}'} e^{-i(\mathbf{q} - \mathbf{q}')\cdot \mathbf{R}} 
    \nonumber \\
    &= \frac{1}{\Omega_{\mathrm{cell}}} \int_{\Omega_{\mathrm{cell}}} d\mathbf{r} \int d\mathbf{r}' \, 
    e^{-i(\mathbf{G} + \mathbf{q}) \cdot \mathbf{r}} f(\mathbf{r}, \mathbf{r}') e^{i(\mathbf{G}' + \mathbf{q}') \cdot \mathbf{r}'} \frac{\Omega_{\mathrm{cell}}}{\Omega} \sum_{\mathbf{R}} e^{-i(\mathbf{q} - \mathbf{q}')\cdot \mathbf{R}}
    \nonumber \\
    &= f(\mathbf{G} + \mathbf{q}, \mathbf{G}' + \mathbf{q}) \frac{(2\pi)^D}{\Omega} \delta(\mathbf{q} - \mathbf{q}').
\end{align}
\end{widetext}
For the periodic two-point functions, the notation $f_{\mathbf{G}\mathbf{G}'}(\mathbf{q}) \equiv f(\mathbf{G}+\mathbf{q}, \mathbf{G}'+\mathbf{q})$ is introduced and referred to as the lattice Fourier transform.

\bibliography{bibliography}

\end{document}